\documentclass[aps,prl,superscriptaddress,twocolumn,nofootinbib]{revtex4-2}

\usepackage[colorlinks=true,citecolor=green,urlcolor=blue]{hyperref}

\usepackage{amssymb}
\usepackage{amsmath,color}
\usepackage{enumitem}
\usepackage{graphicx}
\usepackage{bbm}
\usepackage{subfigure}
\usepackage{verbatim}
\usepackage[T1]{fontenc}
\usepackage[utf8]{inputenc}
\usepackage[normalem]{ulem}
\usepackage{longtable}

\usepackage{url}
\usepackage{orcidlink}

\usepackage{xcolor}
\usepackage{graphicx}
\usepackage{dcolumn}
\usepackage{bm}

\usepackage{newunicodechar}
\newunicodechar{−}{-}

\makeatletter
\renewcommand{\section}[1]{%
  \par\addvspace{0.5ex}%
  \noindent\textit{#1.}\hspace{0.5em}%
}
\makeatother

\begin{document}

\title{Breaking the Dark Sector Degeneracy with Nonparametric Expansion--Growth Reconstruction}

\author{Changyu You\orcidlink{0009-0004-0791-3272}}
\affiliation{School of Physics and Technology, Wuhan University, Wuhan 430072, China}  

\author{Rong-Gen Cai\orcidlink{0000-0002-3539-7103}}
\affiliation{Institute of Fundamental Physics and Quantum Technology, Ningbo University, Ningbo 315211, China}

\author{Tao Yang\orcidlink{0000-0002-2161-0495}}
\email[Corresponding author ]{yangtao@whu.edu.cn}
\affiliation{School of Physics and Technology, Wuhan University, Wuhan 430072, China}

\date{\today}

\begin{abstract}
The dark energy equation of state (EoS) and a possible dark-sector interaction are degenerate at the level of background expansion: the same expansion history may be interpreted as time-varying dark energy, energy exchange with dark matter, or a mixture of both. We introduce a data-driven, nonparametric expansion--growth framework that breaks this degeneracy by simultaneously reconstructing the dark energy EoS and the dark-sector coupling. Allowing the interacting matter density to evolve freely, we incorporate the coupling into the linear matter perturbation equation and reconstruct the expansion and growth histories using Gaussian processes, with hyperparameters marginalized in a Bayesian treatment. Applying this method to Pantheon+, cosmic chronometers, BAO measurements including DESI DR2, and redshift-space-distortion data, we infer both $w_{\rm de}(z)$ and the interaction history over $0\lesssim z\lesssim 2$, without assuming either a parametric EoS or a prescribed interaction form. We find no statistically significant evidence for either a nonzero interaction or dark energy dynamics: the reconstructed coupling and dark energy EoS remain consistent with the $\Lambda$CDM limit. Our results establish an expansion--growth consistency test for coupled dark-sector physics and provide a model-independent route to distinguish genuine dark energy dynamics from effective dark-sector energy exchange.
\end{abstract}

\maketitle

\section{Introduction} \label{sec:intro}
The $\Lambda$CDM model, in which cosmic acceleration is driven by a cosmological constant, provides the simplest and most successful baseline for interpreting observations of the late Universe. It consistently describes a broad range of data, from Type Ia supernovae and baryon acoustic oscillations (BAO) to cosmic microwave background (CMB) anisotropies and large-scale structure~\citep{SupernovaSearchTeam:1998fmf,SupernovaCosmologyProject:1998vns,SDSS:2003eyi,SDSS:2005xqv,WMAP:2012nax,Chuang:2013hya,Planck:2018vyg}. Nevertheless, the physical origin of the cosmological constant remains unknown, and the model faces both theoretical puzzles, such as fine tuning and cosmic coincidence~\citep{Zlatev:1998tr,Sahni:1999gb,Weinberg:2000yb}, and observational tensions, most notably the Hubble tension~\citep{Planck:2018vyg,Riess:2021jrx}. These issues have motivated a broad class of alternatives, including dynamical dark energy (DDE)~\citep{Copeland:2006wr,Tsujikawa:2013fta,Armendariz-Picon:2000ulo,Linder:2007wa,Saridakis:2010mf}, modified gravity~\citep{Kunz:2006ca,Clifton:2011jh,Hu:2007nk}, and interacting dark energy (IDE)~\citep{Amendola:1999er,Shahalam:2015sja,Pourtsidou:2013nha,Costa:2016tpb,Giare:2024smz}.

The question has become especially timely after the recent BAO measurements from the Dark Energy Spectroscopic Instrument (DESI)~\citep{Dinda:2024kjf,Dinda:2024ktd,DESI:2024aqx,Keeley:2025stf,Cortes:2024lgw}. Within the Chevallier--Polarski--Linder (CPL) parametrization, DESI analyses report a preference for dynamical dark energy over $\Lambda$CDM when combined with external datasets~\citep{DESI:2024mwx,DESI:2025zgx,DESI:2025wyn}. This has triggered extensive studies using alternative parametrizations, binning approaches, and Gaussian-process reconstructions to test whether the EoS of dark energy evolves with time~\citep{Giare:2024gpk,DESI:2025fii,Dinda:2024ktd}. However, an apparent deviation from $\Lambda$CDM in the background expansion alone does not identify the underlying physics. The same $H(z)$ can be reproduced by a time-dependent EoS, by an interaction between dark energy and dark matter, or by a combination of both.

This ambiguity is a manifestation of the dark sector degeneracy: gravitational observations of the homogeneous background constrain the total dark-sector evolution, but do not uniquely determine how that total is split into dark matter and dark energy~\citep{Kunz:2007rk,vonMarttens:2019ixw}. In particular, if only background observables are used, the dark energy EoS $w_{\rm de}(z)$ and the interaction history $Q(z)$ enter the Friedmann and continuity equations in a coupled way~\citep{Yang:2015tzc,Yang:2020jze,You:2025uon}. Therefore, reconstructing one of them generally requires assuming the other. Existing studies often fix $w_{\rm de}=-1$, adopt a low-dimensional parametrization such as CPL, or prescribe a specific interaction law before constraining the dark-sector coupling~\citep{Yang:2015tzc,Wang:2015wga,Abedin:2025yru,Yang:2020jze,You:2025uon,Mukherjee:2021ggf,Bonilla:2021dql}. Such assumptions are well motivated for testing specific models, but they do not answer the more basic question of whether a deviation from $\Lambda$CDM originates from dark energy dynamics, dark-sector energy exchange, or both.

The problem is broader than phenomenological interacting dark energy. In scalar--tensor theories, a nonminimal coupling between a scalar field and curvature can be moved by a conformal transformation from a Jordan-frame description to an Einstein-frame description, where the scalar is minimally coupled to curvature but matter fields acquire an effective scalar-dependent coupling. In this sense, a theory that appears as modified gravity in one frame can appear as an effective scalar--matter, and potentially dark-sector interaction in another~\citep{Pettorino:2008ez,Wetterich:2014bma,Postma:2014vaa}. This connection has received renewed attention in the DESI era, where nonminimally coupled scalar fields and broader scalar--tensor modified-gravity models have been investigated as possible physical explanations of evolving or phantom-crossing dark energy~\citep{Chudaykin:2024gol,Ye:2024ywg,Wolf:2024stt,Wolf:2025jed,Ishak:2024jhs,Postolak:2025qmv}. A method that reconstructs both $w_{\rm de}(z)$ and $Q(z)$ without imposing either function therefore provides a direct diagnostic not only of the nature of dark energy, but also of possible effective departures from general relativity.

In this Letter, we present, to our knowledge, the first data-driven and nonparametric expansion--growth reconstruction that simultaneously constrains the dark energy EoS and the DE--DM interaction without assuming a functional form for either. The key idea is to promote the comoving matter density to a free function and to include the corresponding interaction terms in the linear matter perturbation equation. Expansion data determine the background geometry, while redshift-space-distortion (RSD) measurements provide independent information on the growth history. The combination breaks the background-level EoS--coupling degeneracy and turns expansion--growth consistency into a null test of coupled dark-sector physics.

Applying this framework to Pantheon+, cosmic chronometers, BAO data including DESI DR2, and RSD measurements, We find no statistically significant evidence for either a nonzero interaction or dark energy dynamics: the reconstructed coupling and dark energy EoS remain consistent with the $\Lambda$CDM limit. However, the reconstructed EoS exhibits more dataset-dependent behavior than the dark sector interaction. Thus, current data suggest that the expansion-sector deviations are more readily absorbed by dark energy dynamics than by a robust interaction signal in the growth sector, although improved growth measurements will be essential for a decisive test.

\section{Method and data}\label{sec:datamethod}
We consider a spatially flat Friedmann--Lema\^{i}tre--Robertson--Walker (FLRW) Universe in which dark energy may be dynamical, $w_{\rm de}\neq -1$, and may exchange energy with dark matter. Neglecting radiation at late times, the Friedmann and continuity equations are
\begin{eqnarray}
    \bar{\rho}_{\rm de}+\bar{\rho}_{\rm dm}+\bar{\rho}_b&=&3H^2\,, \label{eq:Friedmann}   \\
    \dot{\bar{\rho}}_{\rm dm}+3H\bar{\rho}_{\rm dm}&=&\frac{\dot\rho_f}{a^3} \, ,\label{eq:rhodm}  \\
    \dot{\bar{\rho}}_b+3H\bar{\rho}_b&=&0 \, ,\label{eq:rhob}\\ 
    \dot{\bar{\rho}}_{\rm de}+3H(1+w_{\rm de})\bar{\rho}_{\rm de}&=&-\frac{\dot\rho_f}{a^3}\, .\label{eq:rhode}
\end{eqnarray}
Here $a$ is the scale factor, $H=\dot a/a$ is the Hubble parameter, and dots denote derivatives with respect to cosmic time. We use natural units with $M_{\rm pl}=1$.

The quantity
\begin{equation}
    \bar{\rho}_m=\bar{\rho}_{\rm dm}+\bar{\rho}_b=\frac{\rho_f}{a^3} \label{eq:rhof}
\end{equation}
defines an effective comoving density for the total pressureless matter sector. Since baryons are separately conserved by Eq.~(\ref{eq:rhob}), any redshift variation of $\rho_f$ is sourced by the DE--DM energy exchange. We define
\begin{equation}
    \bar{Q}\equiv \frac{\dot\rho_f}{a^3}\, ,\label{eq:Q}
\end{equation}
so that $\bar{Q}>0$ corresponds to energy transfer from dark energy to dark matter. In the absence of coupling, $\rho_f=3H_0^2\Omega_m$ is constant and $\dot\rho_f=0$. 

At the background level, the EoS is related to the expansion history and the matter evolution through
\begin{equation}
    w_{\rm de}=\frac{-2\dot H-3H^2}{3H^2-\rho_f/a^3} \ . \label{eq:w_de}
\end{equation}
For a fixed $H(z)$, different choices of $\rho_f(z)$ can be exactly compensated by different choices of $w_{\rm de}(z)$. Hence, expansion data alone cannot determine whether a deviation from $\Lambda$CDM originates from dark energy dynamics or from dark-sector coupling. Previous reconstruction studies have therefore typically assumed either the EoS or the interaction in order to infer the other. Our goal is to break this degeneracy by adding the independent growth information encoded in RSD measurements.

In interacting dark-sector scenarios, energy and momentum transfer modify the evolution of density perturbations. The matter perturbation equation has been derived for a variety of interaction models~\citep{SolaPeracaula:2017esw,Sabogal:2024yha,Li:2014cee,Wei:2008vw}. Here, we adopt a general form of the modified equation for the coupled dark sector, following Refs.~\citep{Marcondes:2016reb,Pooya:2025wyd}. On sub-horizon scales, dark energy perturbations are expected to be subdominant for the class of models considered here~\citep{Marcondes:2016reb,Duniya:2013eta,Sabogal:2024yha}. We therefore neglect $\delta_{\rm de}$ relative to $\delta_m$. We assume a vanishing spatial momentum transfer in the dark matter rest frame and an initial condition with no primordial velocity bias between dark matter and baryons, which is required when integrating our model with Redshift-Space Distortion (RSD) measurements~\citep{Motta:2013cwa}.
Under these assumptions, we derive the modified matter perturbation equation (see details in Supplemental Material Section~\ref{sec:modified equation}), which takes the form:
\begin{eqnarray}
    \ddot{\delta}_m &+& A(z)\dot{\delta}_m = B(z)\delta_m \, , \label{eq:delta_mod}\\
     A(z) &=& 2H + \Xi \, , \label{eq:A}\\ 
     B(z) &=& \frac{1}{2}\frac{\rho_f}{a^3} - \dot{\Xi} - 2H\Xi \, , \label{eq:B} 
\end{eqnarray}

where $\Xi = \bar{Q}/\bar{\rho}_m - \delta Q/(\bar{\rho}_m \delta_m)$ denotes the effective perturbative interaction parameter. Here, $\bar{Q}$ and $\delta Q$ represent the background and perturbative components of the energy transfer scalar $Q = \bar{Q} + \delta Q$, respectively.

In this work, we investigate two distinct physical closures for the dark sector interaction, each leading to different observable consequences:
\begin{itemize}
    \item \textbf{Closure I:} The energy transfer rate is proportional to the local dark matter density ($Q \propto \rho_{\rm dm}$). Assuming the density perturbations of dark matter and baryons evolve synchronously ($\delta_{dm}\sim\delta_b\sim\delta_m$), this leads to $\Xi \equiv 0$.
    \item \textbf{Closure II:} The energy transfer rate is proportional to the smooth and homogeneous dark energy density ($Q \propto \rho_{\rm de}$). Assuming dark energy does not cluster, this leads to $\delta Q \approx 0$ and $\Xi = \bar{Q}/\bar{\rho}_m$.
\end{itemize}

Our reconstruction proceeds in three steps. First, we reconstruct $H(z)$ and the matter growth history parameter $f\sigma_8(z)$, together with their derivatives, from the expansion and RSD datasets using Gaussian processes. Second, assuming a specific physical closure, we analytically isolate the effective density parameter $\rho_f$ from Eqs.~\eqref{eq:delta_mod}--\eqref{eq:B} to infer its posterior distribution and then reconstruct the coupling strength $Q$ utilizing Eq.~(\ref{eq:Q}). Finally, we substitute the reconstructed $\rho_f$ into Eq.~\eqref{eq:w_de} to derive the dynamic dark energy equation of state $w_{\rm de}(z)$. Throughout this procedure, all time derivatives are converted to redshift derivatives via $\frac{\mathrm{d}}{\mathrm{d}t} = -H(1+z)\frac{\mathrm{d}}{\mathrm{d}z}$.

Our baseline datasets are as follows:
\begin{itemize}
    \item \textbf{SNe Ia: Pantheon Plus (PP).}
    The Pantheon+ sample contains 1701 light curves from 1550 distinct Type Ia supernovae in the redshift range $0.001\le z \le 2.26$~\citep{Scolnic:2021amr}. To reduce the impact of peculiar velocities at low redshift, we use the 1590 objects with $z>0.01$~\citep{Brout:2022vxf}\footnote{\url{https://github.com/PantheonPlusSH0ES/DataRelease}}. We adopt $M_B=-19.42$ from Table I of Ref.~\citep{Matthewson:2024ffb}.

    \item \textbf{Cosmic chronometers (CC).}
    We use 31 $H(z)$ measurements obtained from the differential-age technique~\citep{Gomez-Valent:2018hwc}.

    \item \textbf{Baryon acoustic oscillations (BAO).}
    We adopt BAO measurements from SDSS, WiggleZ and DESI, summarized in Table III of Ref.~\citep{Yang:2024kdo}. But we replace the DESI DR1 BAO with the recent DESI DR2 BAO data~\citep{DESI:2025zgx}. We use the sound horizon prior $r_d=(147.09\pm0.26)\,{\rm Mpc}$ from Planck 2018~\citep{Planck:2018vyg}.

    \item \textbf{Redshift-space distortions (RSD).}
    We use 20 measurements of $f\sigma_8(z)$ over $0.02<z<1.944$, filtered and summarized in Table 2 of Ref.~\citep{Avila:2022xad}.
\end{itemize}



\section{Reconstructions and results} \label{sec:rec}
In this Letter, we use a nonparametric and data-driven method, namely Gaussian Process(GP), known for its notable ability to predict the evolution of derived cosmological parameters \citep{Seikel:2012uu,Shafieloo:2012ht}. GP provides a framework for reconstructing a function and its derivatives from observational data, whether the data comprise only the function values or are combined with its first-order derivatives, enabling a joint analysis of different datasets \citep{Yang:2025kgc}. From a Bayesian point of view, instead of the optimization method, we adopt in this work a more rigorous approach that marginalizes over the hyperparameter using the MCMC package emcee \citep{Seikel:2013fda,Foreman-Mackey:2012any}. 

In this methodology, background data and structure growth data serve as independent data sources, used respectively to reconstruct $H(z)$ and $f\sigma_8(z)$ via Gaussian Processes, along with their derivatives. However, using these reconstructed quantities to constrain interacting dark sector models requires caution, as the introduction of an interaction parameter  $\Xi(t)$ modifies the matter continuity equation to $\dot{\delta}_m + \theta_m = -\Xi \delta_m$ in the linear regime. Consequently, the standard relation between the velocity divergence and the density contrast is shifted. To interpret redshift-space distortion (\textbf{RSD}) measurements correctly within this framework, we define an effective growth rate:
\begin{equation}
    f_{\text{eff}} \equiv -\frac{\theta_m}{H\delta_m} = \frac{\dot{\delta}_m}{H\delta_m} + \frac{\Xi}{H} \, . \label{eq:f_eff}
\end{equation}
Therefore, depending on the chosen physical closure, the quantity $f\sigma_8(z)$ reconstructed via Gaussian processes carries distinct physical interpretations.

Under \textbf{Closure I}, as detailed in Supplemental Material Section~\ref{sec:modified equation}, both $f\sigma_8(z)$ and the matter density perturbation equation revert to their standard definitions. The energy transfer between the dark sector, $\bar{Q}$, can then be analytically expressed as:
\begin{equation}
\begin{split}
\bar{Q} &=\frac{\dot{\rho}_f}{a^3} 
= 2 \left( \frac{\dddot{\mathcal{D}}+5H\ddot{\mathcal{D}}+2\dot{H}\dot{\mathcal{D}}+6H^2\dot{\mathcal{D}}}{\mathcal{D}} \right. \\
&\quad \left. - \frac{\dot{\mathcal{D}}\ddot{\mathcal{D}}+2H(\dot{\mathcal{D}})^2}{\mathcal{D}^2} \right)
\end{split}
\end{equation}
where $\mathcal{D}(z) \equiv \delta_m(z)/\delta_{m,0}$ denotes the normalized linear growth factor, with $\mathcal{D}(0)=1$.
In this scenario, we treat $\sigma_{8,0}$ as a free parameter governed by the Planck 2018 prior $\sigma_{8,0} = 0.8111 \pm 0.0060$~\citep{Planck:2018vyg}, which enable us to extract $\mathcal{D}(z)$ from $f\sigma_8(z)$ measurements . 

\begin{figure*}
\includegraphics[width=1\textwidth]{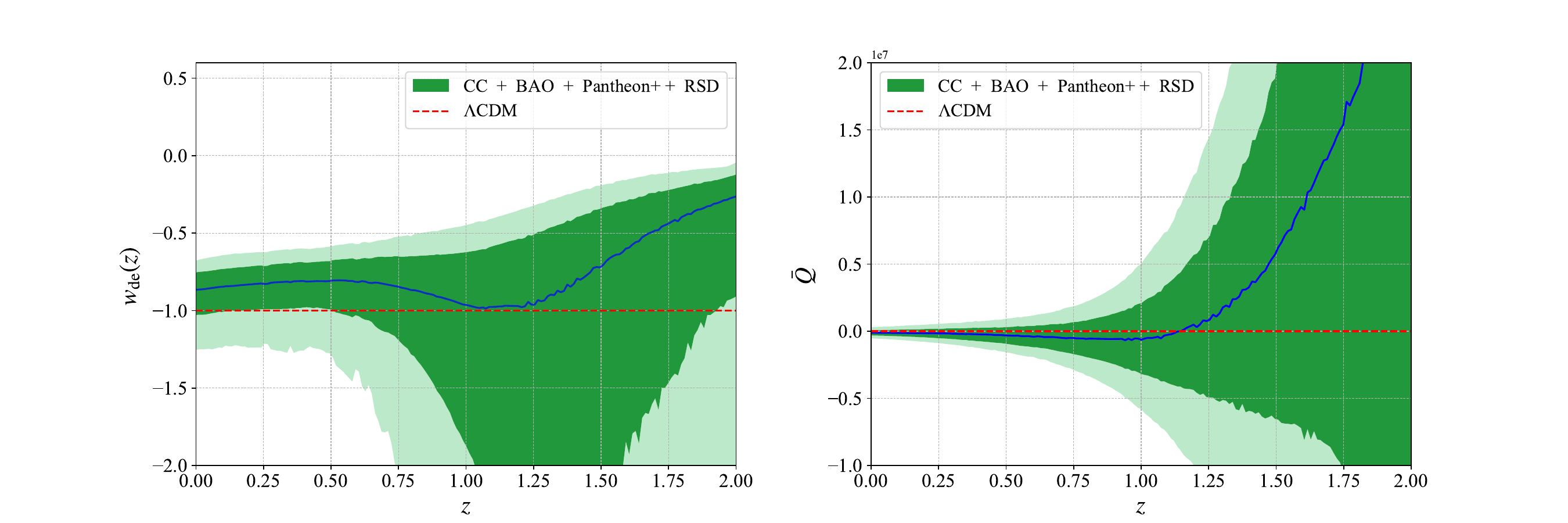}
\caption{Reconstructions of the dark energy equation of state $w_{\rm de}(z)$ and the dark-sector interaction $\bar{Q}$ using CC + BAO + Pantheon+ + RSD measurements under \textbf{Closure I}. Dark and light shaded regions denote the $68\%$ and $95\%$ credible intervals, respectively. Dashed lines represent the $\Lambda$CDM limit.}
\label{fig:single}
\end{figure*}

Fig.~\ref{fig:single} shows the reconstructed dark energy equation of state $w_{\rm de}(z)$ and the dark sector coupling strength $\bar{Q}$ derived from the joint analysis of \textbf{CC}, \textbf{BAO}, \textbf{PP} and \textbf{RSD} datasets. 
We find that both the reconstructed coupling and the EoS are consistent with the $\Lambda$CDM limit within $1\sigma$ uncertainties. 
Therefore, the current data yield no statistically significant evidence for either a nonzero dark-sector interaction or dark energy dynamics in this closure. 
The results for additional combinations of data sets are shown in the Supplemental Material Section~\ref{sec:supp_more_data}.

Under \textbf{Closure II}, where $\Xi(t) = \bar{Q}/\bar{\rho}_m \equiv \Gamma(t)$, the effective growth rate becomes $f_{\rm eff} = \dot{\delta}_m/H\delta_m + \Gamma/H$. To correctly match the RSD measurements, we define the primary observable $L(z) \equiv f_{\rm eff}\sigma_8(z)$. After algebraic manipulations, we can finally express the energy transfer between dark sector $Q$ as:
\begin{equation}
\begin{split}
Q = \bar{Q}=\frac{\dot{\rho}_f}{a^3} 
  &= \frac{2}{L^2H^2} \left[ \dot{M}LH \right. \\
  &\quad \left. +  M(3H^2L - \dot{L}H - L\dot{H}) \right] . \label{eq:Q_closure_two}
\end{split}
\end{equation}
where $K \equiv \dot{L}H + L\dot{H} + 2H^2L$ and $M\equiv \dot K+3HK$.

It is worth noting that under the assumption of \textbf{Closure II} (i.e., the dark sector coupling strength is proportional to the smooth dark energy density), the final reconstructed quantities are highly dependent on the second and higher-order derivatives of the observational parameters, according to Eq.~(\ref{eq:Q_closure_two}). 
Constrained by the statistical precision of current observational data, it is challenging to obtain physical constraints with high statistical significance by directly utilizing real data in this scenario. 

Therefore, to verify the analytical correctness of our methodology under this complex mechanism, we perform a null test using the mock data sets.
Specifically, we generate mock $H(z)$ and self-consistent $f\sigma_8(z)$ data using Planck 2018 $\Lambda$CDM best-fit cosmology as the baseline, with redshifts satisfying a Gamma-distribution peaking at $z\sim0.4$ and a 2$\%$ relative uncertainty matching recent DESI measurement precision.
The results are shown in Fig.~\ref{fig:closure_two_mock} (see Supplemental Material Section~\ref{sec:null_test_closure_two} for more details).
As anticipated, the reconstructions are in excellent agreement with the $\Lambda$CDM predictions at low redshifts, remaining well within the Gaussian uncertainty bands, which proves the correctness of this framework. We observe that at higher redshifts, minor deviations from the $\Lambda$CDM baseline begin to emerge. 
This behavior occurs because both $w_{\rm de}$ and $\bar{Q}$ are highly non-linear functions of the reconstructed observables. These numerical artifacts are further amplified at high redshifts as a direct consequence of the Gamma distribution adopted for our mock data sampling and the base Gaussian Process algorithm.

\begin{figure*}
\includegraphics[width=1\textwidth]{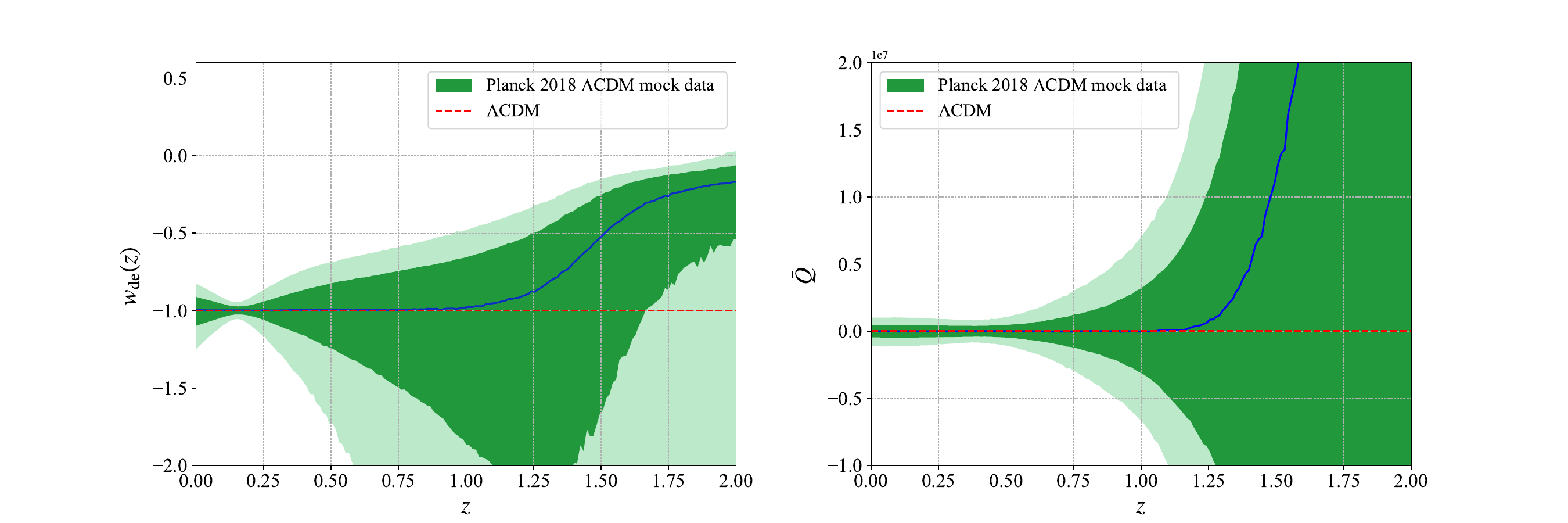}
\caption{Reconstruction of the dark energy equation of state and the dark-sector interaction from the mock data under \textbf{Closure II}. Dark and light shaded bands denote the $68\%$ and $95\%$ credible intervals, respectively. Dashed lines represent the $\Lambda$CDM limit.}
\label{fig:closure_two_mock}
\end{figure*}


\section{Conclusions and Discussions}
We have introduced a data-driven, nonparametric expansion--growth framework that simultaneously reconstructs dark energy dynamics and possible DE--DM interactions. The central result is methodological: by combining background expansion data with growth-of-structure measurements, we break the background-level degeneracy between $w_{\rm de}(z)$ and $Q(z)$ without assuming a functional form for either. To our knowledge, this is the first nonparametric reconstruction framework that treats the dark energy EoS and the dark-sector coupling as two independent unknown functions and infers them jointly from data.

Applying this framework to \textbf{Closure I} with Pantheon+, BAO, CC, combined with RSD measurements, we find no statistically significant evidence for either a nonzero interaction or dark energy dynamics: the reconstructed coupling and dark energy EoS remain consistent within the $\Lambda$CDM limit. However, the reconstructed EoS exhibits more dataset-dependent behavior than the dark sector interaction (See Supplemental Material Section \ref{sec:supp_more_data}).  Current data therefore favor the interpretation that any apparent low-redshift deviation from $\Lambda$CDM is more strongly associated with the background dark energy dynamics than with a confirmed dark-sector energy transfer.

Several methodological points are important for interpreting these results. First, our GP reconstruction marginalizes over kernel hyperparameters rather than fixing them by log-marginal-likelihood optimization.
This is especially relevant for large datasets such as Pantheon+, where multiple likelihood maxima may appear~\citep{Seikel:2013fda}. Second, the choice of mean function in GP reconstruction is also important and should be made carefully.
Following \citet{Hwang:2022hla}, we select $\Lambda$CDM model ($H_0 = 70~\mathrm{km\,s^{-1}\,Mpc^{-1}}$ and $\Omega_m = 0.31$) as our mean function, thereby avoiding nonphysical situations that can arise when a zero mean function is used. 
Third, the priors on $M_B$, $r_d$, and $\sigma_{8,0}$ define the calibration and null-test baseline of the reconstruction (with their individual impacts quantified in the Supplemental Material). Moreover, we emphasize that this data-driven paradigm is inherently flexible, readily accommodating alternative nonparametric techniques--such as Artificial Neural Networks (ANN)~\citep{Wang:2019vxv,Dialektopoulos:2021wde,Abedin:2025yru} and Symbolic Regression (SR)~\citep{Sousa-Neto:2025gpj,Koksbang:2026wvh}--which may offer distinct complementary advantages over standard GP.

While our results show broad consistency with overall trends, we find persistent mild deviations--most notably in the dark energy equation of state, which exhibits greater sensitivity to the background data. 
This indicates a residual tension between current observations and the standard model. 
Future surveys will provide substantially improved expansion and growth measurements, along with tighter control of calibration systematics. 
With such data, the framework developed here can become a decisive expansion--growth consistency test: it can determine whether deviations from $\Lambda$CDM arise from genuine dark energy dynamics, dark-sector coupling, modified gravity effects, or a combination thereof.

\section{Acknowledgements}
T.Y. is supported by the National Natural Science Foundation of China Grant No. 12575063. 
R.-G.C. is supported by the National Natural Science Foundation of China Grant No. 12588101.
C.Y. is supported by the ``Luojia Undergraduate Innovation Research Fund of Wuhan University''. 
The numerical calculations in this paper have been done on the supercomputing system in the Supercomputing Center of Wuhan University.

\bibliography{ref}

\begin{thebibliography}{81}%
\makeatletter
\providecommand \@ifxundefined [1]{%
 \@ifx{#1\undefined}
}%
\providecommand \@ifnum [1]{%
 \ifnum #1\expandafter \@firstoftwo
 \else \expandafter \@secondoftwo
 \fi
}%
\providecommand \@ifx [1]{%
 \ifx #1\expandafter \@firstoftwo
 \else \expandafter \@secondoftwo
 \fi
}%
\providecommand \natexlab [1]{#1}%
\providecommand \enquote  [1]{``#1''}%
\providecommand \bibnamefont  [1]{#1}%
\providecommand \bibfnamefont [1]{#1}%
\providecommand \citenamefont [1]{#1}%
\providecommand \href@noop [0]{\@secondoftwo}%
\providecommand \href [0]{\begingroup \@sanitize@url \@href}%
\providecommand \@href[1]{\@@startlink{#1}\@@href}%
\providecommand \@@href[1]{\endgroup#1\@@endlink}%
\providecommand \@sanitize@url [0]{\catcode `\\12\catcode `\$12\catcode `\&12\catcode `\#12\catcode `\^12\catcode `\_12\catcode `\%12\relax}%
\providecommand \@@startlink[1]{}%
\providecommand \@@endlink[0]{}%
\providecommand \url  [0]{\begingroup\@sanitize@url \@url }%
\providecommand \@url [1]{\endgroup\@href {#1}{\urlprefix }}%
\providecommand \urlprefix  [0]{URL }%
\providecommand \Eprint [0]{\href }%
\providecommand \doibase [0]{https://doi.org/}%
\providecommand \selectlanguage [0]{\@gobble}%
\providecommand \bibinfo  [0]{\@secondoftwo}%
\providecommand \bibfield  [0]{\@secondoftwo}%
\providecommand \translation [1]{[#1]}%
\providecommand \BibitemOpen [0]{}%
\providecommand \bibitemStop [0]{}%
\providecommand \bibitemNoStop [0]{.\EOS\space}%
\providecommand \EOS [0]{\spacefactor3000\relax}%
\providecommand \BibitemShut  [1]{\csname bibitem#1\endcsname}%
\let\auto@bib@innerbib\@empty
\bibitem [{\citenamefont {Riess}\ \emph {et~al.}(1998)\citenamefont {Riess} \emph {et~al.}}]{SupernovaSearchTeam:1998fmf}%
  \BibitemOpen
  \bibfield  {author} {\bibinfo {author} {\bibfnamefont {A.~G.}\ \bibnamefont {Riess}} \emph {et~al.} (\bibinfo {collaboration} {Supernova Search Team}),\ }\bibfield  {title} {\bibinfo {title} {{Observational evidence from supernovae for an accelerating universe and a cosmological constant}},\ }\href {https://doi.org/10.1086/300499} {\bibfield  {journal} {\bibinfo  {journal} {Astron. J.}\ }\textbf {\bibinfo {volume} {116}},\ \bibinfo {pages} {1009} (\bibinfo {year} {1998})},\ \Eprint {https://arxiv.org/abs/astro-ph/9805201} {arXiv:astro-ph/9805201} \BibitemShut {NoStop}%
\bibitem [{\citenamefont {Perlmutter}\ \emph {et~al.}(1999)\citenamefont {Perlmutter} \emph {et~al.}}]{SupernovaCosmologyProject:1998vns}%
  \BibitemOpen
  \bibfield  {author} {\bibinfo {author} {\bibfnamefont {S.}~\bibnamefont {Perlmutter}} \emph {et~al.} (\bibinfo {collaboration} {Supernova Cosmology Project}),\ }\bibfield  {title} {\bibinfo {title} {{Measurements of $\Omega$ and $\Lambda$ from 42 High Redshift Supernovae}},\ }\href {https://doi.org/10.1086/307221} {\bibfield  {journal} {\bibinfo  {journal} {Astrophys. J.}\ }\textbf {\bibinfo {volume} {517}},\ \bibinfo {pages} {565} (\bibinfo {year} {1999})},\ \Eprint {https://arxiv.org/abs/astro-ph/9812133} {arXiv:astro-ph/9812133} \BibitemShut {NoStop}%
\bibitem [{\citenamefont {Tegmark}\ \emph {et~al.}(2004)\citenamefont {Tegmark} \emph {et~al.}}]{SDSS:2003eyi}%
  \BibitemOpen
  \bibfield  {author} {\bibinfo {author} {\bibfnamefont {M.}~\bibnamefont {Tegmark}} \emph {et~al.} (\bibinfo {collaboration} {SDSS}),\ }\bibfield  {title} {\bibinfo {title} {{Cosmological parameters from SDSS and WMAP}},\ }\href {https://doi.org/10.1103/PhysRevD.69.103501} {\bibfield  {journal} {\bibinfo  {journal} {Phys. Rev. D}\ }\textbf {\bibinfo {volume} {69}},\ \bibinfo {pages} {103501} (\bibinfo {year} {2004})},\ \Eprint {https://arxiv.org/abs/astro-ph/0310723} {arXiv:astro-ph/0310723} \BibitemShut {NoStop}%
\bibitem [{\citenamefont {Eisenstein}\ \emph {et~al.}(2005)\citenamefont {Eisenstein} \emph {et~al.}}]{SDSS:2005xqv}%
  \BibitemOpen
  \bibfield  {author} {\bibinfo {author} {\bibfnamefont {D.~J.}\ \bibnamefont {Eisenstein}} \emph {et~al.} (\bibinfo {collaboration} {SDSS}),\ }\bibfield  {title} {\bibinfo {title} {{Detection of the Baryon Acoustic Peak in the Large-Scale Correlation Function of SDSS Luminous Red Galaxies}},\ }\href {https://doi.org/10.1086/466512} {\bibfield  {journal} {\bibinfo  {journal} {Astrophys. J.}\ }\textbf {\bibinfo {volume} {633}},\ \bibinfo {pages} {560} (\bibinfo {year} {2005})},\ \Eprint {https://arxiv.org/abs/astro-ph/0501171} {arXiv:astro-ph/0501171} \BibitemShut {NoStop}%
\bibitem [{\citenamefont {Hinshaw}\ \emph {et~al.}(2013)\citenamefont {Hinshaw} \emph {et~al.}}]{WMAP:2012nax}%
  \BibitemOpen
  \bibfield  {author} {\bibinfo {author} {\bibfnamefont {G.}~\bibnamefont {Hinshaw}} \emph {et~al.} (\bibinfo {collaboration} {WMAP}),\ }\bibfield  {title} {\bibinfo {title} {{Nine-Year Wilkinson Microwave Anisotropy Probe (WMAP) Observations: Cosmological Parameter Results}},\ }\href {https://doi.org/10.1088/0067-0049/208/2/19} {\bibfield  {journal} {\bibinfo  {journal} {Astrophys. J. Suppl.}\ }\textbf {\bibinfo {volume} {208}},\ \bibinfo {pages} {19} (\bibinfo {year} {2013})},\ \Eprint {https://arxiv.org/abs/1212.5226} {arXiv:1212.5226 [astro-ph.CO]} \BibitemShut {NoStop}%
\bibitem [{\citenamefont {Chuang}\ \emph {et~al.}(2013)\citenamefont {Chuang} \emph {et~al.}}]{Chuang:2013hya}%
  \BibitemOpen
  \bibfield  {author} {\bibinfo {author} {\bibfnamefont {C.-H.}\ \bibnamefont {Chuang}} \emph {et~al.},\ }\bibfield  {title} {\bibinfo {title} {{The clustering of galaxies in the SDSS-III Baryon Oscillation Spectroscopic Survey: single-probe measurements and the strong power of normalized growth rate on constraining dark energy}},\ }\href {https://doi.org/10.1093/mnras/stt988} {\bibfield  {journal} {\bibinfo  {journal} {Mon. Not. Roy. Astron. Soc.}\ }\textbf {\bibinfo {volume} {433}},\ \bibinfo {pages} {3559} (\bibinfo {year} {2013})},\ \Eprint {https://arxiv.org/abs/1303.4486} {arXiv:1303.4486 [astro-ph.CO]} \BibitemShut {NoStop}%
\bibitem [{\citenamefont {Aghanim}\ \emph {et~al.}(2020)\citenamefont {Aghanim} \emph {et~al.}}]{Planck:2018vyg}%
  \BibitemOpen
  \bibfield  {author} {\bibinfo {author} {\bibfnamefont {N.}~\bibnamefont {Aghanim}} \emph {et~al.} (\bibinfo {collaboration} {Planck}),\ }\bibfield  {title} {\bibinfo {title} {{Planck 2018 results. VI. Cosmological parameters}},\ }\href {https://doi.org/10.1051/0004-6361/201833910} {\bibfield  {journal} {\bibinfo  {journal} {Astron. Astrophys.}\ }\textbf {\bibinfo {volume} {641}},\ \bibinfo {pages} {A6} (\bibinfo {year} {2020})},\ \bibinfo {note} {[Erratum: Astron.Astrophys. 652, C4 (2021)]},\ \Eprint {https://arxiv.org/abs/1807.06209} {arXiv:1807.06209 [astro-ph.CO]} \BibitemShut {NoStop}%
\bibitem [{\citenamefont {Zlatev}\ \emph {et~al.}(1999)\citenamefont {Zlatev}, \citenamefont {Wang},\ and\ \citenamefont {Steinhardt}}]{Zlatev:1998tr}%
  \BibitemOpen
  \bibfield  {author} {\bibinfo {author} {\bibfnamefont {I.}~\bibnamefont {Zlatev}}, \bibinfo {author} {\bibfnamefont {L.-M.}\ \bibnamefont {Wang}},\ and\ \bibinfo {author} {\bibfnamefont {P.~J.}\ \bibnamefont {Steinhardt}},\ }\bibfield  {title} {\bibinfo {title} {{Quintessence, cosmic coincidence, and the cosmological constant}},\ }\href {https://doi.org/10.1103/PhysRevLett.82.896} {\bibfield  {journal} {\bibinfo  {journal} {Phys. Rev. Lett.}\ }\textbf {\bibinfo {volume} {82}},\ \bibinfo {pages} {896} (\bibinfo {year} {1999})},\ \Eprint {https://arxiv.org/abs/astro-ph/9807002} {arXiv:astro-ph/9807002} \BibitemShut {NoStop}%
\bibitem [{\citenamefont {Sahni}\ and\ \citenamefont {Starobinsky}(2000)}]{Sahni:1999gb}%
  \BibitemOpen
  \bibfield  {author} {\bibinfo {author} {\bibfnamefont {V.}~\bibnamefont {Sahni}}\ and\ \bibinfo {author} {\bibfnamefont {A.~A.}\ \bibnamefont {Starobinsky}},\ }\bibfield  {title} {\bibinfo {title} {{The Case for a positive cosmological Lambda term}},\ }\href {https://doi.org/10.1142/S0218271800000542} {\bibfield  {journal} {\bibinfo  {journal} {Int. J. Mod. Phys. D}\ }\textbf {\bibinfo {volume} {9}},\ \bibinfo {pages} {373} (\bibinfo {year} {2000})},\ \Eprint {https://arxiv.org/abs/astro-ph/9904398} {arXiv:astro-ph/9904398} \BibitemShut {NoStop}%
\bibitem [{\citenamefont {Weinberg}(2000)}]{Weinberg:2000yb}%
  \BibitemOpen
  \bibfield  {author} {\bibinfo {author} {\bibfnamefont {S.}~\bibnamefont {Weinberg}},\ }\bibfield  {title} {\bibinfo {title} {{The Cosmological constant problems}},\ }in\ \href {https://doi.org/10.1007/978-3-662-04587-9_2} {\emph {\bibinfo {booktitle} {{4th International Symposium on Sources and Detection of Dark Matter in the Universe (DM 2000)}}}}\ (\bibinfo {year} {2000})\ pp.\ \bibinfo {pages} {18--26},\ \Eprint {https://arxiv.org/abs/astro-ph/0005265} {arXiv:astro-ph/0005265} \BibitemShut {NoStop}%
\bibitem [{\citenamefont {Riess}\ \emph {et~al.}(2022)\citenamefont {Riess} \emph {et~al.}}]{Riess:2021jrx}%
  \BibitemOpen
  \bibfield  {author} {\bibinfo {author} {\bibfnamefont {A.~G.}\ \bibnamefont {Riess}} \emph {et~al.},\ }\bibfield  {title} {\bibinfo {title} {{A Comprehensive Measurement of the Local Value of the Hubble Constant with 1 km s$^{−1}$ Mpc$^{−1}$ Uncertainty from the Hubble Space Telescope and the SH0ES Team}},\ }\href {https://doi.org/10.3847/2041-8213/ac5c5b} {\bibfield  {journal} {\bibinfo  {journal} {Astrophys. J. Lett.}\ }\textbf {\bibinfo {volume} {934}},\ \bibinfo {pages} {L7} (\bibinfo {year} {2022})},\ \Eprint {https://arxiv.org/abs/2112.04510} {arXiv:2112.04510 [astro-ph.CO]} \BibitemShut {NoStop}%
\bibitem [{\citenamefont {Copeland}\ \emph {et~al.}(2006)\citenamefont {Copeland}, \citenamefont {Sami},\ and\ \citenamefont {Tsujikawa}}]{Copeland:2006wr}%
  \BibitemOpen
  \bibfield  {author} {\bibinfo {author} {\bibfnamefont {E.~J.}\ \bibnamefont {Copeland}}, \bibinfo {author} {\bibfnamefont {M.}~\bibnamefont {Sami}},\ and\ \bibinfo {author} {\bibfnamefont {S.}~\bibnamefont {Tsujikawa}},\ }\bibfield  {title} {\bibinfo {title} {{Dynamics of dark energy}},\ }\href {https://doi.org/10.1142/S021827180600942X} {\bibfield  {journal} {\bibinfo  {journal} {Int. J. Mod. Phys. D}\ }\textbf {\bibinfo {volume} {15}},\ \bibinfo {pages} {1753} (\bibinfo {year} {2006})},\ \Eprint {https://arxiv.org/abs/hep-th/0603057} {arXiv:hep-th/0603057} \BibitemShut {NoStop}%
\bibitem [{\citenamefont {Tsujikawa}(2013)}]{Tsujikawa:2013fta}%
  \BibitemOpen
  \bibfield  {author} {\bibinfo {author} {\bibfnamefont {S.}~\bibnamefont {Tsujikawa}},\ }\bibfield  {title} {\bibinfo {title} {{Quintessence: A Review}},\ }\href {https://doi.org/10.1088/0264-9381/30/21/214003} {\bibfield  {journal} {\bibinfo  {journal} {Class. Quant. Grav.}\ }\textbf {\bibinfo {volume} {30}},\ \bibinfo {pages} {214003} (\bibinfo {year} {2013})},\ \Eprint {https://arxiv.org/abs/1304.1961} {arXiv:1304.1961 [gr-qc]} \BibitemShut {NoStop}%
\bibitem [{\citenamefont {Armendariz-Picon}\ \emph {et~al.}(2001)\citenamefont {Armendariz-Picon}, \citenamefont {Mukhanov},\ and\ \citenamefont {Steinhardt}}]{Armendariz-Picon:2000ulo}%
  \BibitemOpen
  \bibfield  {author} {\bibinfo {author} {\bibfnamefont {C.}~\bibnamefont {Armendariz-Picon}}, \bibinfo {author} {\bibfnamefont {V.~F.}\ \bibnamefont {Mukhanov}},\ and\ \bibinfo {author} {\bibfnamefont {P.~J.}\ \bibnamefont {Steinhardt}},\ }\bibfield  {title} {\bibinfo {title} {{Essentials of k essence}},\ }\href {https://doi.org/10.1103/PhysRevD.63.103510} {\bibfield  {journal} {\bibinfo  {journal} {Phys. Rev. D}\ }\textbf {\bibinfo {volume} {63}},\ \bibinfo {pages} {103510} (\bibinfo {year} {2001})},\ \Eprint {https://arxiv.org/abs/astro-ph/0006373} {arXiv:astro-ph/0006373} \BibitemShut {NoStop}%
\bibitem [{\citenamefont {Linder}(2008)}]{Linder:2007wa}%
  \BibitemOpen
  \bibfield  {author} {\bibinfo {author} {\bibfnamefont {E.~V.}\ \bibnamefont {Linder}},\ }\bibfield  {title} {\bibinfo {title} {{The Dynamics of Quintessence, The Quintessence of Dynamics}},\ }\href {https://doi.org/10.1007/s10714-007-0550-z} {\bibfield  {journal} {\bibinfo  {journal} {Gen. Rel. Grav.}\ }\textbf {\bibinfo {volume} {40}},\ \bibinfo {pages} {329} (\bibinfo {year} {2008})},\ \Eprint {https://arxiv.org/abs/0704.2064} {arXiv:0704.2064 [astro-ph]} \BibitemShut {NoStop}%
\bibitem [{\citenamefont {Saridakis}\ and\ \citenamefont {Sushkov}(2010)}]{Saridakis:2010mf}%
  \BibitemOpen
  \bibfield  {author} {\bibinfo {author} {\bibfnamefont {E.~N.}\ \bibnamefont {Saridakis}}\ and\ \bibinfo {author} {\bibfnamefont {S.~V.}\ \bibnamefont {Sushkov}},\ }\bibfield  {title} {\bibinfo {title} {{Quintessence and phantom cosmology with non-minimal derivative coupling}},\ }\href {https://doi.org/10.1103/PhysRevD.81.083510} {\bibfield  {journal} {\bibinfo  {journal} {Phys. Rev. D}\ }\textbf {\bibinfo {volume} {81}},\ \bibinfo {pages} {083510} (\bibinfo {year} {2010})},\ \Eprint {https://arxiv.org/abs/1002.3478} {arXiv:1002.3478 [gr-qc]} \BibitemShut {NoStop}%
\bibitem [{\citenamefont {Kunz}\ and\ \citenamefont {Sapone}(2007)}]{Kunz:2006ca}%
  \BibitemOpen
  \bibfield  {author} {\bibinfo {author} {\bibfnamefont {M.}~\bibnamefont {Kunz}}\ and\ \bibinfo {author} {\bibfnamefont {D.}~\bibnamefont {Sapone}},\ }\bibfield  {title} {\bibinfo {title} {{Dark Energy versus Modified Gravity}},\ }\href {https://doi.org/10.1103/PhysRevLett.98.121301} {\bibfield  {journal} {\bibinfo  {journal} {Phys. Rev. Lett.}\ }\textbf {\bibinfo {volume} {98}},\ \bibinfo {pages} {121301} (\bibinfo {year} {2007})},\ \Eprint {https://arxiv.org/abs/astro-ph/0612452} {arXiv:astro-ph/0612452} \BibitemShut {NoStop}%
\bibitem [{\citenamefont {Clifton}\ \emph {et~al.}(2012)\citenamefont {Clifton}, \citenamefont {Ferreira}, \citenamefont {Padilla},\ and\ \citenamefont {Skordis}}]{Clifton:2011jh}%
  \BibitemOpen
  \bibfield  {author} {\bibinfo {author} {\bibfnamefont {T.}~\bibnamefont {Clifton}}, \bibinfo {author} {\bibfnamefont {P.~G.}\ \bibnamefont {Ferreira}}, \bibinfo {author} {\bibfnamefont {A.}~\bibnamefont {Padilla}},\ and\ \bibinfo {author} {\bibfnamefont {C.}~\bibnamefont {Skordis}},\ }\bibfield  {title} {\bibinfo {title} {{Modified Gravity and Cosmology}},\ }\href {https://doi.org/10.1016/j.physrep.2012.01.001} {\bibfield  {journal} {\bibinfo  {journal} {Phys. Rept.}\ }\textbf {\bibinfo {volume} {513}},\ \bibinfo {pages} {1} (\bibinfo {year} {2012})},\ \Eprint {https://arxiv.org/abs/1106.2476} {arXiv:1106.2476 [astro-ph.CO]} \BibitemShut {NoStop}%
\bibitem [{\citenamefont {Hu}\ and\ \citenamefont {Sawicki}(2007)}]{Hu:2007nk}%
  \BibitemOpen
  \bibfield  {author} {\bibinfo {author} {\bibfnamefont {W.}~\bibnamefont {Hu}}\ and\ \bibinfo {author} {\bibfnamefont {I.}~\bibnamefont {Sawicki}},\ }\bibfield  {title} {\bibinfo {title} {{Models of f(R) Cosmic Acceleration that Evade Solar-System Tests}},\ }\href {https://doi.org/10.1103/PhysRevD.76.064004} {\bibfield  {journal} {\bibinfo  {journal} {Phys. Rev. D}\ }\textbf {\bibinfo {volume} {76}},\ \bibinfo {pages} {064004} (\bibinfo {year} {2007})},\ \Eprint {https://arxiv.org/abs/0705.1158} {arXiv:0705.1158 [astro-ph]} \BibitemShut {NoStop}%
\bibitem [{\citenamefont {Amendola}(2000)}]{Amendola:1999er}%
  \BibitemOpen
  \bibfield  {author} {\bibinfo {author} {\bibfnamefont {L.}~\bibnamefont {Amendola}},\ }\bibfield  {title} {\bibinfo {title} {{Coupled quintessence}},\ }\href {https://doi.org/10.1103/PhysRevD.62.043511} {\bibfield  {journal} {\bibinfo  {journal} {Phys. Rev. D}\ }\textbf {\bibinfo {volume} {62}},\ \bibinfo {pages} {043511} (\bibinfo {year} {2000})},\ \Eprint {https://arxiv.org/abs/astro-ph/9908023} {arXiv:astro-ph/9908023} \BibitemShut {NoStop}%
\bibitem [{\citenamefont {Shahalam}\ \emph {et~al.}(2015)\citenamefont {Shahalam}, \citenamefont {Pathak}, \citenamefont {Verma}, \citenamefont {Khlopov},\ and\ \citenamefont {Myrzakulov}}]{Shahalam:2015sja}%
  \BibitemOpen
  \bibfield  {author} {\bibinfo {author} {\bibfnamefont {M.}~\bibnamefont {Shahalam}}, \bibinfo {author} {\bibfnamefont {S.~D.}\ \bibnamefont {Pathak}}, \bibinfo {author} {\bibfnamefont {M.~M.}\ \bibnamefont {Verma}}, \bibinfo {author} {\bibfnamefont {M.~Y.}\ \bibnamefont {Khlopov}},\ and\ \bibinfo {author} {\bibfnamefont {R.}~\bibnamefont {Myrzakulov}},\ }\bibfield  {title} {\bibinfo {title} {{Dynamics of interacting quintessence}},\ }\href {https://doi.org/10.1140/epjc/s10052-015-3608-1} {\bibfield  {journal} {\bibinfo  {journal} {Eur. Phys. J. C}\ }\textbf {\bibinfo {volume} {75}},\ \bibinfo {pages} {395} (\bibinfo {year} {2015})},\ \Eprint {https://arxiv.org/abs/1503.08712} {arXiv:1503.08712 [gr-qc]} \BibitemShut {NoStop}%
\bibitem [{\citenamefont {Pourtsidou}\ \emph {et~al.}(2013)\citenamefont {Pourtsidou}, \citenamefont {Skordis},\ and\ \citenamefont {Copeland}}]{Pourtsidou:2013nha}%
  \BibitemOpen
  \bibfield  {author} {\bibinfo {author} {\bibfnamefont {A.}~\bibnamefont {Pourtsidou}}, \bibinfo {author} {\bibfnamefont {C.}~\bibnamefont {Skordis}},\ and\ \bibinfo {author} {\bibfnamefont {E.~J.}\ \bibnamefont {Copeland}},\ }\bibfield  {title} {\bibinfo {title} {{Models of dark matter coupled to dark energy}},\ }\href {https://doi.org/10.1103/PhysRevD.88.083505} {\bibfield  {journal} {\bibinfo  {journal} {Phys. Rev. D}\ }\textbf {\bibinfo {volume} {88}},\ \bibinfo {pages} {083505} (\bibinfo {year} {2013})},\ \Eprint {https://arxiv.org/abs/1307.0458} {arXiv:1307.0458 [astro-ph.CO]} \BibitemShut {NoStop}%
\bibitem [{\citenamefont {Costa}\ \emph {et~al.}(2017)\citenamefont {Costa}, \citenamefont {Xu}, \citenamefont {Wang},\ and\ \citenamefont {Abdalla}}]{Costa:2016tpb}%
  \BibitemOpen
  \bibfield  {author} {\bibinfo {author} {\bibfnamefont {A.~A.}\ \bibnamefont {Costa}}, \bibinfo {author} {\bibfnamefont {X.-D.}\ \bibnamefont {Xu}}, \bibinfo {author} {\bibfnamefont {B.}~\bibnamefont {Wang}},\ and\ \bibinfo {author} {\bibfnamefont {E.}~\bibnamefont {Abdalla}},\ }\bibfield  {title} {\bibinfo {title} {{Constraints on interacting dark energy models from Planck 2015 and redshift-space distortion data}},\ }\href {https://doi.org/10.1088/1475-7516/2017/01/028} {\bibfield  {journal} {\bibinfo  {journal} {JCAP}\ }\textbf {\bibinfo {volume} {01}},\ \bibinfo {pages} {028}},\ \Eprint {https://arxiv.org/abs/1605.04138} {arXiv:1605.04138 [astro-ph.CO]} \BibitemShut {NoStop}%
\bibitem [{\citenamefont {Giar{\`e}}\ \emph {et~al.}(2024{\natexlab{a}})\citenamefont {Giar{\`e}}, \citenamefont {Sabogal}, \citenamefont {Nunes},\ and\ \citenamefont {Di~Valentino}}]{Giare:2024smz}%
  \BibitemOpen
  \bibfield  {author} {\bibinfo {author} {\bibfnamefont {W.}~\bibnamefont {Giar{\`e}}}, \bibinfo {author} {\bibfnamefont {M.~A.}\ \bibnamefont {Sabogal}}, \bibinfo {author} {\bibfnamefont {R.~C.}\ \bibnamefont {Nunes}},\ and\ \bibinfo {author} {\bibfnamefont {E.}~\bibnamefont {Di~Valentino}},\ }\bibfield  {title} {\bibinfo {title} {{Interacting Dark Energy after DESI Baryon Acoustic Oscillation Measurements}},\ }\href {https://doi.org/10.1103/PhysRevLett.133.251003} {\bibfield  {journal} {\bibinfo  {journal} {Phys. Rev. Lett.}\ }\textbf {\bibinfo {volume} {133}},\ \bibinfo {pages} {251003} (\bibinfo {year} {2024}{\natexlab{a}})},\ \Eprint {https://arxiv.org/abs/2404.15232} {arXiv:2404.15232 [astro-ph.CO]} \BibitemShut {NoStop}%
\bibitem [{\citenamefont {Dinda}(2024)}]{Dinda:2024kjf}%
  \BibitemOpen
  \bibfield  {author} {\bibinfo {author} {\bibfnamefont {B.~R.}\ \bibnamefont {Dinda}},\ }\bibfield  {title} {\bibinfo {title} {{A new diagnostic for the null test of dynamical dark energy in light of DESI 2024 and other BAO data}},\ }\href {https://doi.org/10.1088/1475-7516/2024/09/062} {\bibfield  {journal} {\bibinfo  {journal} {JCAP}\ }\textbf {\bibinfo {volume} {09}},\ \bibinfo {pages} {062}},\ \Eprint {https://arxiv.org/abs/2405.06618} {arXiv:2405.06618 [astro-ph.CO]} \BibitemShut {NoStop}%
\bibitem [{\citenamefont {Dinda}\ and\ \citenamefont {Maartens}(2025)}]{Dinda:2024ktd}%
  \BibitemOpen
  \bibfield  {author} {\bibinfo {author} {\bibfnamefont {B.~R.}\ \bibnamefont {Dinda}}\ and\ \bibinfo {author} {\bibfnamefont {R.}~\bibnamefont {Maartens}},\ }\bibfield  {title} {\bibinfo {title} {{Model-agnostic assessment of dark energy after DESI DR1 BAO}},\ }\href {https://doi.org/10.1088/1475-7516/2025/01/120} {\bibfield  {journal} {\bibinfo  {journal} {JCAP}\ }\textbf {\bibinfo {volume} {01}},\ \bibinfo {pages} {120}},\ \Eprint {https://arxiv.org/abs/2407.17252} {arXiv:2407.17252 [astro-ph.CO]} \BibitemShut {NoStop}%
\bibitem [{\citenamefont {Calderon}\ \emph {et~al.}(2024)\citenamefont {Calderon} \emph {et~al.}}]{DESI:2024aqx}%
  \BibitemOpen
  \bibfield  {author} {\bibinfo {author} {\bibfnamefont {R.}~\bibnamefont {Calderon}} \emph {et~al.} (\bibinfo {collaboration} {DESI}),\ }\bibfield  {title} {\bibinfo {title} {{DESI 2024: reconstructing dark energy using crossing statistics with DESI DR1 BAO data}},\ }\href {https://doi.org/10.1088/1475-7516/2024/10/048} {\bibfield  {journal} {\bibinfo  {journal} {JCAP}\ }\textbf {\bibinfo {volume} {10}},\ \bibinfo {pages} {048}},\ \Eprint {https://arxiv.org/abs/2405.04216} {arXiv:2405.04216 [astro-ph.CO]} \BibitemShut {NoStop}%
\bibitem [{\citenamefont {Keeley}\ \emph {et~al.}(2025)\citenamefont {Keeley}, \citenamefont {Abazajian}, \citenamefont {Kaplinghat},\ and\ \citenamefont {Shafieloo}}]{Keeley:2025stf}%
  \BibitemOpen
  \bibfield  {author} {\bibinfo {author} {\bibfnamefont {R.~E.}\ \bibnamefont {Keeley}}, \bibinfo {author} {\bibfnamefont {K.~N.}\ \bibnamefont {Abazajian}}, \bibinfo {author} {\bibfnamefont {M.}~\bibnamefont {Kaplinghat}},\ and\ \bibinfo {author} {\bibfnamefont {A.}~\bibnamefont {Shafieloo}},\ }\bibfield  {title} {\bibinfo {title} {{Preference for evolving dark energy from cosmological distance measurements and possible signatures in the growth rate of perturbations}},\ }\href {https://doi.org/10.1103/sqsv-hy65} {\bibfield  {journal} {\bibinfo  {journal} {Phys. Rev. D}\ }\textbf {\bibinfo {volume} {112}},\ \bibinfo {pages} {043501} (\bibinfo {year} {2025})},\ \Eprint {https://arxiv.org/abs/2502.12667} {arXiv:2502.12667 [astro-ph.CO]} \BibitemShut {NoStop}%
\bibitem [{\citenamefont {Cort{\^e}s}\ and\ \citenamefont {Liddle}(2024)}]{Cortes:2024lgw}%
  \BibitemOpen
  \bibfield  {author} {\bibinfo {author} {\bibfnamefont {M.}~\bibnamefont {Cort{\^e}s}}\ and\ \bibinfo {author} {\bibfnamefont {A.~R.}\ \bibnamefont {Liddle}},\ }\bibfield  {title} {\bibinfo {title} {{Interpreting DESI's evidence for evolving dark energy}},\ }\href {https://doi.org/10.1088/1475-7516/2024/12/007} {\bibfield  {journal} {\bibinfo  {journal} {JCAP}\ }\textbf {\bibinfo {volume} {12}},\ \bibinfo {pages} {007}},\ \Eprint {https://arxiv.org/abs/2404.08056} {arXiv:2404.08056 [astro-ph.CO]} \BibitemShut {NoStop}%
\bibitem [{\citenamefont {Adame}\ \emph {et~al.}(2025)\citenamefont {Adame} \emph {et~al.}}]{DESI:2024mwx}%
  \BibitemOpen
  \bibfield  {author} {\bibinfo {author} {\bibfnamefont {A.~G.}\ \bibnamefont {Adame}} \emph {et~al.} (\bibinfo {collaboration} {DESI}),\ }\bibfield  {title} {\bibinfo {title} {{DESI 2024 VI: cosmological constraints from the measurements of baryon acoustic oscillations}},\ }\href {https://doi.org/10.1088/1475-7516/2025/02/021} {\bibfield  {journal} {\bibinfo  {journal} {JCAP}\ }\textbf {\bibinfo {volume} {02}},\ \bibinfo {pages} {021}},\ \Eprint {https://arxiv.org/abs/2404.03002} {arXiv:2404.03002 [astro-ph.CO]} \BibitemShut {NoStop}%
\bibitem [{\citenamefont {Abdul~Karim}\ \emph {et~al.}(2025)\citenamefont {Abdul~Karim} \emph {et~al.}}]{DESI:2025zgx}%
  \BibitemOpen
  \bibfield  {author} {\bibinfo {author} {\bibfnamefont {M.}~\bibnamefont {Abdul~Karim}} \emph {et~al.} (\bibinfo {collaboration} {DESI}),\ }\bibfield  {title} {\bibinfo {title} {{DESI DR2 results. II. Measurements of baryon acoustic oscillations and cosmological constraints}},\ }\href {https://doi.org/10.1103/tr6y-kpc6} {\bibfield  {journal} {\bibinfo  {journal} {Phys. Rev. D}\ }\textbf {\bibinfo {volume} {112}},\ \bibinfo {pages} {083515} (\bibinfo {year} {2025})},\ \Eprint {https://arxiv.org/abs/2503.14738} {arXiv:2503.14738 [astro-ph.CO]} \BibitemShut {NoStop}%
\bibitem [{\citenamefont {Gu}\ \emph {et~al.}(2025)\citenamefont {Gu} \emph {et~al.}}]{DESI:2025wyn}%
  \BibitemOpen
  \bibfield  {author} {\bibinfo {author} {\bibfnamefont {G.}~\bibnamefont {Gu}} \emph {et~al.} (\bibinfo {collaboration} {DESI}),\ }\bibfield  {title} {\bibinfo {title} {{Dynamical dark energy in light of the DESI DR2 baryonic acoustic oscillations measurements}},\ }\href {https://doi.org/10.1038/s41550-025-02669-6} {\bibfield  {journal} {\bibinfo  {journal} {Nature Astron.}\ }\textbf {\bibinfo {volume} {9}},\ \bibinfo {pages} {1879} (\bibinfo {year} {2025})},\ \bibinfo {note} {[Erratum: Nature Astron. 9, 1898 (2025)]},\ \Eprint {https://arxiv.org/abs/2504.06118} {arXiv:2504.06118 [astro-ph.CO]} \BibitemShut {NoStop}%
\bibitem [{\citenamefont {Giar{\`e}}\ \emph {et~al.}(2024{\natexlab{b}})\citenamefont {Giar{\`e}}, \citenamefont {Najafi}, \citenamefont {Pan}, \citenamefont {Di~Valentino},\ and\ \citenamefont {Firouzjaee}}]{Giare:2024gpk}%
  \BibitemOpen
  \bibfield  {author} {\bibinfo {author} {\bibfnamefont {W.}~\bibnamefont {Giar{\`e}}}, \bibinfo {author} {\bibfnamefont {M.}~\bibnamefont {Najafi}}, \bibinfo {author} {\bibfnamefont {S.}~\bibnamefont {Pan}}, \bibinfo {author} {\bibfnamefont {E.}~\bibnamefont {Di~Valentino}},\ and\ \bibinfo {author} {\bibfnamefont {J.~T.}\ \bibnamefont {Firouzjaee}},\ }\bibfield  {title} {\bibinfo {title} {{Robust preference for Dynamical Dark Energy in DESI BAO and SN measurements}},\ }\href {https://doi.org/10.1088/1475-7516/2024/10/035} {\bibfield  {journal} {\bibinfo  {journal} {JCAP}\ }\textbf {\bibinfo {volume} {10}},\ \bibinfo {pages} {035}},\ \Eprint {https://arxiv.org/abs/2407.16689} {arXiv:2407.16689 [astro-ph.CO]} \BibitemShut {NoStop}%
\bibitem [{\citenamefont {Lodha}\ \emph {et~al.}(2025)\citenamefont {Lodha} \emph {et~al.}}]{DESI:2025fii}%
  \BibitemOpen
  \bibfield  {author} {\bibinfo {author} {\bibfnamefont {K.}~\bibnamefont {Lodha}} \emph {et~al.} (\bibinfo {collaboration} {DESI}),\ }\bibfield  {title} {\bibinfo {title} {{Extended dark energy analysis using DESI DR2 BAO measurements}},\ }\href {https://doi.org/10.1103/w4c6-1r5j} {\bibfield  {journal} {\bibinfo  {journal} {Phys. Rev. D}\ }\textbf {\bibinfo {volume} {112}},\ \bibinfo {pages} {083511} (\bibinfo {year} {2025})},\ \Eprint {https://arxiv.org/abs/2503.14743} {arXiv:2503.14743 [astro-ph.CO]} \BibitemShut {NoStop}%
\bibitem [{\citenamefont {Kunz}(2009)}]{Kunz:2007rk}%
  \BibitemOpen
  \bibfield  {author} {\bibinfo {author} {\bibfnamefont {M.}~\bibnamefont {Kunz}},\ }\bibfield  {title} {\bibinfo {title} {{The dark degeneracy: On the number and nature of dark components}},\ }\href {https://doi.org/10.1103/PhysRevD.80.123001} {\bibfield  {journal} {\bibinfo  {journal} {Phys. Rev. D}\ }\textbf {\bibinfo {volume} {80}},\ \bibinfo {pages} {123001} (\bibinfo {year} {2009})},\ \Eprint {https://arxiv.org/abs/astro-ph/0702615} {arXiv:astro-ph/0702615} \BibitemShut {NoStop}%
\bibitem [{\citenamefont {von Marttens}\ \emph {et~al.}(2020)\citenamefont {von Marttens}, \citenamefont {Lombriser}, \citenamefont {Kunz}, \citenamefont {Marra}, \citenamefont {Casarini},\ and\ \citenamefont {Alcaniz}}]{vonMarttens:2019ixw}%
  \BibitemOpen
  \bibfield  {author} {\bibinfo {author} {\bibfnamefont {R.}~\bibnamefont {von Marttens}}, \bibinfo {author} {\bibfnamefont {L.}~\bibnamefont {Lombriser}}, \bibinfo {author} {\bibfnamefont {M.}~\bibnamefont {Kunz}}, \bibinfo {author} {\bibfnamefont {V.}~\bibnamefont {Marra}}, \bibinfo {author} {\bibfnamefont {L.}~\bibnamefont {Casarini}},\ and\ \bibinfo {author} {\bibfnamefont {J.}~\bibnamefont {Alcaniz}},\ }\bibfield  {title} {\bibinfo {title} {{Dark degeneracy I: Dynamical or interacting dark energy?}},\ }\href {https://doi.org/10.1016/j.dark.2020.100490} {\bibfield  {journal} {\bibinfo  {journal} {Phys. Dark Univ.}\ }\textbf {\bibinfo {volume} {28}},\ \bibinfo {pages} {100490} (\bibinfo {year} {2020})},\ \Eprint {https://arxiv.org/abs/1911.02618} {arXiv:1911.02618 [astro-ph.CO]} \BibitemShut {NoStop}%
\bibitem [{\citenamefont {Yang}\ \emph {et~al.}(2015)\citenamefont {Yang}, \citenamefont {Guo},\ and\ \citenamefont {Cai}}]{Yang:2015tzc}%
  \BibitemOpen
  \bibfield  {author} {\bibinfo {author} {\bibfnamefont {T.}~\bibnamefont {Yang}}, \bibinfo {author} {\bibfnamefont {Z.-K.}\ \bibnamefont {Guo}},\ and\ \bibinfo {author} {\bibfnamefont {R.-G.}\ \bibnamefont {Cai}},\ }\bibfield  {title} {\bibinfo {title} {{Reconstructing the interaction between dark energy and dark matter using Gaussian Processes}},\ }\href {https://doi.org/10.1103/PhysRevD.91.123533} {\bibfield  {journal} {\bibinfo  {journal} {Phys. Rev. D}\ }\textbf {\bibinfo {volume} {91}},\ \bibinfo {pages} {123533} (\bibinfo {year} {2015})},\ \Eprint {https://arxiv.org/abs/1505.04443} {arXiv:1505.04443 [astro-ph.CO]} \BibitemShut {NoStop}%
\bibitem [{\citenamefont {Yang}(2020)}]{Yang:2020jze}%
  \BibitemOpen
  \bibfield  {author} {\bibinfo {author} {\bibfnamefont {T.}~\bibnamefont {Yang}},\ }\bibfield  {title} {\bibinfo {title} {{Model-Independent Perspectives on Coupled Dark Energy and the Swampland}},\ }\href {https://doi.org/10.1103/PhysRevD.102.083511} {\bibfield  {journal} {\bibinfo  {journal} {Phys. Rev. D}\ }\textbf {\bibinfo {volume} {102}},\ \bibinfo {pages} {083511} (\bibinfo {year} {2020})},\ \Eprint {https://arxiv.org/abs/2006.14511} {arXiv:2006.14511 [astro-ph.CO]} \BibitemShut {NoStop}%
\bibitem [{\citenamefont {You}\ \emph {et~al.}(2025)\citenamefont {You}, \citenamefont {Wang},\ and\ \citenamefont {Yang}}]{You:2025uon}%
  \BibitemOpen
  \bibfield  {author} {\bibinfo {author} {\bibfnamefont {C.}~\bibnamefont {You}}, \bibinfo {author} {\bibfnamefont {D.}~\bibnamefont {Wang}},\ and\ \bibinfo {author} {\bibfnamefont {T.}~\bibnamefont {Yang}},\ }\bibfield  {title} {\bibinfo {title} {{Dynamical dark energy implies a coupled dark sector: Insights from DESI DR2 via a data-driven approach}},\ }\href {https://doi.org/10.1103/f6v7-n9fr} {\bibfield  {journal} {\bibinfo  {journal} {Phys. Rev. D}\ }\textbf {\bibinfo {volume} {112}},\ \bibinfo {pages} {043503} (\bibinfo {year} {2025})},\ \Eprint {https://arxiv.org/abs/2504.00985} {arXiv:2504.00985 [astro-ph.CO]} \BibitemShut {NoStop}%
\bibitem [{\citenamefont {Wang}\ \emph {et~al.}(2015)\citenamefont {Wang}, \citenamefont {Zhao}, \citenamefont {Wands}, \citenamefont {Pogosian},\ and\ \citenamefont {Crittenden}}]{Wang:2015wga}%
  \BibitemOpen
  \bibfield  {author} {\bibinfo {author} {\bibfnamefont {Y.}~\bibnamefont {Wang}}, \bibinfo {author} {\bibfnamefont {G.-B.}\ \bibnamefont {Zhao}}, \bibinfo {author} {\bibfnamefont {D.}~\bibnamefont {Wands}}, \bibinfo {author} {\bibfnamefont {L.}~\bibnamefont {Pogosian}},\ and\ \bibinfo {author} {\bibfnamefont {R.~G.}\ \bibnamefont {Crittenden}},\ }\bibfield  {title} {\bibinfo {title} {{Reconstruction of the dark matter{\textendash}vacuum energy interaction}},\ }\href {https://doi.org/10.1103/PhysRevD.92.103005} {\bibfield  {journal} {\bibinfo  {journal} {Phys. Rev. D}\ }\textbf {\bibinfo {volume} {92}},\ \bibinfo {pages} {103005} (\bibinfo {year} {2015})},\ \Eprint {https://arxiv.org/abs/1505.01373} {arXiv:1505.01373 [astro-ph.CO]} \BibitemShut {NoStop}%
\bibitem [{\citenamefont {Abedin}\ \emph {et~al.}(2025)\citenamefont {Abedin}, \citenamefont {Wang}, \citenamefont {Ma},\ and\ \citenamefont {Pan}}]{Abedin:2025yru}%
  \BibitemOpen
  \bibfield  {author} {\bibinfo {author} {\bibfnamefont {M.}~\bibnamefont {Abedin}}, \bibinfo {author} {\bibfnamefont {G.-J.}\ \bibnamefont {Wang}}, \bibinfo {author} {\bibfnamefont {Y.-Z.}\ \bibnamefont {Ma}},\ and\ \bibinfo {author} {\bibfnamefont {S.}~\bibnamefont {Pan}},\ }\bibfield  {title} {\bibinfo {title} {{In search of an interaction in the dark sector through Gaussian Process and ANN approaches}},\ }\href {https://doi.org/10.1093/mnras/staf762} {\bibfield  {journal} {\bibinfo  {journal} {Mon. Not. Roy. Astron. Soc.}\ }\textbf {\bibinfo {volume} {540}},\ \bibinfo {pages} {2253} (\bibinfo {year} {2025})},\ \Eprint {https://arxiv.org/abs/2505.04336} {arXiv:2505.04336 [astro-ph.CO]} \BibitemShut {NoStop}%
\bibitem [{\citenamefont {Mukherjee}\ and\ \citenamefont {Banerjee}(2021)}]{Mukherjee:2021ggf}%
  \BibitemOpen
  \bibfield  {author} {\bibinfo {author} {\bibfnamefont {P.}~\bibnamefont {Mukherjee}}\ and\ \bibinfo {author} {\bibfnamefont {N.}~\bibnamefont {Banerjee}},\ }\bibfield  {title} {\bibinfo {title} {{Nonparametric reconstruction of interaction in the cosmic dark sector}},\ }\href {https://doi.org/10.1103/PhysRevD.103.123530} {\bibfield  {journal} {\bibinfo  {journal} {Phys. Rev. D}\ }\textbf {\bibinfo {volume} {103}},\ \bibinfo {pages} {123530} (\bibinfo {year} {2021})},\ \Eprint {https://arxiv.org/abs/2105.09995} {arXiv:2105.09995 [astro-ph.CO]} \BibitemShut {NoStop}%
\bibitem [{\citenamefont {Bonilla}\ \emph {et~al.}(2022)\citenamefont {Bonilla}, \citenamefont {Kumar}, \citenamefont {Nunes},\ and\ \citenamefont {Pan}}]{Bonilla:2021dql}%
  \BibitemOpen
  \bibfield  {author} {\bibinfo {author} {\bibfnamefont {A.}~\bibnamefont {Bonilla}}, \bibinfo {author} {\bibfnamefont {S.}~\bibnamefont {Kumar}}, \bibinfo {author} {\bibfnamefont {R.~C.}\ \bibnamefont {Nunes}},\ and\ \bibinfo {author} {\bibfnamefont {S.}~\bibnamefont {Pan}},\ }\bibfield  {title} {\bibinfo {title} {{Reconstruction of the dark sectors{\textquoteright} interaction: A model-independent inference and forecast from GW standard sirens}},\ }\href {https://doi.org/10.1093/mnras/stac687} {\bibfield  {journal} {\bibinfo  {journal} {Mon. Not. Roy. Astron. Soc.}\ }\textbf {\bibinfo {volume} {512}},\ \bibinfo {pages} {4231} (\bibinfo {year} {2022})},\ \Eprint {https://arxiv.org/abs/2102.06149} {arXiv:2102.06149 [astro-ph.CO]} \BibitemShut {NoStop}%
\bibitem [{\citenamefont {Pettorino}\ and\ \citenamefont {Baccigalupi}(2008)}]{Pettorino:2008ez}%
  \BibitemOpen
  \bibfield  {author} {\bibinfo {author} {\bibfnamefont {V.}~\bibnamefont {Pettorino}}\ and\ \bibinfo {author} {\bibfnamefont {C.}~\bibnamefont {Baccigalupi}},\ }\bibfield  {title} {\bibinfo {title} {{Coupled and Extended Quintessence: theoretical differences and structure formation}},\ }\href {https://doi.org/10.1103/PhysRevD.77.103003} {\bibfield  {journal} {\bibinfo  {journal} {Phys. Rev. D}\ }\textbf {\bibinfo {volume} {77}},\ \bibinfo {pages} {103003} (\bibinfo {year} {2008})},\ \Eprint {https://arxiv.org/abs/0802.1086} {arXiv:0802.1086 [astro-ph]} \BibitemShut {NoStop}%
\bibitem [{\citenamefont {Wetterich}(2015)}]{Wetterich:2014bma}%
  \BibitemOpen
  \bibfield  {author} {\bibinfo {author} {\bibfnamefont {C.}~\bibnamefont {Wetterich}},\ }\bibfield  {title} {\bibinfo {title} {{Modified gravity and coupled quintessence}},\ }\href {https://doi.org/10.1007/978-3-319-10070-8_3} {\bibfield  {journal} {\bibinfo  {journal} {Lect. Notes Phys.}\ }\textbf {\bibinfo {volume} {892}},\ \bibinfo {pages} {57} (\bibinfo {year} {2015})},\ \Eprint {https://arxiv.org/abs/1402.5031} {arXiv:1402.5031 [astro-ph.CO]} \BibitemShut {NoStop}%
\bibitem [{\citenamefont {Postma}\ and\ \citenamefont {Volponi}(2014)}]{Postma:2014vaa}%
  \BibitemOpen
  \bibfield  {author} {\bibinfo {author} {\bibfnamefont {M.}~\bibnamefont {Postma}}\ and\ \bibinfo {author} {\bibfnamefont {M.}~\bibnamefont {Volponi}},\ }\bibfield  {title} {\bibinfo {title} {{Equivalence of the Einstein and Jordan frames}},\ }\href {https://doi.org/10.1103/PhysRevD.90.103516} {\bibfield  {journal} {\bibinfo  {journal} {Phys. Rev. D}\ }\textbf {\bibinfo {volume} {90}},\ \bibinfo {pages} {103516} (\bibinfo {year} {2014})},\ \Eprint {https://arxiv.org/abs/1407.6874} {arXiv:1407.6874 [astro-ph.CO]} \BibitemShut {NoStop}%
\bibitem [{\citenamefont {Chudaykin}\ and\ \citenamefont {Kunz}(2024)}]{Chudaykin:2024gol}%
  \BibitemOpen
  \bibfield  {author} {\bibinfo {author} {\bibfnamefont {A.}~\bibnamefont {Chudaykin}}\ and\ \bibinfo {author} {\bibfnamefont {M.}~\bibnamefont {Kunz}},\ }\bibfield  {title} {\bibinfo {title} {{Modified gravity interpretation of the evolving dark energy in light of DESI data}},\ }\href {https://doi.org/10.1103/PhysRevD.110.123524} {\bibfield  {journal} {\bibinfo  {journal} {Phys. Rev. D}\ }\textbf {\bibinfo {volume} {110}},\ \bibinfo {pages} {123524} (\bibinfo {year} {2024})},\ \Eprint {https://arxiv.org/abs/2407.02558} {arXiv:2407.02558 [astro-ph.CO]} \BibitemShut {NoStop}%
\bibitem [{\citenamefont {Ye}\ \emph {et~al.}(2025)\citenamefont {Ye}, \citenamefont {Martinelli}, \citenamefont {Hu},\ and\ \citenamefont {Silvestri}}]{Ye:2024ywg}%
  \BibitemOpen
  \bibfield  {author} {\bibinfo {author} {\bibfnamefont {G.}~\bibnamefont {Ye}}, \bibinfo {author} {\bibfnamefont {M.}~\bibnamefont {Martinelli}}, \bibinfo {author} {\bibfnamefont {B.}~\bibnamefont {Hu}},\ and\ \bibinfo {author} {\bibfnamefont {A.}~\bibnamefont {Silvestri}},\ }\bibfield  {title} {\bibinfo {title} {{Hints of Nonminimally Coupled Gravity in DESI 2024 Baryon Acoustic Oscillation Measurements}},\ }\href {https://doi.org/10.1103/PhysRevLett.134.181002} {\bibfield  {journal} {\bibinfo  {journal} {Phys. Rev. Lett.}\ }\textbf {\bibinfo {volume} {134}},\ \bibinfo {pages} {181002} (\bibinfo {year} {2025})},\ \Eprint {https://arxiv.org/abs/2407.15832} {arXiv:2407.15832 [astro-ph.CO]} \BibitemShut {NoStop}%
\bibitem [{\citenamefont {Wolf}\ \emph {et~al.}(2025{\natexlab{a}})\citenamefont {Wolf}, \citenamefont {Ferreira},\ and\ \citenamefont {Garc{\'\i}a-Garc{\'\i}a}}]{Wolf:2024stt}%
  \BibitemOpen
  \bibfield  {author} {\bibinfo {author} {\bibfnamefont {W.~J.}\ \bibnamefont {Wolf}}, \bibinfo {author} {\bibfnamefont {P.~G.}\ \bibnamefont {Ferreira}},\ and\ \bibinfo {author} {\bibfnamefont {C.}~\bibnamefont {Garc{\'\i}a-Garc{\'\i}a}},\ }\bibfield  {title} {\bibinfo {title} {{Matching current observational constraints with nonminimally coupled dark energy}},\ }\href {https://doi.org/10.1103/PhysRevD.111.L041303} {\bibfield  {journal} {\bibinfo  {journal} {Phys. Rev. D}\ }\textbf {\bibinfo {volume} {111}},\ \bibinfo {pages} {L041303} (\bibinfo {year} {2025}{\natexlab{a}})},\ \Eprint {https://arxiv.org/abs/2409.17019} {arXiv:2409.17019 [astro-ph.CO]} \BibitemShut {NoStop}%
\bibitem [{\citenamefont {Wolf}\ \emph {et~al.}(2025{\natexlab{b}})\citenamefont {Wolf}, \citenamefont {Garc{\'\i}a-Garc{\'\i}a}, \citenamefont {Anton},\ and\ \citenamefont {Ferreira}}]{Wolf:2025jed}%
  \BibitemOpen
  \bibfield  {author} {\bibinfo {author} {\bibfnamefont {W.~J.}\ \bibnamefont {Wolf}}, \bibinfo {author} {\bibfnamefont {C.}~\bibnamefont {Garc{\'\i}a-Garc{\'\i}a}}, \bibinfo {author} {\bibfnamefont {T.}~\bibnamefont {Anton}},\ and\ \bibinfo {author} {\bibfnamefont {P.~G.}\ \bibnamefont {Ferreira}},\ }\bibfield  {title} {\bibinfo {title} {{Assessing Cosmological Evidence for Nonminimal Coupling}},\ }\href {https://doi.org/10.1103/jysf-k72m} {\bibfield  {journal} {\bibinfo  {journal} {Phys. Rev. Lett.}\ }\textbf {\bibinfo {volume} {135}},\ \bibinfo {pages} {081001} (\bibinfo {year} {2025}{\natexlab{b}})},\ \Eprint {https://arxiv.org/abs/2504.07679} {arXiv:2504.07679 [astro-ph.CO]} \BibitemShut {NoStop}%
\bibitem [{\citenamefont {Ishak}\ \emph {et~al.}(2025)\citenamefont {Ishak} \emph {et~al.}}]{Ishak:2024jhs}%
  \BibitemOpen
  \bibfield  {author} {\bibinfo {author} {\bibfnamefont {M.}~\bibnamefont {Ishak}} \emph {et~al.},\ }\bibfield  {title} {\bibinfo {title} {{Modified gravity constraints from the full shape modeling of clustering measurements from DESI 2024}},\ }\href {https://doi.org/10.1088/1475-7516/2025/09/053} {\bibfield  {journal} {\bibinfo  {journal} {JCAP}\ }\textbf {\bibinfo {volume} {09}},\ \bibinfo {pages} {053}},\ \Eprint {https://arxiv.org/abs/2411.12026} {arXiv:2411.12026 [astro-ph.CO]} \BibitemShut {NoStop}%
\bibitem [{\citenamefont {Postolak}(2025)}]{Postolak:2025qmv}%
  \BibitemOpen
  \bibfield  {author} {\bibinfo {author} {\bibfnamefont {M.}~\bibnamefont {Postolak}},\ }\bibfield  {title} {\bibinfo {title} {{Non-minimally coupled scalar field dark sector of the universe: in-depth (Einstein frame) case study}},\ }\Eprint {https://arxiv.org/abs/2505.07456} {arXiv:2505.07456 [gr-qc]}  (\bibinfo {year} {2025})\BibitemShut {NoStop}%
\bibitem [{\citenamefont {Sol{\`a}~Peracaula}\ \emph {et~al.}(2018)\citenamefont {Sol{\`a}~Peracaula}, \citenamefont {de~Cruz~P{\'e}rez},\ and\ \citenamefont {Gomez-Valent}}]{SolaPeracaula:2017esw}%
  \BibitemOpen
  \bibfield  {author} {\bibinfo {author} {\bibfnamefont {J.}~\bibnamefont {Sol{\`a}~Peracaula}}, \bibinfo {author} {\bibfnamefont {J.}~\bibnamefont {de~Cruz~P{\'e}rez}},\ and\ \bibinfo {author} {\bibfnamefont {A.}~\bibnamefont {Gomez-Valent}},\ }\bibfield  {title} {\bibinfo {title} {{Possible signals of vacuum dynamics in the Universe}},\ }\href {https://doi.org/10.1093/mnras/sty1253} {\bibfield  {journal} {\bibinfo  {journal} {Mon. Not. Roy. Astron. Soc.}\ }\textbf {\bibinfo {volume} {478}},\ \bibinfo {pages} {4357} (\bibinfo {year} {2018})},\ \Eprint {https://arxiv.org/abs/1703.08218} {arXiv:1703.08218 [astro-ph.CO]} \BibitemShut {NoStop}%
\bibitem [{\citenamefont {Sabogal}\ \emph {et~al.}(2024)\citenamefont {Sabogal}, \citenamefont {Silva}, \citenamefont {Nunes}, \citenamefont {Kumar}, \citenamefont {Di~Valentino},\ and\ \citenamefont {Giar{\`e}}}]{Sabogal:2024yha}%
  \BibitemOpen
  \bibfield  {author} {\bibinfo {author} {\bibfnamefont {M.~A.}\ \bibnamefont {Sabogal}}, \bibinfo {author} {\bibfnamefont {E.}~\bibnamefont {Silva}}, \bibinfo {author} {\bibfnamefont {R.~C.}\ \bibnamefont {Nunes}}, \bibinfo {author} {\bibfnamefont {S.}~\bibnamefont {Kumar}}, \bibinfo {author} {\bibfnamefont {E.}~\bibnamefont {Di~Valentino}},\ and\ \bibinfo {author} {\bibfnamefont {W.}~\bibnamefont {Giar{\`e}}},\ }\bibfield  {title} {\bibinfo {title} {{Quantifying the S8 tension and evidence for interacting dark energy from redshift-space distortion measurements}},\ }\href {https://doi.org/10.1103/PhysRevD.110.123508} {\bibfield  {journal} {\bibinfo  {journal} {Phys. Rev. D}\ }\textbf {\bibinfo {volume} {110}},\ \bibinfo {pages} {123508} (\bibinfo {year} {2024})},\ \Eprint {https://arxiv.org/abs/2408.12403} {arXiv:2408.12403 [astro-ph.CO]} \BibitemShut {NoStop}%
\bibitem [{\citenamefont {Li}\ \emph {et~al.}(2014)\citenamefont {Li}, \citenamefont {Zhang},\ and\ \citenamefont {Zhang}}]{Li:2014cee}%
  \BibitemOpen
  \bibfield  {author} {\bibinfo {author} {\bibfnamefont {Y.-H.}\ \bibnamefont {Li}}, \bibinfo {author} {\bibfnamefont {J.-F.}\ \bibnamefont {Zhang}},\ and\ \bibinfo {author} {\bibfnamefont {X.}~\bibnamefont {Zhang}},\ }\bibfield  {title} {\bibinfo {title} {{Exploring the full parameter space for an interacting dark energy model with recent observations including redshift-space distortions: Application of the parametrized post-Friedmann approach}},\ }\href {https://doi.org/10.1103/PhysRevD.90.123007} {\bibfield  {journal} {\bibinfo  {journal} {Phys. Rev. D}\ }\textbf {\bibinfo {volume} {90}},\ \bibinfo {pages} {123007} (\bibinfo {year} {2014})},\ \Eprint {https://arxiv.org/abs/1409.7205} {arXiv:1409.7205 [astro-ph.CO]} \BibitemShut {NoStop}%
\bibitem [{\citenamefont {Wei}\ and\ \citenamefont {Zhang}(2008)}]{Wei:2008vw}%
  \BibitemOpen
  \bibfield  {author} {\bibinfo {author} {\bibfnamefont {H.}~\bibnamefont {Wei}}\ and\ \bibinfo {author} {\bibfnamefont {S.~N.}\ \bibnamefont {Zhang}},\ }\bibfield  {title} {\bibinfo {title} {{How to Distinguish Dark Energy and Modified Gravity?}},\ }\href {https://doi.org/10.1103/PhysRevD.78.023011} {\bibfield  {journal} {\bibinfo  {journal} {Phys. Rev. D}\ }\textbf {\bibinfo {volume} {78}},\ \bibinfo {pages} {023011} (\bibinfo {year} {2008})},\ \Eprint {https://arxiv.org/abs/0803.3292} {arXiv:0803.3292 [astro-ph]} \BibitemShut {NoStop}%
\bibitem [{\citenamefont {Marcondes}\ \emph {et~al.}(2016)\citenamefont {Marcondes}, \citenamefont {Landim}, \citenamefont {Costa}, \citenamefont {Wang},\ and\ \citenamefont {Abdalla}}]{Marcondes:2016reb}%
  \BibitemOpen
  \bibfield  {author} {\bibinfo {author} {\bibfnamefont {R.~J.~F.}\ \bibnamefont {Marcondes}}, \bibinfo {author} {\bibfnamefont {R.~C.~G.}\ \bibnamefont {Landim}}, \bibinfo {author} {\bibfnamefont {A.~A.}\ \bibnamefont {Costa}}, \bibinfo {author} {\bibfnamefont {B.}~\bibnamefont {Wang}},\ and\ \bibinfo {author} {\bibfnamefont {E.}~\bibnamefont {Abdalla}},\ }\bibfield  {title} {\bibinfo {title} {{Analytic study of the effect of dark energy-dark matter interaction on the growth of structures}},\ }\href {https://doi.org/10.1088/1475-7516/2016/12/009} {\bibfield  {journal} {\bibinfo  {journal} {JCAP}\ }\textbf {\bibinfo {volume} {12}},\ \bibinfo {pages} {009}},\ \Eprint {https://arxiv.org/abs/1605.05264} {arXiv:1605.05264 [astro-ph.CO]} \BibitemShut {NoStop}%
\bibitem [{\citenamefont {Pooya}(2025)}]{Pooya:2025wyd}%
  \BibitemOpen
  \bibfield  {author} {\bibinfo {author} {\bibfnamefont {N.~N.}\ \bibnamefont {Pooya}},\ }\bibfield  {title} {\bibinfo {title} {{Constraints on interacting holographic dark energy models: Implications from background and perturbations data}},\ }\href {https://doi.org/10.1103/bxpb-tnpl} {\bibfield  {journal} {\bibinfo  {journal} {Phys. Rev. D}\ }\textbf {\bibinfo {volume} {112}},\ \bibinfo {pages} {103506} (\bibinfo {year} {2025})},\ \Eprint {https://arxiv.org/abs/2510.08875} {arXiv:2510.08875 [astro-ph.CO]} \BibitemShut {NoStop}%
\bibitem [{\citenamefont {Duniya}\ \emph {et~al.}(2013)\citenamefont {Duniya}, \citenamefont {Bertacca},\ and\ \citenamefont {Maartens}}]{Duniya:2013eta}%
  \BibitemOpen
  \bibfield  {author} {\bibinfo {author} {\bibfnamefont {D.}~\bibnamefont {Duniya}}, \bibinfo {author} {\bibfnamefont {D.}~\bibnamefont {Bertacca}},\ and\ \bibinfo {author} {\bibfnamefont {R.}~\bibnamefont {Maartens}},\ }\bibfield  {title} {\bibinfo {title} {{Clustering of quintessence on horizon scales and its imprint on HI intensity mapping}},\ }\href {https://doi.org/10.1088/1475-7516/2013/10/015} {\bibfield  {journal} {\bibinfo  {journal} {JCAP}\ }\textbf {\bibinfo {volume} {10}},\ \bibinfo {pages} {015}},\ \Eprint {https://arxiv.org/abs/1305.4509} {arXiv:1305.4509 [astro-ph.CO]} \BibitemShut {NoStop}%
\bibitem [{\citenamefont {Motta}\ \emph {et~al.}(2013)\citenamefont {Motta}, \citenamefont {Sawicki}, \citenamefont {Saltas}, \citenamefont {Amendola},\ and\ \citenamefont {Kunz}}]{Motta:2013cwa}%
  \BibitemOpen
  \bibfield  {author} {\bibinfo {author} {\bibfnamefont {M.}~\bibnamefont {Motta}}, \bibinfo {author} {\bibfnamefont {I.}~\bibnamefont {Sawicki}}, \bibinfo {author} {\bibfnamefont {I.~D.}\ \bibnamefont {Saltas}}, \bibinfo {author} {\bibfnamefont {L.}~\bibnamefont {Amendola}},\ and\ \bibinfo {author} {\bibfnamefont {M.}~\bibnamefont {Kunz}},\ }\bibfield  {title} {\bibinfo {title} {{Probing Dark Energy through Scale Dependence}},\ }\href {https://doi.org/10.1103/PhysRevD.88.124035} {\bibfield  {journal} {\bibinfo  {journal} {Phys. Rev. D}\ }\textbf {\bibinfo {volume} {88}},\ \bibinfo {pages} {124035} (\bibinfo {year} {2013})},\ \Eprint {https://arxiv.org/abs/1305.0008} {arXiv:1305.0008 [astro-ph.CO]} \BibitemShut {NoStop}%
\bibitem [{\citenamefont {Scolnic}\ \emph {et~al.}(2022)\citenamefont {Scolnic} \emph {et~al.}}]{Scolnic:2021amr}%
  \BibitemOpen
  \bibfield  {author} {\bibinfo {author} {\bibfnamefont {D.}~\bibnamefont {Scolnic}} \emph {et~al.},\ }\bibfield  {title} {\bibinfo {title} {{The Pantheon+ Analysis: The Full Data Set and Light-curve Release}},\ }\href {https://doi.org/10.3847/1538-4357/ac8b7a} {\bibfield  {journal} {\bibinfo  {journal} {Astrophys. J.}\ }\textbf {\bibinfo {volume} {938}},\ \bibinfo {pages} {113} (\bibinfo {year} {2022})},\ \Eprint {https://arxiv.org/abs/2112.03863} {arXiv:2112.03863 [astro-ph.CO]} \BibitemShut {NoStop}%
\bibitem [{\citenamefont {Brout}\ \emph {et~al.}(2022)\citenamefont {Brout} \emph {et~al.}}]{Brout:2022vxf}%
  \BibitemOpen
  \bibfield  {author} {\bibinfo {author} {\bibfnamefont {D.}~\bibnamefont {Brout}} \emph {et~al.},\ }\bibfield  {title} {\bibinfo {title} {{The Pantheon+ Analysis: Cosmological Constraints}},\ }\href {https://doi.org/10.3847/1538-4357/ac8e04} {\bibfield  {journal} {\bibinfo  {journal} {Astrophys. J.}\ }\textbf {\bibinfo {volume} {938}},\ \bibinfo {pages} {110} (\bibinfo {year} {2022})},\ \Eprint {https://arxiv.org/abs/2202.04077} {arXiv:2202.04077 [astro-ph.CO]} \BibitemShut {NoStop}%
\bibitem [{\citenamefont {Matthewson}\ and\ \citenamefont {Shafieloo}(2025)}]{Matthewson:2024ffb}%
  \BibitemOpen
  \bibfield  {author} {\bibinfo {author} {\bibfnamefont {W.~L.}\ \bibnamefont {Matthewson}}\ and\ \bibinfo {author} {\bibfnamefont {A.}~\bibnamefont {Shafieloo}},\ }\bibfield  {title} {\bibinfo {title} {{Star-crossed labours: checking consistency between current supernovae compilations}},\ }\href {https://doi.org/10.1088/1475-7516/2025/01/064} {\bibfield  {journal} {\bibinfo  {journal} {JCAP}\ }\textbf {\bibinfo {volume} {01}},\ \bibinfo {pages} {064}},\ \Eprint {https://arxiv.org/abs/2409.02550} {arXiv:2409.02550 [astro-ph.CO]} \BibitemShut {NoStop}%
\bibitem [{\citenamefont {G{\'o}mez-Valent}\ and\ \citenamefont {Amendola}(2018)}]{Gomez-Valent:2018hwc}%
  \BibitemOpen
  \bibfield  {author} {\bibinfo {author} {\bibfnamefont {A.}~\bibnamefont {G{\'o}mez-Valent}}\ and\ \bibinfo {author} {\bibfnamefont {L.}~\bibnamefont {Amendola}},\ }\bibfield  {title} {\bibinfo {title} {{$H_0$ from cosmic chronometers and Type Ia supernovae, with Gaussian Processes and the novel Weighted Polynomial Regression method}},\ }\href {https://doi.org/10.1088/1475-7516/2018/04/051} {\bibfield  {journal} {\bibinfo  {journal} {JCAP}\ }\textbf {\bibinfo {volume} {04}},\ \bibinfo {pages} {051}},\ \Eprint {https://arxiv.org/abs/1802.01505} {arXiv:1802.01505 [astro-ph.CO]} \BibitemShut {NoStop}%
\bibitem [{\citenamefont {Yang}\ \emph {et~al.}(2024)\citenamefont {Yang}, \citenamefont {Ren}, \citenamefont {Wang}, \citenamefont {Lu}, \citenamefont {Zhang}, \citenamefont {Cai},\ and\ \citenamefont {Saridakis}}]{Yang:2024kdo}%
  \BibitemOpen
  \bibfield  {author} {\bibinfo {author} {\bibfnamefont {Y.}~\bibnamefont {Yang}}, \bibinfo {author} {\bibfnamefont {X.}~\bibnamefont {Ren}}, \bibinfo {author} {\bibfnamefont {Q.}~\bibnamefont {Wang}}, \bibinfo {author} {\bibfnamefont {Z.}~\bibnamefont {Lu}}, \bibinfo {author} {\bibfnamefont {D.}~\bibnamefont {Zhang}}, \bibinfo {author} {\bibfnamefont {Y.-F.}\ \bibnamefont {Cai}},\ and\ \bibinfo {author} {\bibfnamefont {E.~N.}\ \bibnamefont {Saridakis}},\ }\bibfield  {title} {\bibinfo {title} {{Quintom cosmology and modified gravity after DESI 2024}},\ }\href {https://doi.org/10.1016/j.scib.2024.07.029} {\bibfield  {journal} {\bibinfo  {journal} {Sci. Bull.}\ }\textbf {\bibinfo {volume} {69}},\ \bibinfo {pages} {2698} (\bibinfo {year} {2024})},\ \Eprint {https://arxiv.org/abs/2404.19437} {arXiv:2404.19437 [astro-ph.CO]} \BibitemShut {NoStop}%
\bibitem [{\citenamefont {Avila}\ \emph {et~al.}(2022)\citenamefont {Avila}, \citenamefont {Bernui}, \citenamefont {Bonilla},\ and\ \citenamefont {Nunes}}]{Avila:2022xad}%
  \BibitemOpen
  \bibfield  {author} {\bibinfo {author} {\bibfnamefont {F.}~\bibnamefont {Avila}}, \bibinfo {author} {\bibfnamefont {A.}~\bibnamefont {Bernui}}, \bibinfo {author} {\bibfnamefont {A.}~\bibnamefont {Bonilla}},\ and\ \bibinfo {author} {\bibfnamefont {R.~C.}\ \bibnamefont {Nunes}},\ }\bibfield  {title} {\bibinfo {title} {{Inferring $S_8(z)$ and $\gamma (z)$ with cosmic growth rate measurements using machine learning}},\ }\href {https://doi.org/10.1140/epjc/s10052-022-10561-0} {\bibfield  {journal} {\bibinfo  {journal} {Eur. Phys. J. C}\ }\textbf {\bibinfo {volume} {82}},\ \bibinfo {pages} {594} (\bibinfo {year} {2022})},\ \Eprint {https://arxiv.org/abs/2201.07829} {arXiv:2201.07829 [astro-ph.CO]} \BibitemShut {NoStop}%
\bibitem [{\citenamefont {Seikel}\ \emph {et~al.}(2012)\citenamefont {Seikel}, \citenamefont {Clarkson},\ and\ \citenamefont {Smith}}]{Seikel:2012uu}%
  \BibitemOpen
  \bibfield  {author} {\bibinfo {author} {\bibfnamefont {M.}~\bibnamefont {Seikel}}, \bibinfo {author} {\bibfnamefont {C.}~\bibnamefont {Clarkson}},\ and\ \bibinfo {author} {\bibfnamefont {M.}~\bibnamefont {Smith}},\ }\bibfield  {title} {\bibinfo {title} {{Reconstruction of dark energy and expansion dynamics using Gaussian processes}},\ }\href {https://doi.org/10.1088/1475-7516/2012/06/036} {\bibfield  {journal} {\bibinfo  {journal} {JCAP}\ }\textbf {\bibinfo {volume} {06}},\ \bibinfo {pages} {036}},\ \Eprint {https://arxiv.org/abs/1204.2832} {arXiv:1204.2832 [astro-ph.CO]} \BibitemShut {NoStop}%
\bibitem [{\citenamefont {Shafieloo}\ \emph {et~al.}(2012)\citenamefont {Shafieloo}, \citenamefont {Kim},\ and\ \citenamefont {Linder}}]{Shafieloo:2012ht}%
  \BibitemOpen
  \bibfield  {author} {\bibinfo {author} {\bibfnamefont {A.}~\bibnamefont {Shafieloo}}, \bibinfo {author} {\bibfnamefont {A.~G.}\ \bibnamefont {Kim}},\ and\ \bibinfo {author} {\bibfnamefont {E.~V.}\ \bibnamefont {Linder}},\ }\bibfield  {title} {\bibinfo {title} {{Gaussian Process Cosmography}},\ }\href {https://doi.org/10.1103/PhysRevD.85.123530} {\bibfield  {journal} {\bibinfo  {journal} {Phys. Rev. D}\ }\textbf {\bibinfo {volume} {85}},\ \bibinfo {pages} {123530} (\bibinfo {year} {2012})},\ \Eprint {https://arxiv.org/abs/1204.2272} {arXiv:1204.2272 [astro-ph.CO]} \BibitemShut {NoStop}%
\bibitem [{\citenamefont {Yang}\ \emph {et~al.}(2025)\citenamefont {Yang}, \citenamefont {Wang}, \citenamefont {Li}, \citenamefont {Yuan}, \citenamefont {Ren}, \citenamefont {Saridakis},\ and\ \citenamefont {Cai}}]{Yang:2025kgc}%
  \BibitemOpen
  \bibfield  {author} {\bibinfo {author} {\bibfnamefont {Y.}~\bibnamefont {Yang}}, \bibinfo {author} {\bibfnamefont {Q.}~\bibnamefont {Wang}}, \bibinfo {author} {\bibfnamefont {C.}~\bibnamefont {Li}}, \bibinfo {author} {\bibfnamefont {P.}~\bibnamefont {Yuan}}, \bibinfo {author} {\bibfnamefont {X.}~\bibnamefont {Ren}}, \bibinfo {author} {\bibfnamefont {E.~N.}\ \bibnamefont {Saridakis}},\ and\ \bibinfo {author} {\bibfnamefont {Y.-F.}\ \bibnamefont {Cai}},\ }\bibfield  {title} {\bibinfo {title} {{Gaussian process reconstructions and model building of quintom dark energy from latest cosmological observations}},\ }\href {https://doi.org/10.1088/1475-7516/2025/08/050} {\bibfield  {journal} {\bibinfo  {journal} {JCAP}\ }\textbf {\bibinfo {volume} {08}},\ \bibinfo {pages} {050}},\ \Eprint {https://arxiv.org/abs/2501.18336} {arXiv:2501.18336 [astro-ph.CO]} \BibitemShut {NoStop}%
\bibitem [{\citenamefont {Seikel}\ and\ \citenamefont {Clarkson}(2013)}]{Seikel:2013fda}%
  \BibitemOpen
  \bibfield  {author} {\bibinfo {author} {\bibfnamefont {M.}~\bibnamefont {Seikel}}\ and\ \bibinfo {author} {\bibfnamefont {C.}~\bibnamefont {Clarkson}},\ }\bibfield  {title} {\bibinfo {title} {{Optimising Gaussian processes for reconstructing dark energy dynamics from supernovae}},\ }\Eprint {https://arxiv.org/abs/1311.6678} {arXiv:1311.6678 [astro-ph.CO]}  (\bibinfo {year} {2013})\BibitemShut {NoStop}%
\bibitem [{\citenamefont {Foreman-Mackey}\ \emph {et~al.}(2013)\citenamefont {Foreman-Mackey}, \citenamefont {Hogg}, \citenamefont {Lang},\ and\ \citenamefont {Goodman}}]{Foreman-Mackey:2012any}%
  \BibitemOpen
  \bibfield  {author} {\bibinfo {author} {\bibfnamefont {D.}~\bibnamefont {Foreman-Mackey}}, \bibinfo {author} {\bibfnamefont {D.~W.}\ \bibnamefont {Hogg}}, \bibinfo {author} {\bibfnamefont {D.}~\bibnamefont {Lang}},\ and\ \bibinfo {author} {\bibfnamefont {J.}~\bibnamefont {Goodman}},\ }\bibfield  {title} {\bibinfo {title} {{emcee: The MCMC Hammer}},\ }\href {https://doi.org/10.1086/670067} {\bibfield  {journal} {\bibinfo  {journal} {Publ. Astron. Soc. Pac.}\ }\textbf {\bibinfo {volume} {125}},\ \bibinfo {pages} {306} (\bibinfo {year} {2013})},\ \Eprint {https://arxiv.org/abs/1202.3665} {arXiv:1202.3665 [astro-ph.IM]} \BibitemShut {NoStop}%
\bibitem [{\citenamefont {Hwang}\ \emph {et~al.}(2023)\citenamefont {Hwang}, \citenamefont {L'Huillier}, \citenamefont {Keeley}, \citenamefont {Jee},\ and\ \citenamefont {Shafieloo}}]{Hwang:2022hla}%
  \BibitemOpen
  \bibfield  {author} {\bibinfo {author} {\bibfnamefont {S.-g.}\ \bibnamefont {Hwang}}, \bibinfo {author} {\bibfnamefont {B.}~\bibnamefont {L'Huillier}}, \bibinfo {author} {\bibfnamefont {R.~E.}\ \bibnamefont {Keeley}}, \bibinfo {author} {\bibfnamefont {M.~J.}\ \bibnamefont {Jee}},\ and\ \bibinfo {author} {\bibfnamefont {A.}~\bibnamefont {Shafieloo}},\ }\bibfield  {title} {\bibinfo {title} {{How to use GP: effects of the mean function and hyperparameter selection on Gaussian process regression}},\ }\href {https://doi.org/10.1088/1475-7516/2023/02/014} {\bibfield  {journal} {\bibinfo  {journal} {JCAP}\ }\textbf {\bibinfo {volume} {02}},\ \bibinfo {pages} {014}},\ \Eprint {https://arxiv.org/abs/2206.15081} {arXiv:2206.15081 [astro-ph.CO]} \BibitemShut {NoStop}%
\bibitem [{\citenamefont {Wang}\ \emph {et~al.}(2020)\citenamefont {Wang}, \citenamefont {Ma}, \citenamefont {Li},\ and\ \citenamefont {Xia}}]{Wang:2019vxv}%
  \BibitemOpen
  \bibfield  {author} {\bibinfo {author} {\bibfnamefont {G.-J.}\ \bibnamefont {Wang}}, \bibinfo {author} {\bibfnamefont {X.-J.}\ \bibnamefont {Ma}}, \bibinfo {author} {\bibfnamefont {S.-Y.}\ \bibnamefont {Li}},\ and\ \bibinfo {author} {\bibfnamefont {J.-Q.}\ \bibnamefont {Xia}},\ }\bibfield  {title} {\bibinfo {title} {{Reconstructing Functions and Estimating Parameters with Artificial Neural Networks: A Test with a Hubble Parameter and SNe Ia}},\ }\href {https://doi.org/10.3847/1538-4365/ab620b} {\bibfield  {journal} {\bibinfo  {journal} {Astrophys. J. Suppl.}\ }\textbf {\bibinfo {volume} {246}},\ \bibinfo {pages} {13} (\bibinfo {year} {2020})},\ \Eprint {https://arxiv.org/abs/1910.03636} {arXiv:1910.03636 [astro-ph.CO]} \BibitemShut {NoStop}%
\bibitem [{\citenamefont {Dialektopoulos}\ \emph {et~al.}(2022)\citenamefont {Dialektopoulos}, \citenamefont {Said}, \citenamefont {Mifsud}, \citenamefont {Sultana},\ and\ \citenamefont {Adami}}]{Dialektopoulos:2021wde}%
  \BibitemOpen
  \bibfield  {author} {\bibinfo {author} {\bibfnamefont {K.}~\bibnamefont {Dialektopoulos}}, \bibinfo {author} {\bibfnamefont {J.~L.}\ \bibnamefont {Said}}, \bibinfo {author} {\bibfnamefont {J.}~\bibnamefont {Mifsud}}, \bibinfo {author} {\bibfnamefont {J.}~\bibnamefont {Sultana}},\ and\ \bibinfo {author} {\bibfnamefont {K.~Z.}\ \bibnamefont {Adami}},\ }\bibfield  {title} {\bibinfo {title} {{Neural network reconstruction of late-time cosmology and null tests}},\ }\href {https://doi.org/10.1088/1475-7516/2022/02/023} {\bibfield  {journal} {\bibinfo  {journal} {JCAP}\ }\textbf {\bibinfo {volume} {02}}\bibfield  {number} {\bibinfo  {number} { (02)},\ \bibinfo {pages} {023}},\ }\Eprint {https://arxiv.org/abs/2111.11462} {arXiv:2111.11462 [astro-ph.CO]} \BibitemShut {NoStop}%
\bibitem [{\citenamefont {Sousa-Neto}\ \emph {et~al.}(2025)\citenamefont {Sousa-Neto}, \citenamefont {Bengaly}, \citenamefont {Gonzalez},\ and\ \citenamefont {Alcaniz}}]{Sousa-Neto:2025gpj}%
  \BibitemOpen
  \bibfield  {author} {\bibinfo {author} {\bibfnamefont {A.}~\bibnamefont {Sousa-Neto}}, \bibinfo {author} {\bibfnamefont {C.}~\bibnamefont {Bengaly}}, \bibinfo {author} {\bibfnamefont {J.~E.}\ \bibnamefont {Gonzalez}},\ and\ \bibinfo {author} {\bibfnamefont {J.}~\bibnamefont {Alcaniz}},\ }\bibfield  {title} {\bibinfo {title} {{Symbolic regression analysis of dynamical dark energy with DESI-DR2 and SN data}},\ }\href {https://doi.org/10.1016/j.dark.2025.102108} {\bibfield  {journal} {\bibinfo  {journal} {Phys. Dark Univ.}\ }\textbf {\bibinfo {volume} {50}},\ \bibinfo {pages} {102108} (\bibinfo {year} {2025})},\ \Eprint {https://arxiv.org/abs/2502.10506} {arXiv:2502.10506 [astro-ph.CO]} \BibitemShut {NoStop}%
\bibitem [{\citenamefont {Koksbang}\ and\ \citenamefont {Heinesen}(2026)}]{Koksbang:2026wvh}%
  \BibitemOpen
  \bibfield  {author} {\bibinfo {author} {\bibfnamefont {S.~M.}\ \bibnamefont {Koksbang}}\ and\ \bibinfo {author} {\bibfnamefont {A.}~\bibnamefont {Heinesen}},\ }\bibfield  {title} {\bibinfo {title} {{Model-independent constraints on generalized FLRW consistency relations with bootstrap-based symbolic regression}},\ }\Eprint {https://arxiv.org/abs/2604.05822} {arXiv:2604.05822 [astro-ph.CO]}  (\bibinfo {year} {2026})\BibitemShut {NoStop}%
\bibitem [{\citenamefont {Escobal}\ \emph {et~al.}(2026)\citenamefont {Escobal}, \citenamefont {Macedo}, \citenamefont {Jesus}, \citenamefont {Nunes},\ and\ \citenamefont {Lima}}]{Escobal:2026lnp}%
  \BibitemOpen
  \bibfield  {author} {\bibinfo {author} {\bibfnamefont {A.~A.}\ \bibnamefont {Escobal}}, \bibinfo {author} {\bibfnamefont {H.~A.~P.}\ \bibnamefont {Macedo}}, \bibinfo {author} {\bibfnamefont {J.~F.}\ \bibnamefont {Jesus}}, \bibinfo {author} {\bibfnamefont {R.~C.}\ \bibnamefont {Nunes}},\ and\ \bibinfo {author} {\bibfnamefont {J.~A.~S.}\ \bibnamefont {Lima}},\ }\bibfield  {title} {\bibinfo {title} {{Linear Growth of Matter Perturbations Probed by Redshift-Space Distortions in Interacting $\Lambda(t)$CDM Cosmologies}},\ }\Eprint {https://arxiv.org/abs/2602.11310} {arXiv:2602.11310 [astro-ph.CO]}  (\bibinfo {year} {2026})\BibitemShut {NoStop}%
\bibitem [{\citenamefont {Barua}\ and\ \citenamefont {Desai}(2025)}]{Barua:2024gei}%
  \BibitemOpen
  \bibfield  {author} {\bibinfo {author} {\bibfnamefont {S.}~\bibnamefont {Barua}}\ and\ \bibinfo {author} {\bibfnamefont {S.}~\bibnamefont {Desai}},\ }\bibfield  {title} {\bibinfo {title} {{Effect of peak absolute magnitude of Type Ia supernovae and sound horizon values on the Hubble constant using DESI Data Release 1 results}},\ }\href {https://doi.org/10.1140/epjc/s10052-025-14219-5} {\bibfield  {journal} {\bibinfo  {journal} {Eur. Phys. J. C}\ }\textbf {\bibinfo {volume} {85}},\ \bibinfo {pages} {470} (\bibinfo {year} {2025})},\ \Eprint {https://arxiv.org/abs/2412.19240} {arXiv:2412.19240 [astro-ph.CO]} \BibitemShut {NoStop}%
\bibitem [{\citenamefont {Benisty}\ \emph {et~al.}(2023)\citenamefont {Benisty}, \citenamefont {Mifsud}, \citenamefont {Levi~Said},\ and\ \citenamefont {Staicova}}]{Benisty:2022psx}%
  \BibitemOpen
  \bibfield  {author} {\bibinfo {author} {\bibfnamefont {D.}~\bibnamefont {Benisty}}, \bibinfo {author} {\bibfnamefont {J.}~\bibnamefont {Mifsud}}, \bibinfo {author} {\bibfnamefont {J.}~\bibnamefont {Levi~Said}},\ and\ \bibinfo {author} {\bibfnamefont {D.}~\bibnamefont {Staicova}},\ }\bibfield  {title} {\bibinfo {title} {{On the robustness of the constancy of the Supernova absolute magnitude: Non-parametric reconstruction {\&} Bayesian approaches}},\ }\href {https://doi.org/10.1016/j.dark.2022.101160} {\bibfield  {journal} {\bibinfo  {journal} {Phys. Dark Univ.}\ }\textbf {\bibinfo {volume} {39}},\ \bibinfo {pages} {101160} (\bibinfo {year} {2023})},\ \Eprint {https://arxiv.org/abs/2202.04677} {arXiv:2202.04677 [astro-ph.CO]} \BibitemShut {NoStop}%
\bibitem [{\citenamefont {Nesseris}\ \emph {et~al.}(2017)\citenamefont {Nesseris}, \citenamefont {Pantazis},\ and\ \citenamefont {Perivolaropoulos}}]{Nesseris:2017vor}%
  \BibitemOpen
  \bibfield  {author} {\bibinfo {author} {\bibfnamefont {S.}~\bibnamefont {Nesseris}}, \bibinfo {author} {\bibfnamefont {G.}~\bibnamefont {Pantazis}},\ and\ \bibinfo {author} {\bibfnamefont {L.}~\bibnamefont {Perivolaropoulos}},\ }\bibfield  {title} {\bibinfo {title} {{Tension and constraints on modified gravity parametrizations of $G_{\textrm{eff}}(z)$ from growth rate and Planck data}},\ }\href {https://doi.org/10.1103/PhysRevD.96.023542} {\bibfield  {journal} {\bibinfo  {journal} {Phys. Rev. D}\ }\textbf {\bibinfo {volume} {96}},\ \bibinfo {pages} {023542} (\bibinfo {year} {2017})},\ \Eprint {https://arxiv.org/abs/1703.10538} {arXiv:1703.10538 [astro-ph.CO]} \BibitemShut {NoStop}%
\bibitem [{\citenamefont {Skara}\ and\ \citenamefont {Perivolaropoulos}(2020)}]{Skara:2019usd}%
  \BibitemOpen
  \bibfield  {author} {\bibinfo {author} {\bibfnamefont {F.}~\bibnamefont {Skara}}\ and\ \bibinfo {author} {\bibfnamefont {L.}~\bibnamefont {Perivolaropoulos}},\ }\bibfield  {title} {\bibinfo {title} {{Tension of the $E_G$ statistic and redshift space distortion data with the Planck - $\Lambda CDM$ model and implications for weakening gravity}},\ }\href {https://doi.org/10.1103/PhysRevD.101.063521} {\bibfield  {journal} {\bibinfo  {journal} {Phys. Rev. D}\ }\textbf {\bibinfo {volume} {101}},\ \bibinfo {pages} {063521} (\bibinfo {year} {2020})},\ \Eprint {https://arxiv.org/abs/1911.10609} {arXiv:1911.10609 [astro-ph.CO]} \BibitemShut {NoStop}%
\end{thebibliography}%

\appendix
\onecolumngrid 
\clearpage

\makeatletter
\setcounter{secnumdepth}{2}

\renewcommand{\thesubsection}{\Roman{subsection}}

\def\section{\@startsection{section}{1}{\z@}
    {0.8cm}{0.5cm} 
    {\bfseries\centering\selectfont}}
\makeatother

\section*{Supplemental Material}

\subsection{Matter Perturbation Equation for RSD with Interacting Dark Sector}\label{sec:modified equation}

We conduct our analysis within the framework of linear perturbation theory in the Newtonian gauge. The spacetime line element is expressed in terms of the cosmic physical time $t$:
\begin{equation}\label{eq:metric}
    ds^2 = -(1+2\Psi)dt^2 + a^2(t)(1-2\Phi)\delta_{ij}dx^i dx^j
\end{equation}
where $\Psi$ and $\Phi$ denote the gravitational potential and spatial curvature perturbations, respectively.

In our work, we consider an Interacting Dark Energy (IDE) framework. The baryons do not participate in the dark sector interactions and thus obey the standard conservation laws:
\begin{equation}\label{eq:baryon_con}
    \dot{\delta}_b = -\theta_b
\end{equation}
\begin{equation}\label{eq:baryon_euler}
    \dot{\theta}_b = -2H\theta_b + \frac{k^2}{a^2}\Psi
\end{equation}
where $\delta_x \equiv \delta\rho_x/\bar{\rho}_x$ is the density contrast and $\theta_x \equiv \frac{1}{a}\partial_i v^i_x$ represents the velocity divergence, where $v^i = a\frac{dx^i}{dt}$ is the peculiar velocity. 

Crucially, we treat dark matter as a pressureless fluid, similar to baryons, which implies $p_{dm}=\delta p_{dm}=0$. We assume a vanishing spatial momentum transfer in the dark matter rest frame and an adiabatic initial condition with no primordial velocity bias between dark matter and baryons, which ensures $\theta_b \equiv \theta_{dm}$ at all subsequent times. This condition is strictly required when integrating our model with Redshift-Space Distortion (RSD) measurements~\citep{Motta:2013cwa}. Under this assumption, the Euler equation for dark matter perturbations retains its standard form:
\begin{equation}\label{eq:cdm_euler}
    \dot{\theta}_{dm} = -2H\theta_{dm} + \frac{k^2}{a^2}\Psi.
\end{equation}

The condition of zero spatial momentum transfer in the dark matter rest frame dictates that the energy-momentum transfer four-vector is strictly proportional to the dark matter four-velocity, expressed as $Q^\nu = Q u^\nu_{dm}$, where $Q$ denotes the energy transfer scalar. Under this kinematic restriction, the modified continuity equation for the dark matter density perturbation takes the form \citep{Marcondes:2016reb}:
\begin{equation}\label{eq:cdm_con}
    \dot{\delta}_{dm} = -\theta_{dm} - \frac{\bar{Q}}{\bar{\rho}_{dm}}\delta_{dm} + \frac{\delta Q}{\bar{\rho}_{dm}} \ .
\end{equation}
where $\bar{Q}$ is the background energy transfer, and $\delta Q$ is the local interaction perturbation with $Q=\bar{Q}+\delta Q$. And $\bar{\rho}_{dm}$ obeys the background continuity equation described in Eq.(\ref{eq:rhodm}).

We define the effective macroscopic variables for the total matter ($m$) as the density-weighted linear combinations of the individual components:
\begin{align}
    \bar{\rho}_m &= \bar{\rho}_{dm} + \bar{\rho}_b \label{eq:rhom} \\
    \bar{\rho}_m \delta_m &= \bar{\rho}_{dm}\delta_{dm} + \bar{\rho}_b\delta_b \label{eq:deltam} \\
    \bar{\rho}_m \theta_m &= \bar{\rho}_{dm}\theta_{dm} + \bar{\rho}_b\theta_b \label{eq:thetam}
\end{align}

Based on the background evolution of both components described in Eqs.(\ref{eq:rhodm}-\ref{eq:rhof}), we obtain:
\begin{equation}\label{eq:rhom_dot}
    \dot{\bar{\rho}}_m+3H\bar{\rho}_m = \bar{Q} = \frac{\dot{\rho}_f}{a^3}
\end{equation}
where $\bar{\rho}_m=\rho_f/a^3$.

Differentiating Eq.(\ref{eq:deltam}) with respect to the cosmic time $t$ yields:
\begin{equation}\label{eq:dt_deltam}
    \frac{d}{dt}(\bar{\rho}_m \delta_m) = \dot{\bar{\rho}}_m \delta_m + \bar{\rho}_m \dot{\delta}_m = (-3H\bar{\rho}_m + \bar{Q})\delta_m + \bar{\rho}_m \dot{\delta}_m
\end{equation}

Expanding the derivative using the component-level equations, we find:
\begin{align}\label{eq:expand_deltam}
    \frac{d}{dt}(\bar{\rho}_{dm}\delta_{dm} + \bar{\rho}_b\delta_b) &= \dot{\bar{\rho}}_{dm}\delta_{dm} + \bar{\rho}_{dm}\dot{\delta}_{dm} + \dot{\bar{\rho}}_b\delta_b + \bar{\rho}_b\dot{\delta}_b \nonumber \\
    &= (-3H\bar{\rho}_{dm} + \bar{Q})\delta_{dm} + \bar{\rho}_{dm}\left(-\theta_{dm} - \frac{\bar{Q}}{\bar{\rho}_{dm}}\delta_{dm}+\frac{\delta Q}{\bar{\rho}_{dm}}\right) - 3H\bar{\rho}_b\delta_b - \bar{\rho}_b\theta_b \nonumber \\
    &= -3H(\bar{\rho}_{dm}\delta_{dm} + \bar{\rho}_b\delta_b) - (\bar{\rho}_{dm}\theta_{dm} + \bar{\rho}_b\theta_b) + (\bar{Q}\delta_{dm} - \bar{Q}\delta_{dm})+\delta Q \nonumber \\
    &= -3H\bar{\rho}_m\delta_m - \bar{\rho}_m\theta_m +\delta Q
\end{align}

Equating this result with Eq.(\ref{eq:dt_deltam}) and dividing by $\bar{\rho}_m$, we establish the \textbf{total matter continuity equation}:
\begin{equation}\label{eq:total_con}
    \dot{\delta}_m = -\theta_m - \frac{\bar{Q}}{\bar{\rho}_m}\delta_m+\frac{\delta Q}{\bar{\rho}_m}
\end{equation}

Given $\theta_b=\theta_{dm}$, Eq.(\ref{eq:thetam}) directly implies that $\theta_m=\theta_b=\theta_{dm}$. Therefore, the \textbf{total matter Euler equation} preserves the standard form:
\begin{equation}\label{eq:total_euler}
    \dot{\theta}_m = -2H\theta_m + \frac{k^2}{a^2}\Psi
\end{equation}

To unify the mathematical treatment of different interaction closures, we introduce an effective perturbative interaction parameter $\Xi(t)$:
\begin{equation}
    \Xi(t) \equiv \frac{\bar{Q}}{\bar{\rho}_m} - \frac{\delta Q}{\bar{\rho}_m \delta_m} \ .
\end{equation}

This allows us to compact the total matter continuity equation into a generalized form:
\begin{equation}\label{eq:theta_isolate}
    \theta_m = -\dot{\delta}_m - \Xi\delta_m \ .
\end{equation}

Taking the time derivative of Eq.(\ref{eq:theta_isolate}):
\begin{equation}\label{eq:dt_theta}
    \dot{\theta}_m = -\ddot{\delta}_m - \dot{\Xi}\delta_m - \Xi\dot{\delta}_m
\end{equation}

Substituting Eq.(\ref{eq:theta_isolate}) and Eq.(\ref{eq:dt_theta}) into the total Euler equation Eq.(\ref{eq:total_euler}), we obtain:
\begin{equation}\label{eq:combined_mid}
    -\ddot{\delta}_m - \dot{\Xi}\delta_m - \Xi\dot{\delta}_m = -2H(-\dot{\delta}_m - \Xi\delta_m) + \frac{k^2}{a^2}\Psi
\end{equation}

Rearranging the terms yields:
\begin{equation}\label{eq:ode_step2}
    \ddot{\delta}_m + (2H + \Xi)\dot{\delta}_m + (\dot{\Xi} + 2H\Xi)\delta_m = -\frac{k^2}{a^2}\Psi
\end{equation}

By invoking the cosmological Poisson equation in the sub-horizon limit ($k \gg aH$) and neglecting the dark energy clustering ($\delta_{de}\approx 0$)~\citep{Marcondes:2016reb,Duniya:2013eta,Sabogal:2024yha}, we apply the relation:
\begin{equation}\label{eq:poisson}
    \frac{k^2}{a^2}\Psi = -4\pi G \bar{\rho}_m \delta_m=-\frac{1}{2}\bar{\rho}_m \delta_m
\end{equation}
where we take natural units with $M_{\rm pl}=1$.

Substituting Eq.(\ref{eq:poisson}) into Eq.(\ref{eq:ode_step2}), we arrive at the final modified second-order perturbation equation for the total matter density contrast under the IDE framework:
\begin{equation}\label{eq:ode_final}
    \ddot{\delta}_m + (2H + \Xi)\dot{\delta}_m + (\dot{\Xi} + 2H\Xi - \frac{1}{2} \bar{\rho}_m)\delta_m = 0
\end{equation}

\subsection{Interaction Closures and Reconstruction Parameters} \label{sec:closures}

In the linear regime, the matter density contrast $\delta_m$ evolves according to the continuity equation. 
In standard $\Lambda$CDM, the velocity divergence $\theta_m \equiv \frac{1}{a} \nabla \cdot \vec{v}$ satisfies $\theta_m = -f H \delta_m$, with $f \equiv d \ln \delta_m / d \ln a$.
When a non-zero interaction rate $\Xi(t)$ between dark matter and dark energy is introduced, the continuity equation is generalized to:
\begin{equation}
    \dot{\delta}_m + \theta_m = -\Xi \delta_m \ .
\end{equation}

To properly interpret the Redshift-Space Distortion (RSD) measurements within this interacting framework, we define an effective growth rate:
\begin{equation}
    f_{\text{eff}} = \frac{\dot{\delta}_m}{H\delta_m} + \frac{\Xi}{H} = f_{\delta} + \frac{\Xi}{H} \ .
\end{equation}
In this work, we investigate two distinct physical closures for the dark sector interaction, each leading to different observable consequences.

\subsubsection{Closure I: Interaction Proportional to Dark Matter Density}

In the first scenario, we assume that the energy transfer is strictly proportional to the local dark matter density, yielding $Q \propto \rho_{dm}$ and $\delta Q = \bar{Q}\delta_{dm}$. Hence, the modified continuity equation for dark matter reduces to the standard form, identical to that of baryons (see Eqs. (\ref{eq:baryon_con}) and (\ref{eq:cdm_con})). By further adopting adiabatic initial conditions ($\delta_{dm,i} = \delta_{b,i}$) and assuming no primordial velocity bias, the density perturbations of dark matter and baryons evolve synchronously. Therefore, the equivalence $\delta_{dm} \approx \delta_b \approx \delta_m$ holds dynamically in the linear regime. 
Under this local matter-proportional transfer assumption, the effective perturbative source term perfectly cancels out ($\Xi(t) = 0$). Consequently, the matter density perturbation equation reverts to its standard-form growth equation:
\begin{equation}
    \ddot{\delta}_m + 2H\dot{\delta}_m = \frac{1}{2}\bar{\rho}_m\delta_m \ .
\end{equation}
In this scenario, the interaction manifests solely in the background continuity equation, thereby altering the expansion history $H(z)$ without directly sourcing the local clustering.

Because the velocity divergence remains kinematically coupled to the density contrast in the standard way, the effective growth rate reduces to the standard definition, $f_{\text{eff}} = f_{\delta}$. For reference, the standard matter density growth rate $f_\delta(z)$ and the RSD observable $F(z)$ are defined as:
\begin{eqnarray}
    f_\delta(z) &=& \frac{d\ln\delta_m}{d\ln a} = -(1+z)\frac{\delta_m'}{\delta_m} \ , \\
    F(z) \equiv f\sigma_8(z) &=& -(1+z)\frac{\sigma_{8,0}}{\delta_{m,0}}\delta_m' \ . \label{eq:F_def}
\end{eqnarray}
Here, a prime denotes a derivative with respect to redshift $z$, and $\sigma_8(z) = \sigma_{8,0}\frac{\delta_m(z)}{\delta_{m,0}}$ is the fluctuation amplitude defined on spheres of radius $8h^{-1}{\rm Mpc}$.

In the reconstruction, we define the normalized growth function
\begin{equation}
    \mathcal{D}(z) \equiv \frac{\delta_m(z)}{\delta_{m,0}} \ , \qquad \mathcal{D}(0)=1 \ .
\end{equation}
Because the perturbation equation is linear in $\delta_m$, the absolute normalization of $\delta_m$ is arbitrary, meaning $\mathcal{D}(z)$ can be used equivalently. From Eq.~(\ref{eq:F_def}), we obtain:
\begin{eqnarray}
    \mathcal{D}'(z) &=& -\frac{1}{\sigma_{8,0}}\frac{F(z)}{1+z} \ , \\
    \mathcal{D}(z) &=& 1 - \frac{1}{\sigma_{8,0}} \int^z_0 \frac{F(\tilde{z})}{1+\tilde{z}} d\tilde{z} \ .
\end{eqnarray}

Replacing $\delta_m$ with $\mathcal{D}$ allows us to express the interaction variable directly in terms of the integrated observables:
\begin{equation}
    \rho_f(z) = \frac{2a^3}{ \mathcal{D}}(\ddot{\mathcal{D}} + 2H\dot{\mathcal{D}}) \ ,
\end{equation}

\begin{equation}
    \bar{Q}=\frac{\dot{\rho}_f}{a^3}=2 \left(\frac{\dddot{\mathcal{D}}+5H\ddot{\mathcal{D}}+2\dot{H}\dot{\mathcal{D}}+6H^2\dot{\mathcal{D}}}{\mathcal{D}}-\frac{\dot{\mathcal{D}}\ddot{\mathcal{D}}+2H(\dot{\mathcal{D}})^2}{\mathcal{D}^2}\right)
\end{equation}

In this closure, a non-interacting $\Lambda$CDM universe corresponds strictly to  $\bar{Q}\equiv 0$. Therefore, any reconstructed temporal deviation from those serves as a direct, model-independent signature of a non-zero dark sector interaction. The reconstruction of $\rho_f$ also permits us to break the dark degeneracy and reconstruct the dark energy equation of state via:
\begin{equation}
    w_{de} = \frac{-2\dot H-3H^2}{3H^2-\rho_f/a^3} \ .
\end{equation}

\subsubsection{Closure II: Interaction Proportional to Dark Energy Density}

The second scenario assumes that the energy transfer is entirely controlled by the smooth, homogeneous dark energy background. Physically, because dark energy does not cluster significantly on sub-horizon scales ($\delta_{de} \approx 0$), the absolute rate of energy transfer is spatially uniform, which dictates that the local perturbation of the interaction term vanishes ($\delta Q \sim 0$). Under this smooth-source assumption, the effective source parameter reduces strictly to the background interaction rate: $\Xi(t) = \bar{Q}/\bar{\rho}_m \equiv \Gamma(t)$. 

Consequently, the continuity equation is explicitly modified to $\dot{\delta}_m + \theta_m = -\Gamma\delta_m$. To correctly interpret RSD data in this regime, the effective growth rate must absorb this background transfer: $f_{\text{eff}} = f_\delta + \Gamma/H$. 

For analytical convenience in reconstructing this highly coupled system, we set $D(z) \equiv \sigma_8(z) = \sigma_{8,0}\frac{\delta_m(z)}{\delta_{m,0}}$, which bypasses the requirement for an explicit $\sigma_{8,0}$ prior. We also introduce a normalized, dimensionless variable $y \equiv \rho_f/\rho_0$, which allows us to rewrite the interaction rate kinematically as $\Gamma(t) = \dot{y}/y$, where $\rho_0$ is the normalization parameter. 

The primary observable $L$ from RSD surveys is then formulated as the product of the effective growth rate and the fluctuation amplitude:
\begin{equation}
    L \equiv f_{\text{eff}} \sigma_8(z) = \left( \frac{\dot{D}}{HD} + \frac{\dot{y}}{Hy} \right) D \ .
\end{equation}

By distributing the amplitude $D$ into the brackets, we arrive at the following fundamental identity:

\begin{equation}\label{eq:L_identity}
L = \frac{\dot{D}y + D\dot{y}}{Hy} = \frac{\dot{(Dy)}}{Hy}.
\end{equation}

For algebraic convenience, we introduce the composite variable $u \equiv Dy$. 
Under the IDE framework, the evolution of $D(t)$ satisfies Eq.~(\ref{eq:ode_final})~\citep{Escobal:2026lnp}:

\begin{equation}\label{eq:D_ODE}
\ddot{D} + A(t)\dot{D} - B(t)D = 0 \ ,
\end{equation}
where the coefficients are given by:
\begin{equation}
A(t) = 2H + \frac{\dot{y}}{y} \ , \quad B(t) = \frac{\rho_0 y}{2a^3} - \frac{\ddot{y}}{y} - 2H\frac{\dot{y}}{y} + \left(\frac{\dot{y}}{y}\right)^2 \ .
\end{equation}

The second time derivative of $u$ expands as $\ddot{u} = \ddot{D}y + 2\dot{D}\dot{y} + D\ddot{y}$. 
Substituting $\ddot{D}$ from Eq.~\eqref{eq:D_ODE}, we obtain:

\begin{align}
\ddot{u} &= \left[ -\left( 2H + \frac{\dot{y}}{y} \right)\dot{D} + \left( \frac{\rho_0 y}{2a^3} - \frac{\ddot{y}}{y} - 2H\frac{\dot{y}}{y} + \left(\frac{\dot{y}}{y}\right)^2 \right)D \right] y + 2\dot{D}\dot{y} + D\ddot{y} \nonumber \\
&= -2H\dot{D}y - \dot{D}\dot{y} + \frac{\rho_0 y}{2a^3}u - D\ddot{y} - 2HD\dot{y} + D\frac{\dot{y}^2}{y} + 2\dot{D}\dot{y} + D\ddot{y} \nonumber \\
&= -2H\dot{D}y + \dot{D}\dot{y} - 2HD\dot{y} + D\frac{\dot{y}^2}{y} + \frac{\rho_0 y}{2a^3}u \ .
\end{align}

To express this entirely in terms of $u$ and $\dot{u}$, we substitute the relation $\dot{D}y = \dot{u} - D\dot{y}$:
\begin{align}
\ddot{u} &= -2H(\dot{u} - D\dot{y}) + (\dot{u} - D\dot{y})\frac{\dot{y}}{y} - 2HD\dot{y} + D\frac{\dot{y}^2}{y} + \frac{\rho_0 y}{2a^3}u \nonumber \\
&= -2H\dot{u} + 2HD\dot{y} + \frac{\dot{y}}{y}\dot{u} - D\frac{\dot{y}^2}{y} - 2HD\dot{y} + D\frac{\dot{y}^2}{y} + \frac{\rho_0 y}{2a^3}u \ .
\end{align}

Remarkably, all non-linear kinematic terms involving $D\dot{y}$ and $D\dot{y}^2/y$ cancel out exactly. This mathematical simplification collapses the expression into a clean second-order differential equation for $u$:
\begin{equation}\label{eq:u_clean}
\ddot{u} + \left( 2H - \frac{\dot{y}}{y} \right)\dot{u} - \frac{\rho_0 y}{2a^3} u = 0 \ .
\end{equation}

To determine the interaction variable $y$, we differentiate the fundamental identity $\dot{u} = LHy$ with respect to cosmic time:
\begin{equation}\label{eq:u_ddot_L}
\ddot{u} = (\dot{L}H + L\dot{H})y + LH\dot{y} \ .
\end{equation}

Substituting $\dot{u}$ and $\ddot{u}$ back into Eq.~\eqref{eq:u_clean} yields:
\begin{equation}
(\dot{L}H + L\dot{H})y + LH\dot{y} + \left( 2H - \frac{\dot{y}}{y} \right)LHy - \frac{\rho_0 y}{2a^3} u = 0 \ .
\end{equation}

Defining the observable composite variable $K(t) \equiv \dot{L}H + L\dot{H} + 2H^2 L$, the equation reduces to:
\begin{equation}
K(t)y - \frac{\rho_0 y}{2a^3}u = 0 \ .
\end{equation}

Since $y \neq 0$, we divide by $y$ to obtain an explicit, purely algebraic relation for $u(t)$:
\begin{equation}
u = \frac{2a^3}{\rho_0} K(t) \ .
\end{equation}

To find the exact solution for the normalized comoving matter-density variable $y$, we enforce the relation $\frac{du}{dt} = \dot{u} = LHy$. Taking the time derivative of $u(t)$ and utilizing $\dot{a} = aH$:
\begin{equation}
\dot{u} = \frac{2}{\rho_0} \left( 3a^2 \dot{a} K(t) + a^3 \dot{K}(t) \right) = \frac{2a^3}{\rho_0} \left( 3HK + \dot{K} \right) \ .
\end{equation}

Equating this to $LHy$ and isolating $y$, we find that the interaction variable $y$ is analytically determined strictly by the background and perturbation observables. We are also able to recover the $\rho_f$ form without normalization:
\begin{equation}
    \rho_f(t) = \frac{2a^3}{LH} \left( \dot{K} + 3HK \right) \ .\label{eq:closure_two_y}
\end{equation}

To recover the explicit physical quantities characterizing the dark sector, we map the normalized variable back to its standard form. For notational simplicity, we introduce a new composite variable $M(t) \equiv \dot{K} + 3HK$. The total matter energy density can then be expressed as:
\begin{equation}
    \rho_m = \frac{\rho_f}{a^3} = \frac{2M}{LH} \ .
\end{equation}

The physical energy exchange rate $Q$ is obtained by taking the cosmic time derivative of $\rho_f$:
\begin{equation}
    Q = \bar{Q}=\frac{\dot{\rho}_f}{a^3} = \frac{2}{L^2H^2} \left[ \dot{M}LH + M(3H^2L - \dot{L}H - L\dot{H}) \right] \ . \label{eq:de_Q_closure_two}
\end{equation}

Furthermore, the reconstructed dark energy equation of state $w_{de}$ is given by:
\begin{equation}
    w_{de} = \frac{-2\dot{H} - 3H^2}{3H^2 - 2M/(LH)} \ .\label{eq:closure_two_w}
\end{equation}

This exact analytical solution provides a remarkably powerful theoretical tool, allowing us to reconstruct the dark energy equation of state and the dark sector interaction history directly from background expansion and structure growth measurements, without relying on a priori parameterizations.

\subsection{Gaussian Process for combined dataset analysis}
\label{sec:Gaussian Process}

Gaussian Process provides a framework for reconstructing a function and its derivatives from observational data, whether the data comprise only the function values or are combined with its first-order derivatives, enabling a joint analysis of different datasets.

For the \textbf{CC} and \textbf{BAO} datasets, the Hubble parameter $H(z)$ and its derivatives can be directly reconstructed using GP. However, only the comoving distance $D_M(z)$ can be reconstructed from the \textbf{PP} sample via:
\begin{equation}
    m = M_B + 5\log_{10}\!\left[(1+z)D_M(z)\right] + 25 \, .
\end{equation}

Fortunately, by applying the relation:
\begin{equation}
    H(z) = \frac{c}{D'_M(z)} \, ,\label{H to dD}
\end{equation}
we are able to derive the distribution of $H(z)$ and related derivatives from the reconstructed $D_M(z)$ and its derivatives. Here, the prime denotes the redshift derivative. 

GP allows for a joint analysis of different datasets, such as the combination of \textbf{PP}+\textbf{BAO} datasets. To this end, we incorporate additional $D_M(z)$ data from \textbf{BAO} into the \textbf{PP} sample, forming a combined input for $D_M(z)$. The $H(z)$ values obtained from \textbf{BAO} are converted into $D'_M(z)$ via Eq.~(\ref{H to dD}), which then serve as the input for the first derivative data in GP.

\subsection{Additional data combinations}
\label{sec:supp_more_data}

Throughout this Supplemental Material, PP denotes Pantheon+, CC denotes cosmic chronometers, and BAO denotes the BAO compilation described in the main text, including DESI DR2. Unless otherwise stated, each expansion-data combination is combined with the RSD growth dataset used in the main analysis.

\begin{figure*}
    \centering
    \includegraphics[width=1\textwidth]{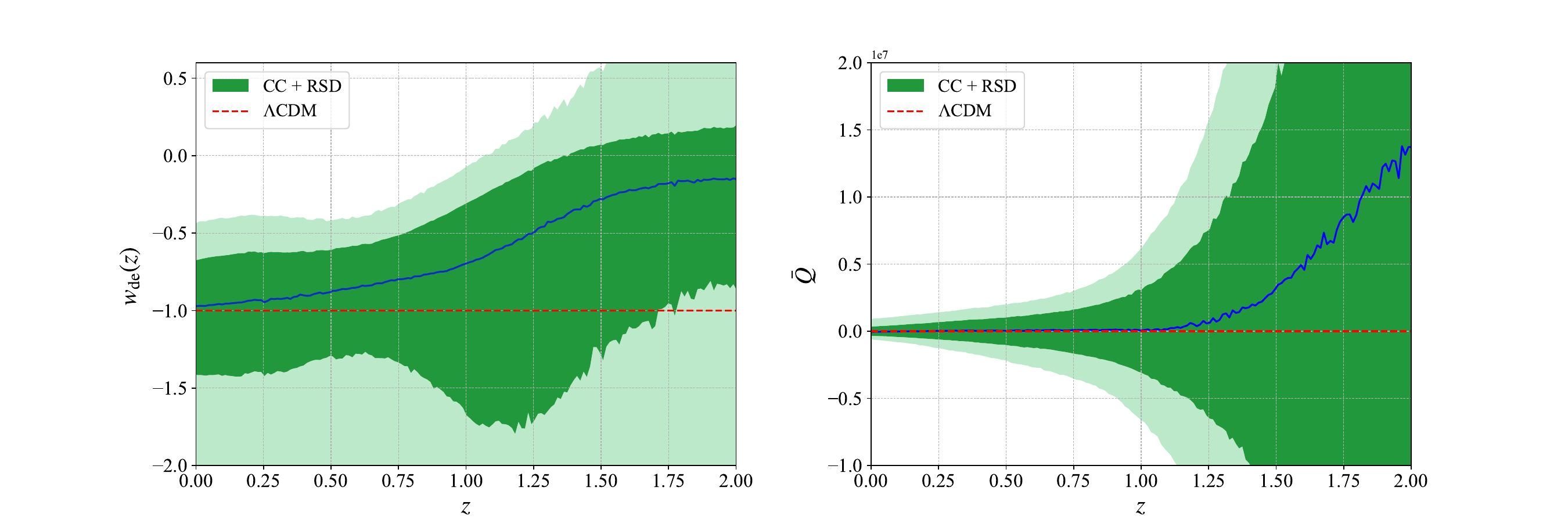}\par
    \includegraphics[width=1\textwidth]{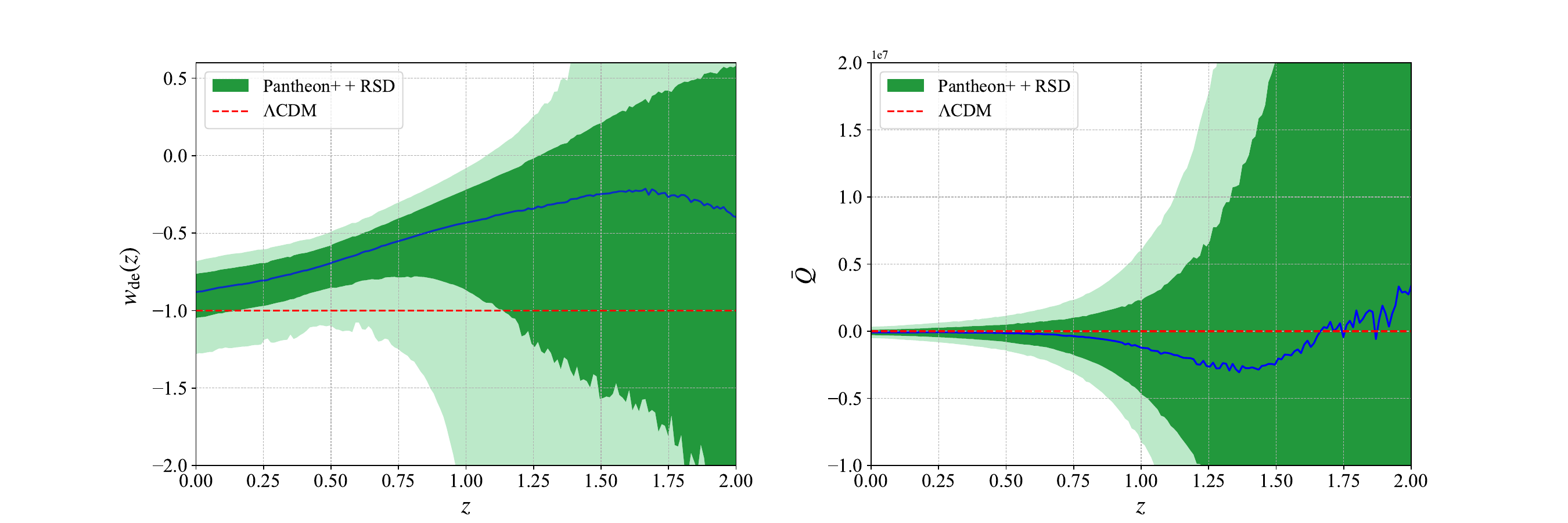}\par
    \includegraphics[width=1\textwidth]{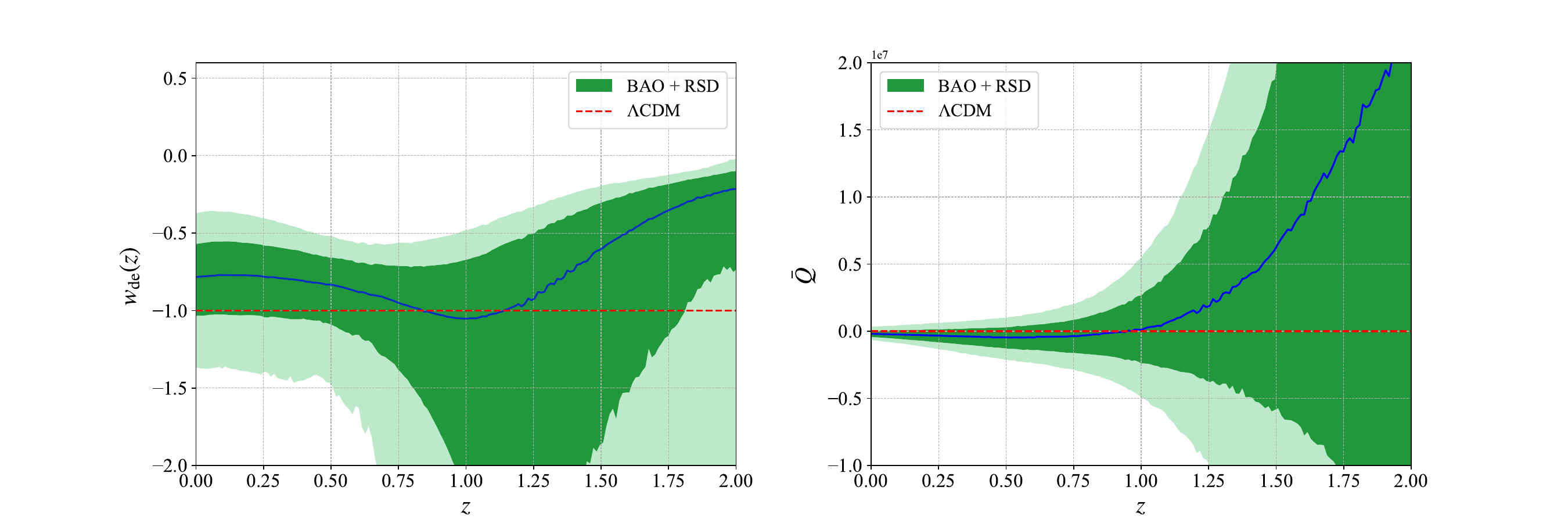}
    
    \caption{Same as figure~\ref{fig:single}, but for single expansion datasets.}
    \label{fig:single_data}
\end{figure*}

\begin{figure*}
    \centering
    \includegraphics[width=1\textwidth]{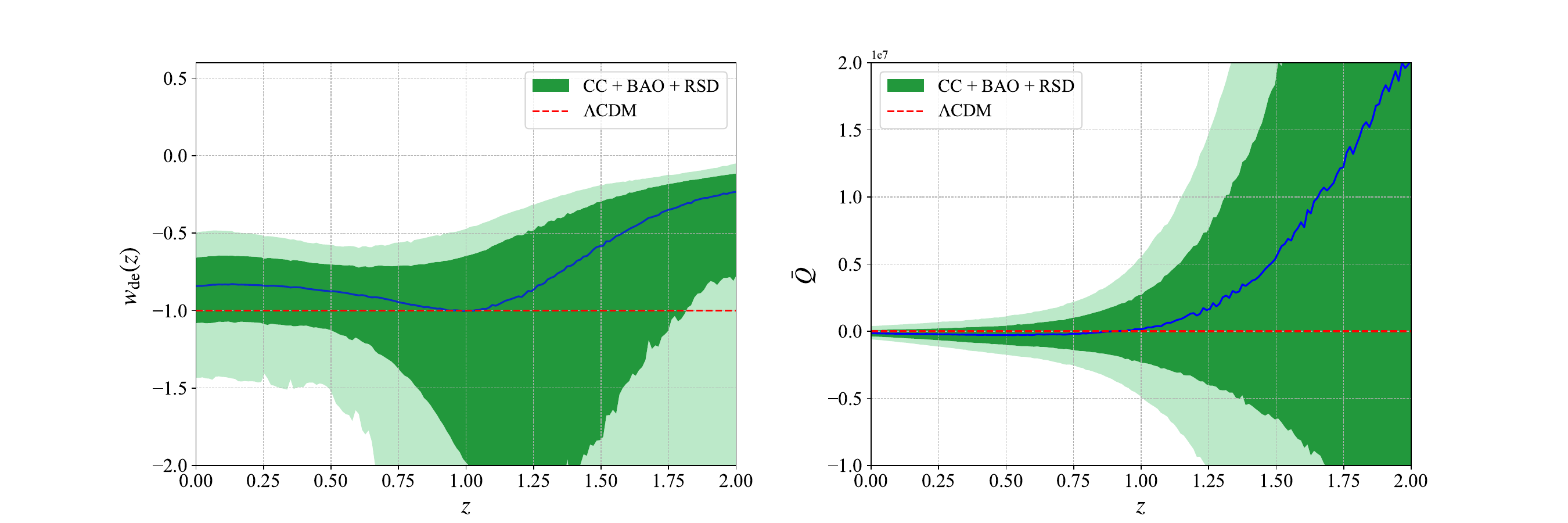}\par
    \includegraphics[width=1\textwidth]{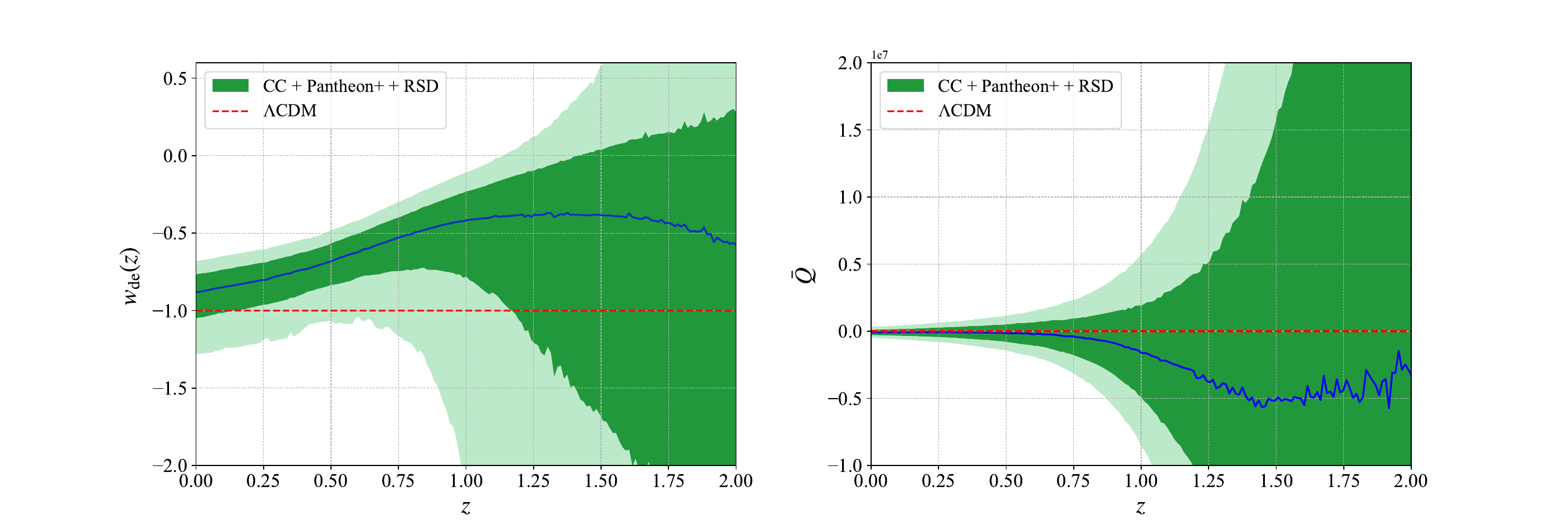}\par
    \includegraphics[width=1\textwidth]{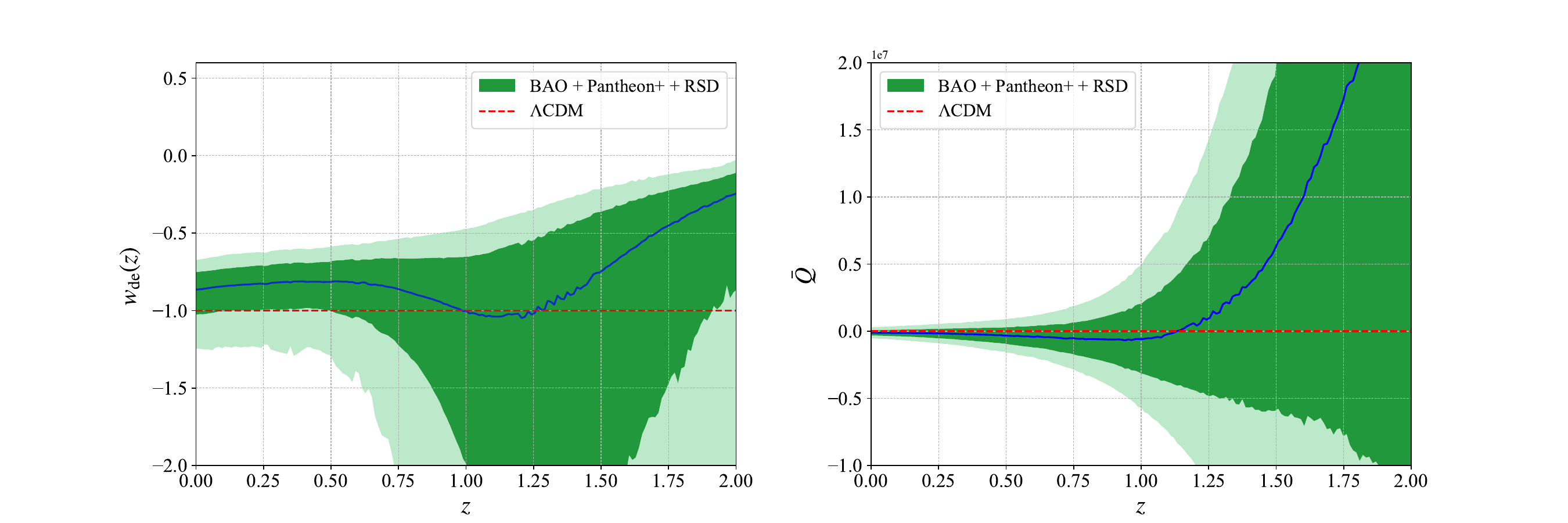}
    
    \caption{Same as figure~\ref{fig:single_data}, but for multiple expansion-data combinations.}
    \label{fig:combined_data}
\end{figure*}

In this section, we present additional reconstructions of the dark energy equation of state and the dark-sector coupling for several expansion-data combinations under \textbf{Closure I}. The results are shown in Figs.~\ref{fig:single_data} and \ref{fig:combined_data} .

For all additional combinations, the reconstructed interaction remains statistically consistent with the no-coupling limit, within the quoted credible intervals. This confirms that the null detection of a dark-sector interaction reported in the main text is not driven by a particular choice of expansion dataset. By contrast, the reconstructed equation of state exhibits visibly different redshift evolution for different background probes. For instance, when constrained by the PP sample, the EoS exhibits quintessence-like dynamical behavior. Such dataset-dependent evolution of the dark energy EoS is expected from Eq.~(\ref{eq:w_de}), because $w_{\rm de}(z)$ depends directly on the reconstructed expansion history once $\rho_f(z)$ is inferred from the growth sector.

The PP sample provides strong leverage at low redshift because of its large number of supernovae, while CC and BAO measurements provide complementary information over a wider redshift range. In particular, the inclusion of BAO data tends to enhance the possibility of a phantom-crossing behavior in the reconstructed $w_{\rm de}(z)$ when $z\sim 0.9$. This feature motivates continued scrutiny with forthcoming higher-precision BAO measurements, especially future DESI releases.

\subsection{$M_B$--$r_d$ calibration degeneracy and the Hubble tension}
\label{sec:supp_MBrd}

The absolute magnitude $M_B$ of Type Ia supernovae fixes the absolute calibration of luminosity distances and is therefore degenerate with the Hubble constant $H_0$. An analogous calibration degeneracy appears in BAO analyses, where the sound horizon at the drag epoch, $r_d$, sets the absolute scale of the standard ruler and is likewise degenerate with $H_0$. In our nonparametric reconstruction, $H_0$ only plays an implicit role; nevertheless, when SNe Ia and BAO are combined, the underlying $H_0$ tension is inherited as a calibration consistency condition in the $M_B$--$r_d$ plane~\citep{Barua:2024gei,Benisty:2022psx}.

This can be seen from the zero-redshift limits
\begin{align}
    H_{0}^{\mathrm{SN}} &= \frac{c}{D'_M(0)}, \label{eq:supp_H0_SN}\\
    H_{0}^{\mathrm{BAO}} &= \frac{c}{r_d \widetilde{D}_H(0)}. \label{eq:supp_H0_BAO}
\end{align}
Here, $\widetilde{D}_H(z)\equiv D_H(z)/r_d$ denotes the line-of-sight BAO observable. Therefore, combining SNe Ia and BAO in a model-independent reconstruction requires a calibration-consistent pair of priors for $M_B$ and $r_d$. Otherwise, one effectively imposes two incompatible absolute distance scales on the same reconstructed expansion history, which can artificially affect the inferred evolution of $w_{\rm de}(z)$ and the interaction history.

\begin{figure*}
\includegraphics[width=1\textwidth]{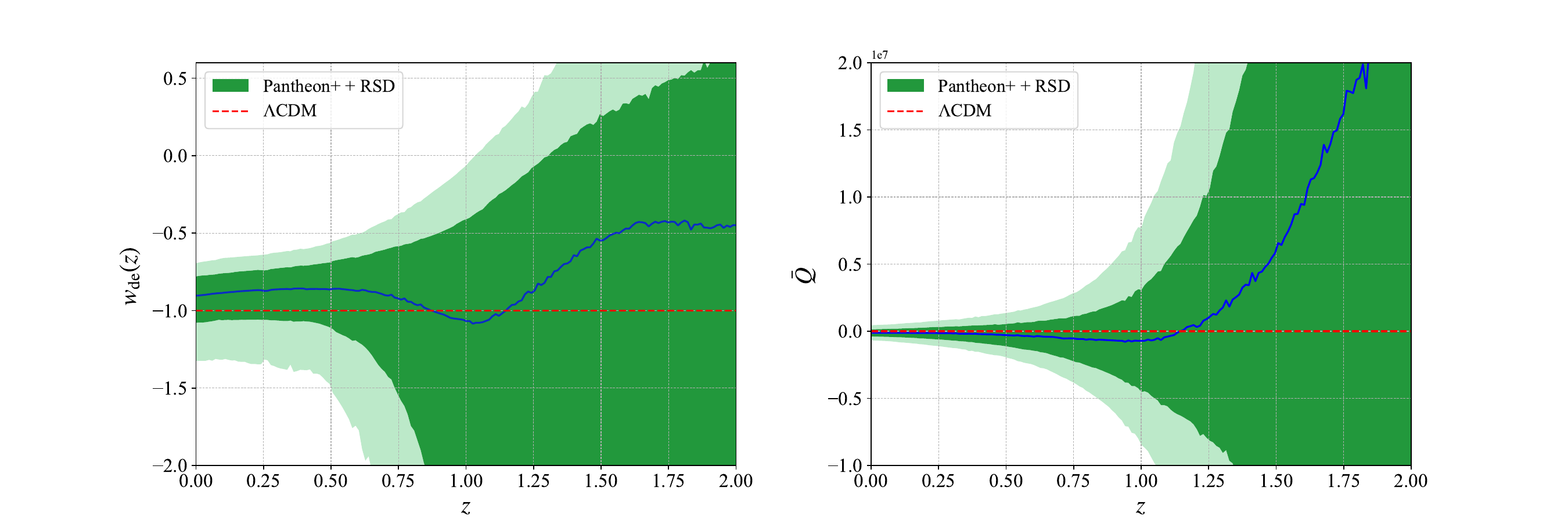}\par
\includegraphics[width=1\textwidth]{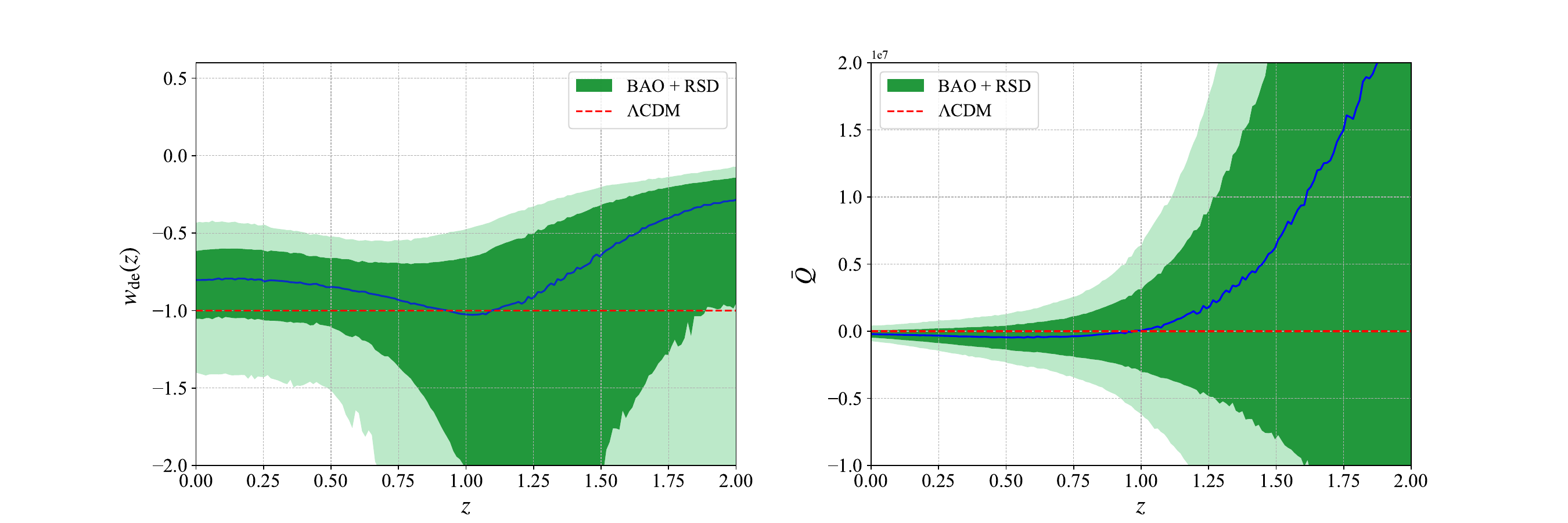}

\caption{Reconstructions of the dark energy equation of state and the dark-sector interaction for single expansion datasets under the alternative calibration $M_B=-19.253$ and $r_d=138.3\pm1.9~{\rm Mpc}$. Dark and light shaded bands denote the $68\%$ and $95\%$ credible intervals, respectively. Dashed lines represent the $\Lambda$CDM limit.}
\label{fig:combine_19.253_single}
\end{figure*}

\begin{figure*}
\includegraphics[width=1\textwidth]{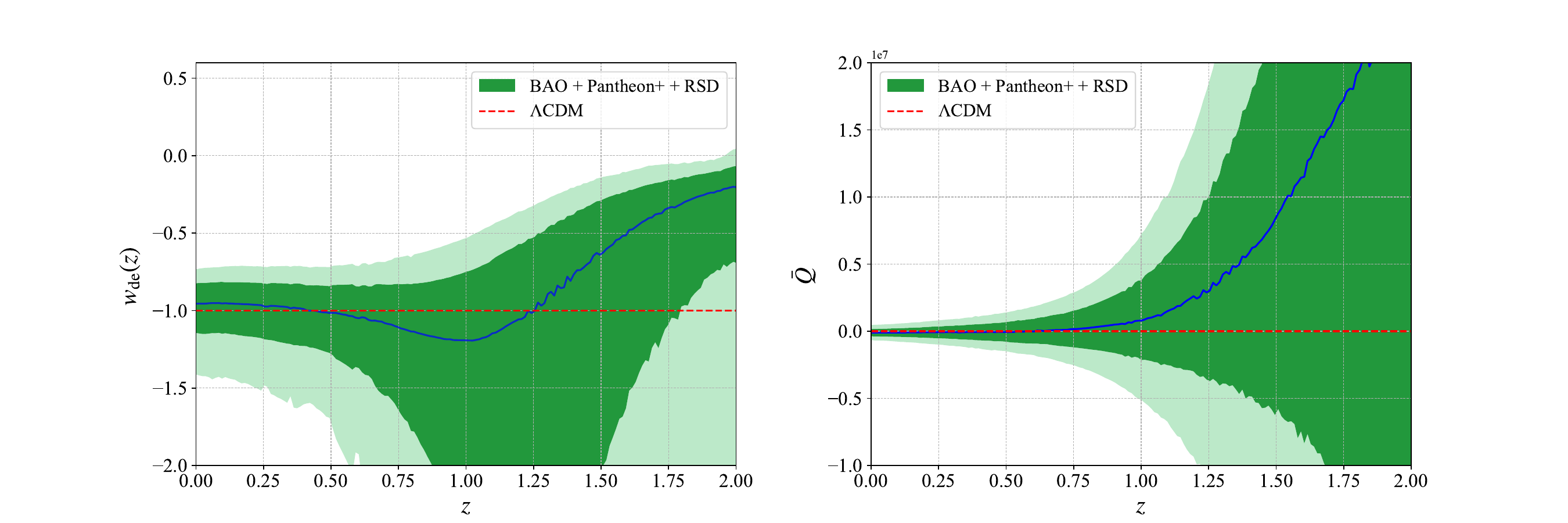}\par
\includegraphics[width=1\textwidth]{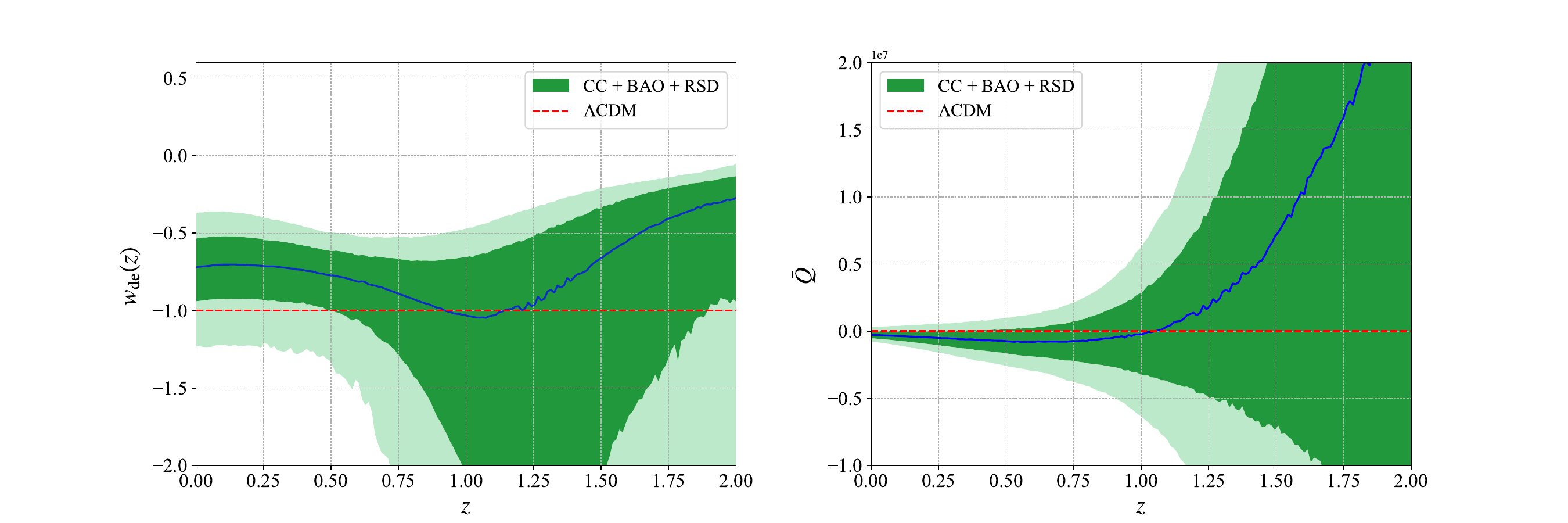}\par
\includegraphics[width=1\textwidth]{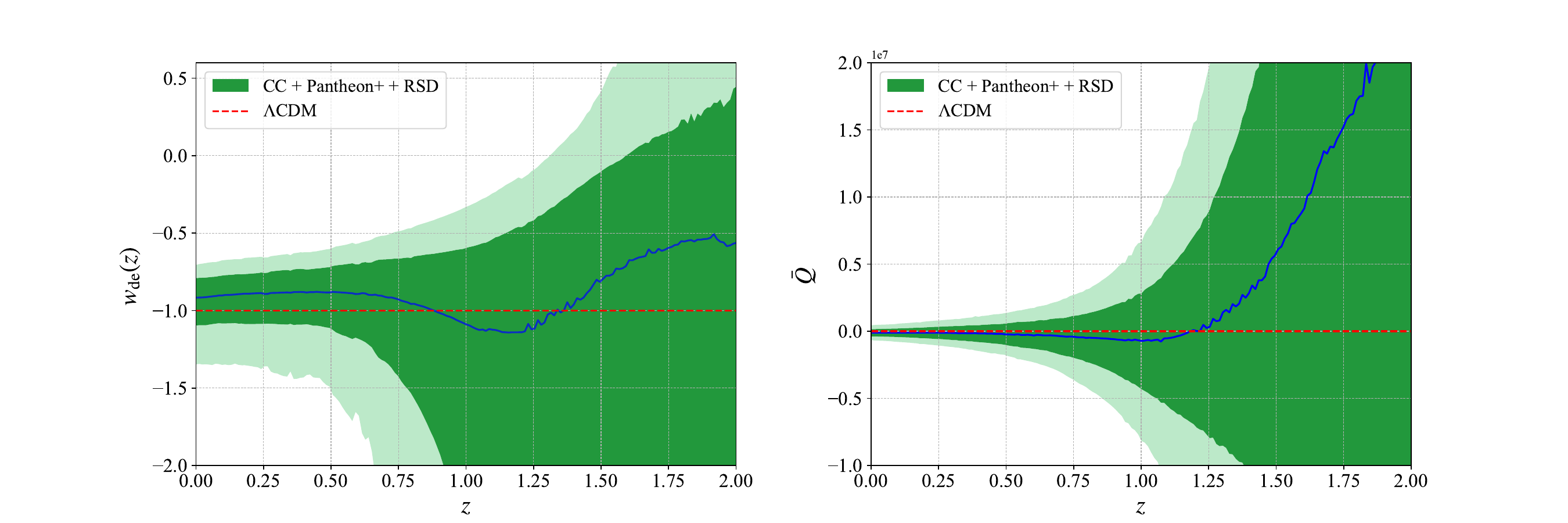}\par
\includegraphics[width=1\textwidth]{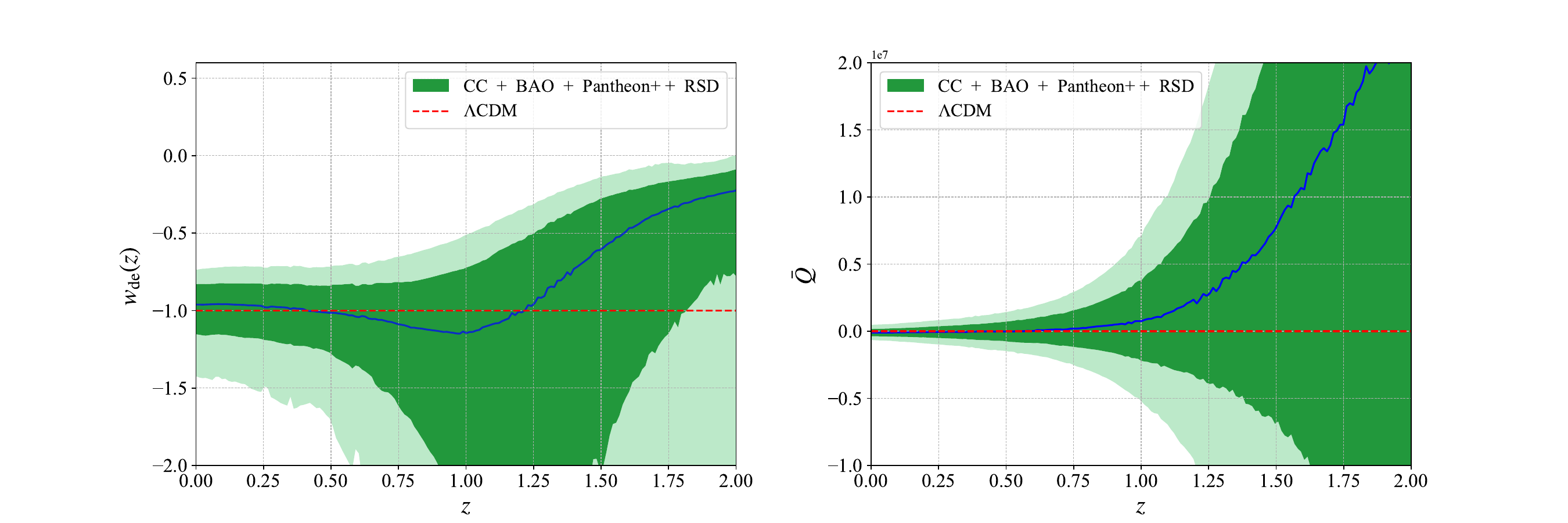}

\caption{Same as figure~\ref{fig:combine_19.253_single}, but for multiple expansion-data combinations.}
\label{fig:combine_19.253_mutiple}
\end{figure*}

In the main text, we use the Planck-calibrated sound horizon and the corresponding supernova calibration as a $\Lambda$CDM-based null-test baseline. As a robustness check, we repeat the analysis using an alternative calibration motivated by the local distance ladder: $M_B=-19.253$ from the Cepheid calibration of Type Ia SNe based on SH0ES observations~\citep{Riess:2021jrx}, together with the corresponding sound horizon $r_d=138.3\pm1.9~{\rm Mpc}$ from Table IV of Ref.~\citep{Barua:2024gei}. The resulting reconstructions are shown in Figs.~\ref{fig:combine_19.253_single} and \ref{fig:combine_19.253_mutiple}.

Changing the calibration priors modifies the detailed reconstructed expansion history and therefore affects the detailed shape of $w_{\rm de}(z)$. However, the qualitative conclusion for the coupling is unchanged: the reconstructed interaction remains consistent with the no-coupling limit within the uncertainties.

The reconstruction of the dark energy EoS exhibits significant sensitivity to the choice of calibration priors. Specifically, adopting a late-universe prior for the absolute magnitude $M_B$ broadens the reconstructed EoS into the Quintom regime, in contrast to the strictly quintessence-like behavior favored by early-universe calibrations, when including \textbf{PP} sample. Similarly, applying an alternative, smaller $r_d$ prior reduces the statistical preference for the phantom-crossing behavior typically induced by the \textbf{BAO} data when calibrated with the standard CMB sound horizon. These sensitivities follow directly from Eq.~(\ref{eq:w_de}): once the growth sector fixes $\rho_f(z)$ within its uncertainty, changes in the reconstructed $H(z)$ propagate directly into the inferred EoS.

These tests show that the main coupling result is robust against reasonable changes in the background calibration priors: current data do not provide statistically significant evidence for nonzero dark-sector energy transfer. The EoS reconstruction, in contrast, is calibration-sensitive because it is directly tied to the absolute expansion history. For this reason, the Planck-calibrated priors used in the main text should be understood as a well-defined $\Lambda$CDM null-test baseline. Alternative calibrations can be consistently incorporated within the same framework and provide a useful way to diagnose how background-scale choices propagate into the reconstructed dark energy dynamics.

\subsection{Redshift-space distortions and $\sigma_{8,0}$}\label{sec:RSD}
Our method uses redshift-space-distortion (RSD) measurements to reconstruct the growth history independently of the background expansion. RSD observations constrain either the growth rate $f(z)$ or, more robustly, the combination $f\sigma_8(z)$, which is less sensitive to the galaxy-bias degeneracy~\citep{Nesseris:2017vor}. These observables are generated by galaxy peculiar velocities sourced by the growth of large-scale-structure perturbations.

Under \textbf{Closure I} (see Section~\ref{sec:closures}), the matter density perturbation and the observable growth rate $F(z) \equiv f\sigma_8(z)$ retain their standard definitions:
\begin{align}
    \ddot{\delta}_m &+ 2H\dot{\delta}_m = \frac{1}{2}\rho_m\delta_m \, , \label{eq:supp_perturbation} \\
    F(z) &= -(1+z)\frac{\sigma_{8,0}}{\delta_{m,0}}\delta_m' \, . \label{eq:supp_fsigma8}
\end{align}

Since Eq.~\eqref{eq:supp_perturbation} is linear in $\delta_m$, its absolute normalization is arbitrary. Introducing the normalized growth function $\mathcal{D}(z) \equiv \delta_m(z)/\delta_{m,0}$ (with $\mathcal{D}(0)=1$), we can relate it directly to the reconstructed observable $F(z)$:
\begin{align}
    \mathcal{D}'(z) &= -\frac{1}{\sigma_{8,0}}\frac{F(z)}{(1+z)} \, ,\label{eq:supp_Dprime} \\
    \mathcal{D}(z) &= 1 - \frac{1}{\sigma_{8,0}} \int^z_0 \frac{F(\tilde{z})}{1+\tilde{z}} \mathrm{d}\tilde{z} \, . \label{eq:supp_D}
\end{align}

By applying Gaussian process (GP) regression directly to the RSD $f\sigma_8(z)$ dataset, we obtain $\mathcal{D}(z)$ and its derivatives via Eqs.~\eqref{eq:supp_Dprime} and \eqref{eq:supp_D}. Because the effective matter density $\rho_f(z)$ is analytically inferred from these growth quantities, the resulting coupling is highly sensitive to both the chosen RSD compilation and the $\sigma_{8,0}$ prior.  This sensitivity is useful diagnostically, but it also means that correlated or internally inconsistent growth data can mimic an apparent interaction signal.

\begin{figure*}
\includegraphics[width=0.8\textwidth]{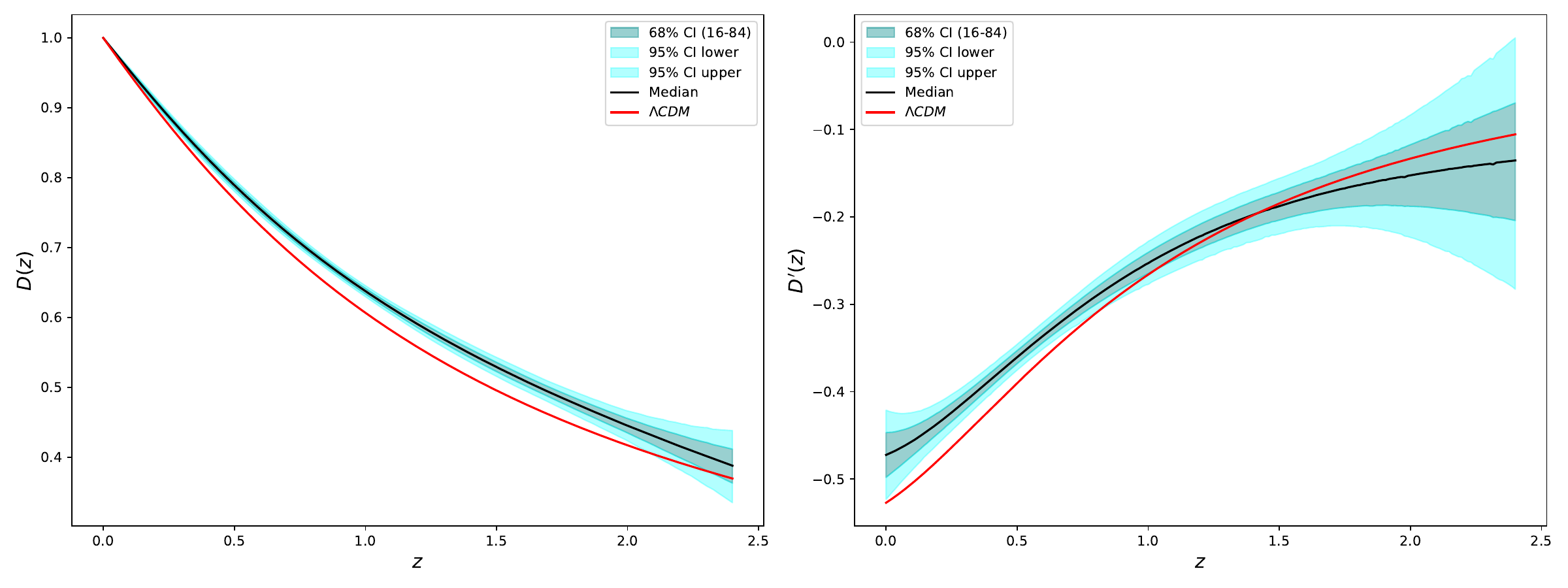}

\caption{Reconstructions of $D(z)$ and $D'(z)$ using the 66-point RSD compilation with $\sigma_{8,0}=0.8111$. The shaded bands denote the $68\%$ credible intervals. Red curves represent the Planck $\Lambda$CDM prediction.}
\label{fig:D_66}
\end{figure*}

As a stress test, we consider the 66-point $f\sigma_8$ compilation of Ref.~\citep{Skara:2019usd}, denoted as \textbf{RSD66}. Many entries in this compilation are correlated or share overlapping survey information. If they are treated as independent inputs in a GP reconstruction, they can overweight particular redshift ranges and distort the inferred growth history. Figure~\ref{fig:D_66} shows that the resulting reconstruction of $\mathcal{D}'(z)$ deviates noticeably from the Planck $\Lambda$CDM baseline at low redshift.

\begin{figure*}
\includegraphics[width=1\textwidth]{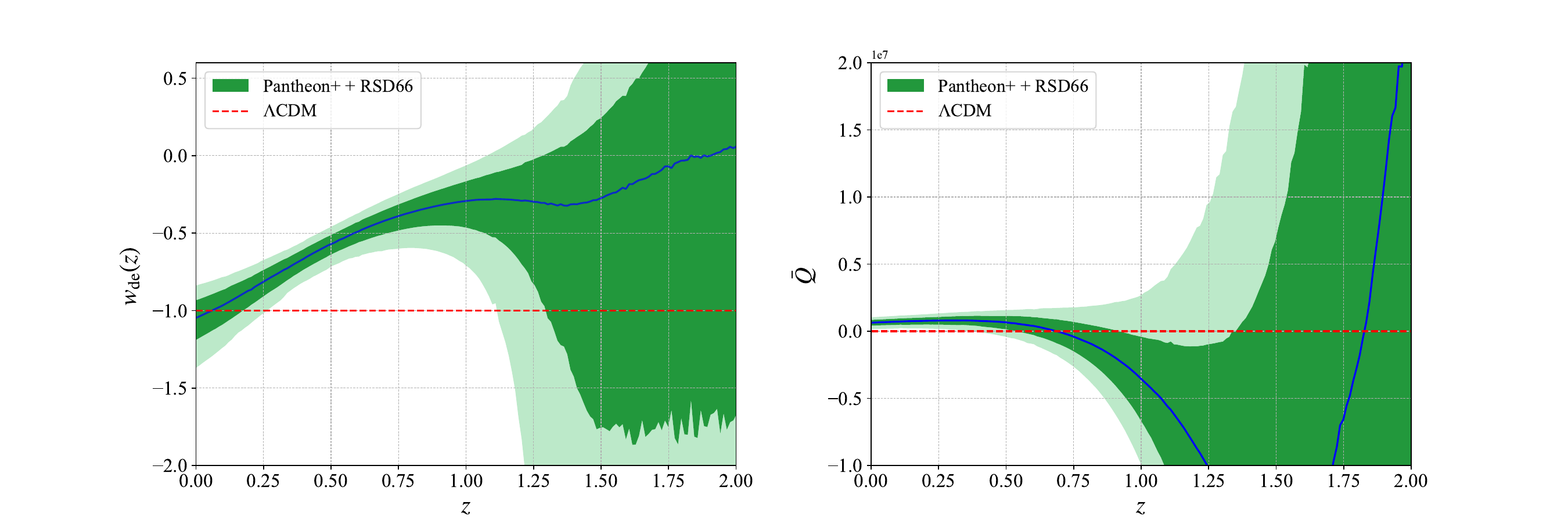}\par
\includegraphics[width=1\textwidth]{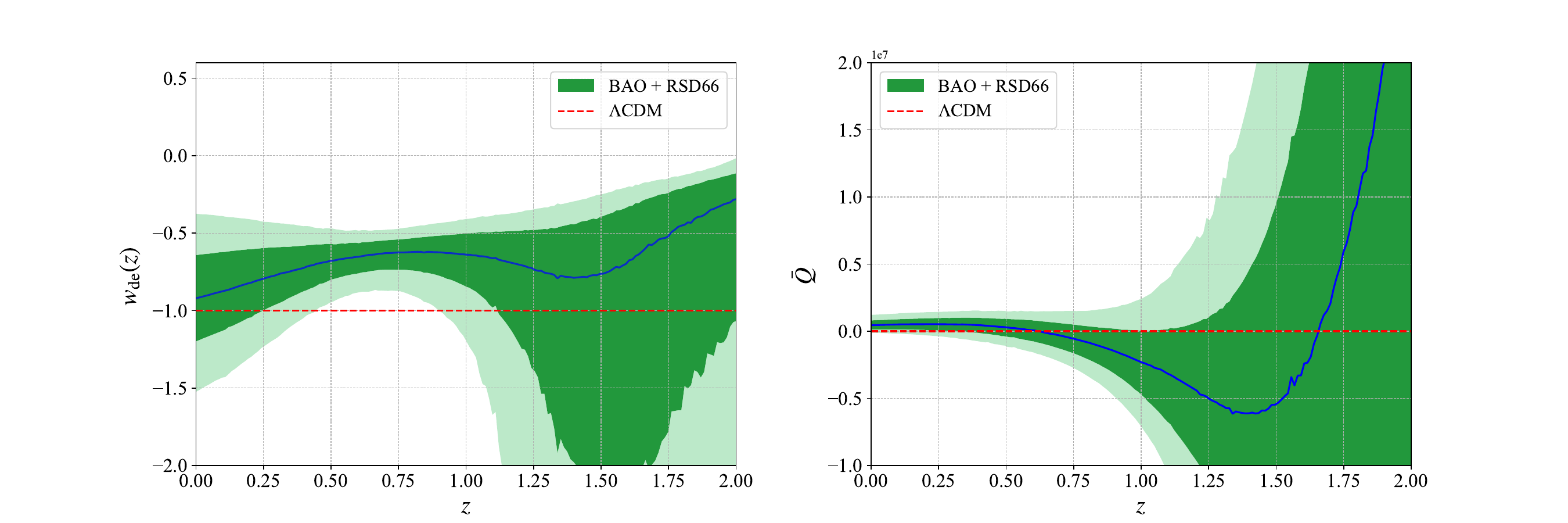}

\caption{Reconstructions of the dark energy equation of state and the dark-sector interaction for single expansion datasets using the \textbf{RSD66} compilation as a growth-sector stress test. Dark and light shaded bands denote the $68\%$ and $95\%$ credible intervals, respectively. Dashed lines represent the $\Lambda$CDM limit.}
\label{fig:combine_66_single}
\end{figure*}

\begin{figure*}
\includegraphics[width=1\textwidth]{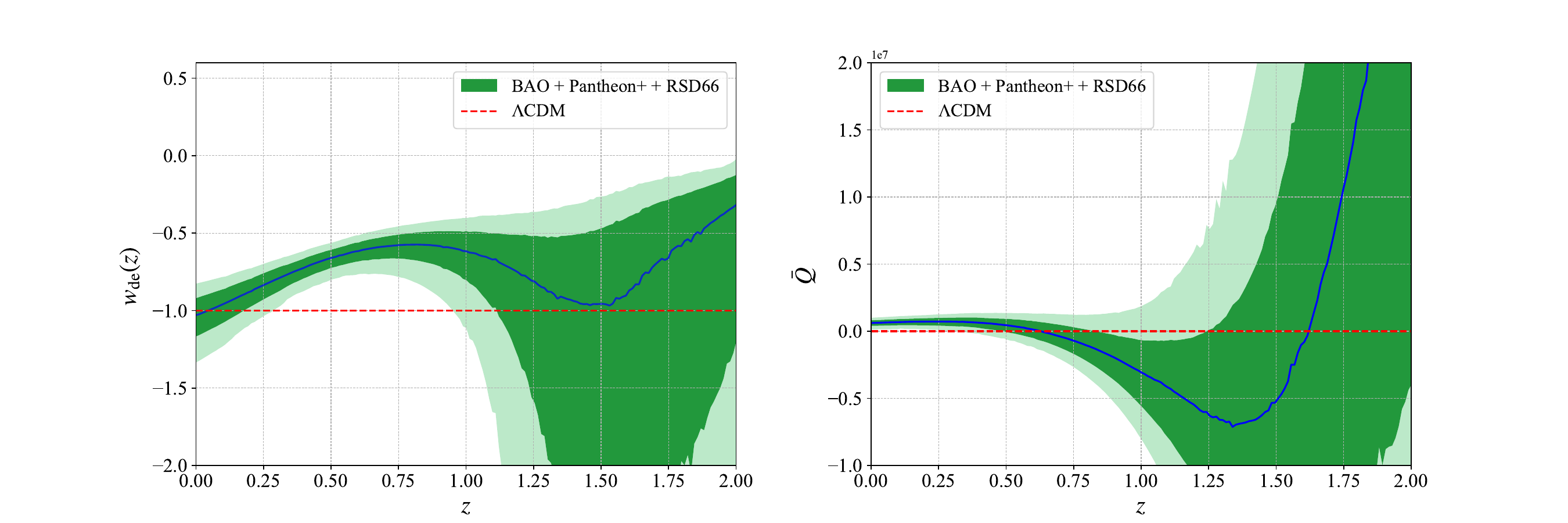}\par
\includegraphics[width=1\textwidth]{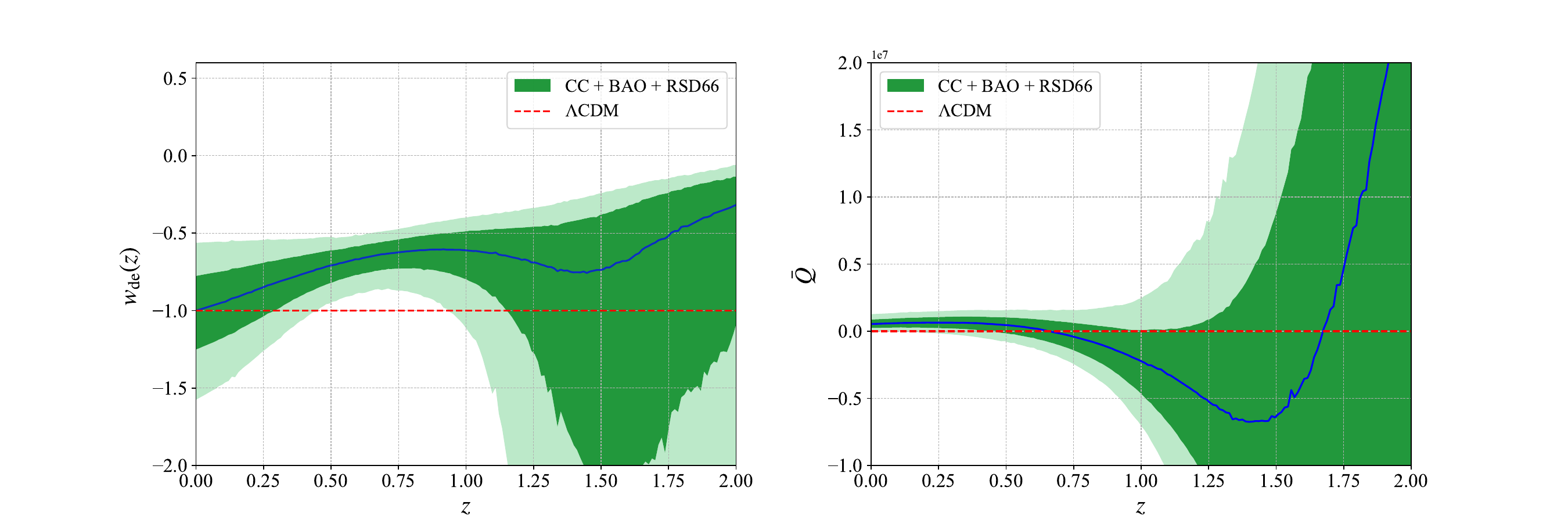}\par
\includegraphics[width=1\textwidth]{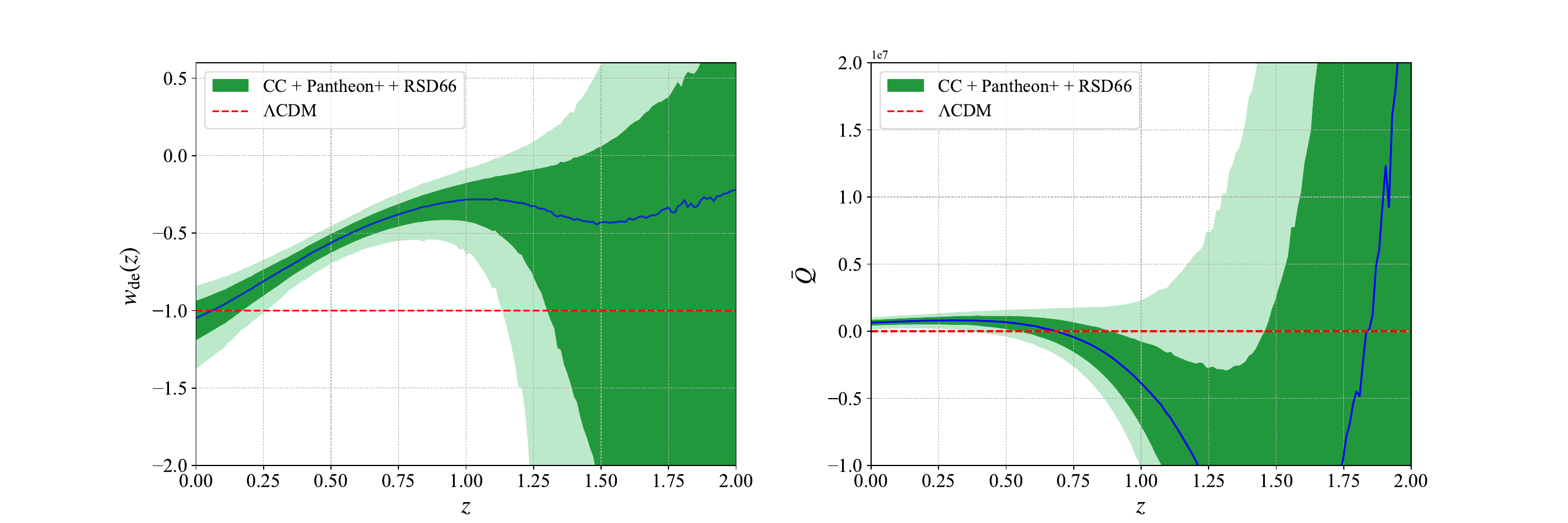}\par
\includegraphics[width=1\textwidth]{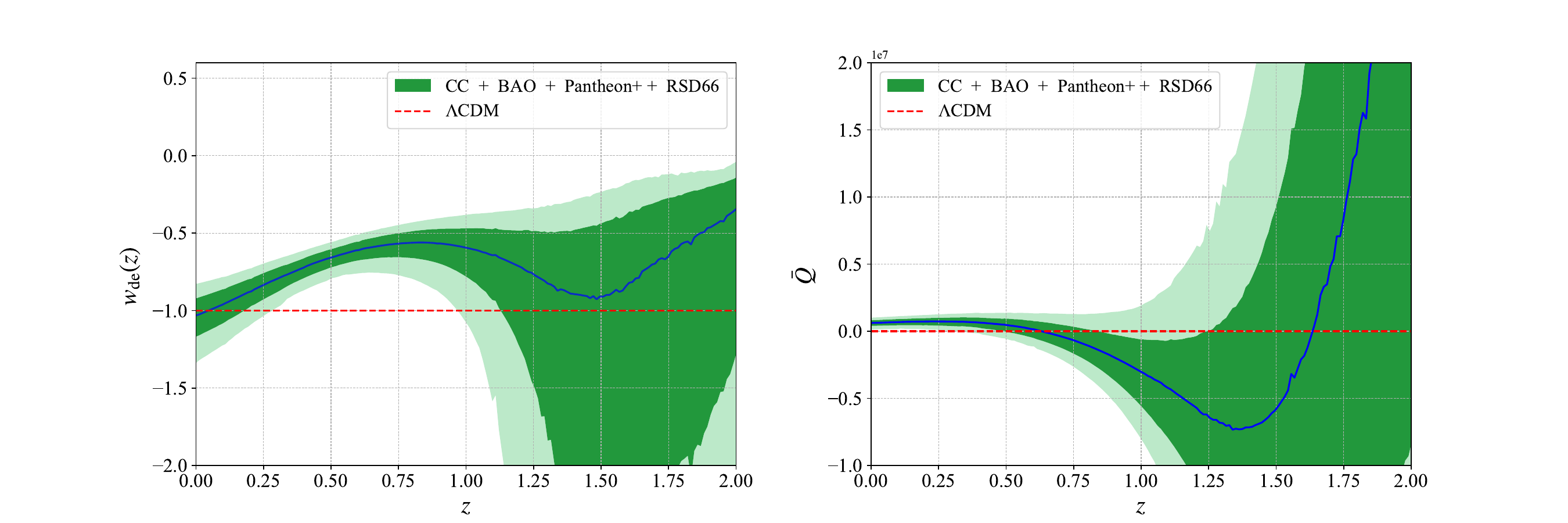}

\caption{Same as figure~\ref{fig:combine_66_single}, but for multiple expansion-data combinations.}
\label{fig:combine_66_mutiple}
\end{figure*}

We intentionally apply our full framework to this \textbf{RSD66} compilation to test whether a growth-sector anomaly propagates into the inferred interaction. As shown in Figs.~\ref{fig:combine_66_single} and \ref{fig:combine_66_mutiple}, a stronger coupling signal at the $1$--$2\sigma$ level emerges in several combinations. This should not be interpreted as evidence for a physical DE--DM interaction, because the correlations in the input growth compilation are not fully controlled. Instead, it demonstrates that the proposed expansion--growth reconstruction is sensitive to anomalies in the perturbation history, as expected from the structure of the method. For the main analysis, we therefore use the 20-point RSD compilation of Ref.~\citep{Avila:2022xad}, which removes strongly correlated redshift bins.

The prior on $\sigma_{8,0}$ also affects the reconstruction through Eqs.~(\ref{eq:supp_Dprime}) and (\ref{eq:supp_D}). In the main text, we adopt the precise Planck 2018 prior $\sigma_{8,0}=0.8111\pm0.0060$~\citep{Planck:2018vyg}, which defines a sharp $\Lambda$CDM null-test baseline. A more late-Universe-oriented prior, such as $\sigma_{8,0}=0.766\pm0.116$ from Ref.~\citep{Avila:2022xad}, is attractive from a model-independent perspective, but its current uncertainty is too large to provide strong diagnostic power. The broad prior substantially enlarges the posterior uncertainty in $\mathcal{D}(z)$ and $\mathcal{D}'(z)$ and therefore washes out potential interaction features.

In summary, the limiting factor for detecting dark-sector interactions with the present framework is the precision and internal consistency of the growth data. Current RSD measurements are not yet sufficient to establish a statistically significant nonzero interaction. Nevertheless, the sensitivity tests above show that the method responds correctly to changes in the perturbation history, making it well suited for future high-precision growth measurements.

\subsection{Null test of Closure II}\label{sec:null_test_closure_two}

In this section, we perform a null test for the \textbf{Closure II} framework, where the dark-sector interaction is assumed to be proportional to the smooth dark energy density. 

As indicated by Eqs.~\eqref{eq:closure_two_y} and \eqref{eq:closure_two_w}, the analytical decoupling in this scenario relies on high-order derivatives of the observables. The propagation of errors through these derivatives severely amplifies the statistical noise, rendering physically meaningful constraints unfeasible with current observational data, particularly the RSD measurements. Consequently, we adopt the Planck 2018 $\Lambda$CDM best-fit cosmology~\citep{Planck:2018vyg} to generate mock datasets for both the Hubble parameter $H(z)$ and the growth observable $f\sigma_8(z)$, utilizing the following priors:
\begin{equation}
    H_0 = 67.4 \pm 0.5~\mathrm{km\,s^{-1}\,Mpc^{-1}} \, , \quad \Omega_m = 0.315 \pm 0.007 \, , \quad \sigma_{8,0} = 0.8111 \pm 0.006 \, .
\end{equation}

To mimic realistic observational distributions, the redshifts of the mock $H(z)$ data are sampled from a Gamma distribution peaking at $z \sim 0.4$. We assign a relative uncertainty of $\approx 2\%$ to these measurements, matching the precision of recent DESI observations. The corresponding $f\sigma_{8}(z)$ mock values are self-consistently derived from the background $H(z)$ under the standard $\Lambda$CDM assumption.

To rigorously validate our entire analytical and numerical pipeline, we first apply Gaussian process regression to the mock datasets to extract the smoothed posteriors of the observables and their derivatives. Subsequently, we substitute these into Eqs.~\eqref{eq:de_Q_closure_two} and \eqref{eq:closure_two_w} to reconstruct the dark-sector interaction and the dark energy equation of state. The results of this null test are presented in Fig.~\ref{fig:closure_two_mock}.

As anticipated, the reconstructions are in excellent agreement with the $\Lambda$CDM predictions at low redshifts, remaining well within the Gaussian uncertainty bands. 
The enlarged uncertainty near $z \sim 0$ emerges fundamentally from the absence of $z<0$ data - an inevitable boundary effect that naturally degrades the constraints on the first and second derivatives in this regime. 
Furthermore, we observe that at higher redshifts, minor deviations from the $\Lambda$CDM baseline begin to emerge. 
This behavior occurs because both $w_{\rm de}$ and $\bar{Q}$ are highly non-linear functions of the reconstructed observables. These numerical artifacts are further amplified at high redshifts as a direct consequence of the Gamma distribution adopted for our mock data sampling and the base Gaussian Process algorithm.

In summary, the analytical framework of \textbf{Closure II} is strictly self-consistent, faithfully recovering the $\Lambda$CDM baseline when supplied with standard-model observables $H(z)$ and $f\sigma_{8}(z)$. Nevertheless, the extreme sensitivity of the decoupled equations to high-order derivatives dictates that this framework can only yield physically meaningful constraints when applied to observational data with high accuracy. Looking ahead, unlocking the full potential of this approach will require not only the unprecedented precision of next-generation cosmological surveys but also the deployment of advanced, noise-resilient reconstruction algorithms to overcome the current limitations of standard Gaussian processes.

\end{document}